\documentclass[iop]{emulateapj}

\newcommand{\kms}{\,\hbox{\hbox{km}\,\hbox{s}$^{-1}$}}

\newcommand{\spi}{{\it Spitzer}}

\newcommand{\mum}   {$\mu$m}

\shorttitle{Far-infrared line spectra of Seyfert galaxies}
\shortauthors{Spinoglio et al.}

\usepackage{graphicx}
\usepackage{lscape}

\usepackage{amsmath}

\begin{document}

\DeclareGraphicsExtensions{.pdf,.gif,.jpg}

\title{Far-infrared line spectra of Seyfert galaxies from the \textit{Herschel}-PACS Spectrometer
}

\author{Luigi Spinoglio\altaffilmark{1}, Miguel Pereira-Santaella\altaffilmark{1},  Kalliopi M. Dasyra\altaffilmark{2}, Luca Calzoletti\altaffilmark{3}, Matthew A. Malkan\altaffilmark{4}, Silvia Tommasin\altaffilmark {5}, Gemma Busquet\altaffilmark{1,6}  
}
\altaffiltext{1}{Istituto di Astrofisica e Planetologia Spaziali, INAF, Via Fosso del Cavaliere 100, I-00133 Roma, Italy}
\email{luigi.spinoglio@iaps.inaf.it}
\altaffiltext{2}{Observatoire de Paris, LERMA (CNRS:UMR8112), 61 Av. de l\'\ Observatoire, F-75014, Paris, France }
\altaffiltext{3}{Agenzia Spaziale Italiana (ASI) Science Data Center, I-00044 Frascati (Roma), Italy}
\altaffiltext{4}{Astronomy Division, University of California, Los Angeles, CA 90095-1547, USA}
\altaffiltext{5}{Weizmann Institute of Science, Department of Neurobiology, Rehovot 76100, Israel}
\altaffiltext{6}{Instituto de Astrof\'isica de Andalucia, CSIC, Glorieta de la Astronom\'ia, s/n, E-18008 Granada, Spain}
\clearpage

\begin{abstract}
We present spectroscopic observations of far-IR fine-structure lines of 26 Seyfert galaxies obtained with the PACS spectrometer onboard  {\it Herschel}. 
These observations are complemented by spectroscopy with both the \spi~ IRS and 
the {\it Herschel} SPIRE spectrometers. The ratios of the [OIII], [NII], [SIII] and [NeV] lines have been used to determine 
electron densities in the ionised gas regions. The [CI] lines, observed with SPIRE,  have been used to measure 
the densities in the neutral gas, while the [OI] lines provide a measure of the gas temperature, 
at densities below $\sim$ 10$^4$ cm$^{-3}$. Using the  [OI]145$\mu$m/63$\mu$m and [SIII]33/18.7$\mu$m line ratios 
we find an anti-correlation of the temperature with the gas density.
 
Using various fine-structure line ratios, we find that density stratiÞcation is common in these active galaxies.
On average, the electron densities increase with the ionisation potential of the ions producing the [NII], [SIII] and [NeV] emission.
The infrared emission lines arise partly in the Narrow Line Region (NLR) photoionised by the AGN central engine,
partly in HII regions photoionised by hot stars, and partly in neutral gas in photo-dissociated regions (PDRs).

We attempt to separate the contributions to the line emission produced in these different regions by comparing our observed emission line ratios
to empirical and theoretical values.
 
In particular, we tried to separate the contribution of AGN and star formation by using a combination of \spi~  and {\it Herschel} lines, 
and we found that, besides the well known mid-IR line ratios, the mixed mid-IR/far-IR line ratio of [OIII]88$\mu$m/[OIV]26$\mu$m can reliably
discriminate the two emission regions, while the total far-IR line ratio of [CII]157$\mu$m/[OI]63$\mu$m is only able to mildly separate the two regimes.

By comparing the observed [CII]157µm/[NII]205µm ratio with photoionisation models, we also found that most
of the [CII] emission in the galaxies we examined is due to PDRs.

\end{abstract}

\keywords{- Galaxies: ISM, nuclei, active, starburst, Seyfert - Techniques: infrared spectroscopy}

\section{Introduction}
\label{intro}

The luminosity of a galaxy may ultimately be produced by young or old stars, or accretion onto a central massive black hole.
The latter two sources emit large portions of ionising photons, which are usually absorbed by gas, producing
emission lines.  
These are particularly strong in the
far-infrared because of their ease of excitation. The strongest lines can exceed a percent of the total far-IR luminosity. 
Through the observation of various line ratios, it is possible to infer what has ionised the gas, and 
to characterise its physical properties, such as the nature and strength of the
interstellar radiation fields, chemical abundances, local temperatures and gas densities. 
The observed mid- and far-IR fine structure lines from different species, with a wide range
different ionisation potentials and excitation conditions \citep[see, e.g.][]{sm92} 
trace different physical conditions and phases of the ISM, all the way from the neutral atomic
gas in interstellar PDRs,
to the extremely ionised  ÒcoronalÓ gas powered by X-rays from the AGN.

Hot (young) stars and black hole accretion discs have strongly different
ionising continuum spectra. 
However this continuum, 
which produces a considerable fraction of the bolometric luminosity in both processes,  
is not observable directly, due to absorption by HI and, at longer wavelengths,
by dust.  Emission lines ratios from the photoionised gas are the best tracers and
discriminators of accretion and star formation processes \citep[see, e.g.,][]{sm92,of06}. 
In order to overcome heavy extinction, observations in the mid- to far-IR are needed to probe obscured regions.
The rest-frame mid- to far-IR spectral region contains several extinction-free emission lines  
which measure the contributions from AGN and star formation to the overall energy budget. 
Since the spectroscopic observations of the
Infrared Space Observatory (ISO) \citep{kes96} Short Wavelength Spectrometer (SWS) \citep{deg96} 
a useful family of spectroscopic diagnostics has been found, such as the
ratios of emission lines tracing the hard UV field found in the narrow line region of AGN (e.g., [Ne V], [OIV]) 
to those tracing stellar HII regions (e.g., [S III], [Ne II]), as well as the strength of the PAH emission features
-- indicators of star formation.
These spectroscopic tools are able to separate the star-forming from AGN-dominated galaxies
because of the large differences in ionisation potential of the observed mid-IR emission lines \citep[e.g.,][]{sm92, gen98}. 
The {\it Spitzer} Space Telescope \citep{wer04} IR Spectrograph  (IRS) \citep{hou04} observations have further extended these 
results for ultra luminous IR galaxies, Seyfert galaxies and starburst galaxies
\citep{arm07, smi07, tom08, vei09, tom10, bra06, bs09}. 

At longer wavelengths, 
spectroscopy of some of the brighter nearby star-forming galaxies and AGN with 
the ISO Long Wavelength Spectrometer (LWS) \citep{cle96} 
has demonstrated the importance and diagnostic power of the rest-frame far-IR \citep{fis99, bh99, s05}.
\citet{fis99} observed in a sample of highly obscured ultraluminous IR galaxies a variety of 
line ratios indicative of contributions from a wide range of physical regions, and even some
peculiar spectra (e.g in Arp220) showing some atomic lines in absorption. 
ISO spectroscopy also showed, for example, that far-IR fine structure
line ratios (e.g. the diagram of [C II]/[OI] vs. [OIII]/[OI]) can be used to separate starburst and AGN
\citep{spi00}. A full decomposition of the primary ionising continua from the AGN and the starburst
in the Seyfert 2 galaxy NGC1068 has been performed through photoionisation models thanks to the
detection of tens of mid- and far-IR emission lines \citep{s05}. 

The mid-IR spectroscopic diagnostics of active and starburst galaxies has been shown in the last decade using results by  
{\it Spitzer} \citep{wer04} IRS \citep{hou04}. 
In a previous study, \citet{abe09} have compared \spi~ and {ISO} spectroscopic data of ULIRG
galaxies with photoionisation models, focusing on the predictions of far-IR line ratios and line to continuum ratios,
with models with high incident photon to particle densities. They found that these models can reproduce many
ULIRG observational characteristics, such as the [CII] line deficit \citep[see, e.g.][]{lum03}.  
Our goal here is to model the Seyfert galaxies. 

We have now extended the spectral range up to
the far-IR, using the results from the {\it Herschel} \citep{pil10} Photodetector Array Camera and Spectrometer (PACS) \citep{pog10}.
We present here the {\it Herschel} PACS spectroscopic observations of 26 Seyfert galaxies.  
These objects have been chosen because they are among the brightest local active galaxies at 60--100 $\mu$m, 
most of which belong to the 12$\mu$m galaxy sample \citep{rms}, they have been observed by 
{\it Spitzer} IRS and their luminosity has been separated between AGN and Starburst components through mid-IR spectroscopy. 
The immediate goals are to obtain a large set of FIR tracers to be able to classify and model the various levels 
of non-thermal and starburst activity in local Seyfert's, using fine-structure ionic 
and neutral transitions that have been observed 
with  {\it Herschel} and  {\it Spitzer}. Most of the galaxies (2/3) have been selected from the Seyfert galaxy 
catalog of the 12$\mu$m Galaxy Sample \citep{rms}, those not in this latter catalog have been selected because 
they are the brightest accretion-powered galaxies locally, as measured by the [OIV]26$\mu$m line with {\it Spitzer}. 
Following the mid-IR spectroscopic classification of \citet{tom10}, the selected sample includes 
10 Seyfert 1's, 5  Hidden Broad Line Region (HBLR) Seyfert 2's  (for a total of 13 AGN1's), 
9 "pure" AGN2's and three lower activity objects, classified as non-Seyfert's (Table \ref{tbl-1}).
Following the classifications given in \citet{tom10}, we have divided the narrow-line
Seyfert type 2 galaxy population into two classes, based on spectropolarimetry.
The ``HBLR" and ``pure type 2's"
are those Seyfert nuclei either having, or not having broad emission
line wings detected only in polarized light \citep[see, e.g.][]{tra01,tra03}. 

\section{Observations}
\label{sec:observations}

\subsection{PACS Spectroscopy}

We obtained far-infrared spectroscopic observations of a sample of 26 Seyfert nuclei, made with the PACS instrument \citep{pog10} 
onboard the ESA \textit{Herschel} Space Observatory \citep{pil10}. 
The selected sources belong to the two observational programmes in guaranteed time \textit{Bright Seyfert Nuclei: PACS spectroscopy} 
(GT1 lspinogl\_4, PI: L. Spinoglio) and \textit{PACS spectroscopy of bright Seyfert galaxies in the [OIV]26$\mu$m line} 
(GT2 lspinogl\_6, PI: L. Spinoglio), with a total time investment of 32 hours  of {\it Herschel}.
In this article we present the analysis of seven fine-structure lines 
(Table \ref{tbl-2}), while leaving for a forthcoming article the molecular line observations (four high-J CO lines and three OH doublets).
Whenever observations of the atomic lines of interest were carried out by other programs, we complemented our data
with PACS archival data. 
The final sample comprises observations acquired in single-pointing mode, carried out in chopping/nodding mode with a throw of 6'. 
Most of the observations were obtained in the range spectroscopy mode, whereas--for enlarging the sample--line spectroscopy observations were included. 
Standard rebinned cubes were used for the scientific analysis. 
Thus the spectral maps are composed by $5\times5$ squared spatial pixels (spaxels), 
with a size of 9.''7, covering a field of view of $47''\times47''$. Our observations were conducted in staring mode, so our maps are not 
Nyquist-sampled with respect to the Point Spread Function (PSF), which ranges from $4.1''$ at 57~$\mu$m\ to $11.6''$ at 163~$\mu$m, 
but the beam size is dominated by the spaxel size below $\lambda$ $\sim$120~$\mu$m (see Table \ref{tbl-2}).

The \spi~ IRAC 3.6$\mu$m images of each galaxy in our sample,  superposed to the frame of the PACS spectrometer of 
47$\arcsec$ $\times$ 47$\arcsec$ and the slit of 22.3$\arcsec$ $\times$ 11.1$\arcsec$ of the long-wavelength high-resolution 
mode (LH) of the IRS spectrometer of \spi, are shown in Appendix \ref{app.c}. 

Data reduction has been performed with standard recipes within HIPE\footnote{HIPE is a joint development by the \textit{Herschel} Science 
Ground Segment Consortium, consisting of ESA, the NASA \textit{Herschel} Science Center, and the HIFI, PACS and SPIRE consortia.} 
v10, starting from Level~0 data, and including the specFlatFieldLine task for applying the flat-field correction at every observed line.  
An external package, developed in the IDL language, was built for performing the spectral analysis over the wide sample of spectral profiles.
The software looks for spectral lines in every spaxel by adopting a threshold approach and, if spectral features are identified, a polynomial 
continuum plus a gaussian profile was fitted by using the IDL GAUSSFIT routine. 
The estimations provided by the fitting procedure were used to build maps of physical (continuum level, line intensity, line peak emission) 
and kinematical (peak velocity, velocity dispersion) parameters.

The observations were based on the assumption that the galactic nuclei are point-like sources located at the central spaxel position.
However, taking into account small telescope mispointings  
(less than 1/2 spaxel) and the large, non-Gaussian point spread function  
(especially at the longer wavelengths), we integrated the line intensity 
measured over the central 3$\times$3 spaxels. 
To recover the absolute flux level of the nuclear emission, 
we applied an aperture correction factor consistent with the spectra extracted from the central 9 spaxels.
The point-source aperture correction factors were obtained by using the calibration product pointSpuceLoss (version 4), 
provided within the HIPE environment for the PACS spectrometer. 

The obtained line profiles are shown in Appendix \ref{app.a} and the measured fluxes are reported in Table \ref{tbl-3}. 
Upper limits were provided by computing the integrated flux within an unresolved Gaussian profile with peak equal to 
3$\times$RMS and sigma derived by the spectral resolution at the considered wavelength. 
The spectral profiles reveal that often the lines are not well represented by single Gaussians;  thus the line intensities 
have been obtained with a numerical integration across the line profiles. 
The velocity limits of the lines were inferred by computing the RMS and adopting a threshold approach. 
A comparison of the two flux estimates for lines with Gaussian-shape profiles results in a difference of less than of the flux 
calibration accuracy, i.e. less than 10\%.  

We point out that the OIII]52$\mu$m line lies at the edge of the spectral coverage of the PACS
BLUE detector, where the relative spectral response function is not well characterised.
As a result, the 52$\mu$m ßuxes measured in the galaxies UGC2608, 3C120, MRK3, NGC5256, IC4329A, NGC5506
and NGC7582 can be affected by a calibration error that exceeds the noise of the spectrum,
as it is shown in Table \ref{tbl-3}. This error is hard to quantify, because few calibration data are available for
it. As an example, in NGC4151, this line was detected by ISO-LWS with a flux of (103$\pm$22)$\times$10$^{-17}$W m$^{-2}$ \citep{spi97}, 
whereas the same line was detected by PACS with a flux of (40.85$\pm$6.30)$\times$10$^{-17}$W m$^{-2}$.
This indicates a factor of 2.5 uncertainty in the flux measurement, which cannot be accounted for by aperture
effects, that are negligible for this high-ionisation line. This uncertainty does not account for cross calibration
errors between the two instruments. Nevertheless, the measured fluxes are not affected by leakage from other
wavelengths because the observations have been made with the second-order grating.
In order to explore the discrepancies found between the ISO-LWS and the Herschel-PACS observations,
we present in Appendix \ref{app.b} the observations of Seyfert galaxies obtained with both spectrometers. 

Table \ref{tbl-2} also contains the SPIRE FTS fluxes of the [NII]205$\mu$m, the [CI]370$\mu$m and the [CI]609$\mu$m lines. 
Most detected objects have been published in \citet{p-s13},  
while for UGC2608, we have reduced the FTS spectrum.
We downloaded SPIRE/FTS high resolution (1.45 GHz) spectroscopic observations of UGC2608 from the {\it Herschel} archive. 
The data were reduced as described by \citet{p-s13}, using the standard pipeline provided by the {\it Herschel} interactive pipeline environment 
software (HIPE) version 11 \citep{ott10}. 
We assumed the point-like corrections for the flux calibrations and extracted the spectra from the central bolometer detector. 
The line fluxes were measured by fitting a \textit{sinc} function to the emission profiles.  

\subsection{Ancillary data: {\it Spitzer} IRS Spectroscopy}

As complementary data, that we will use in the analysis, we also present the {\it Spitzer} IRS high-resolution spectroscopy 
of the 26 galaxies of our sample. Most of the data have been taken from \citet{tom08, tom10}, 
while for the galaxies not observed there, we have reduced the IRS data from the
 {\it Spitzer} Archive. The {\it Spitzer} IRS spectroscopy is presented in Table \ref{tbl-4}.

\section{Results}

\subsection{Density diagnostics} \label{dens_diag}

\begin{figure*}
\centering
\includegraphics[width=8cm]{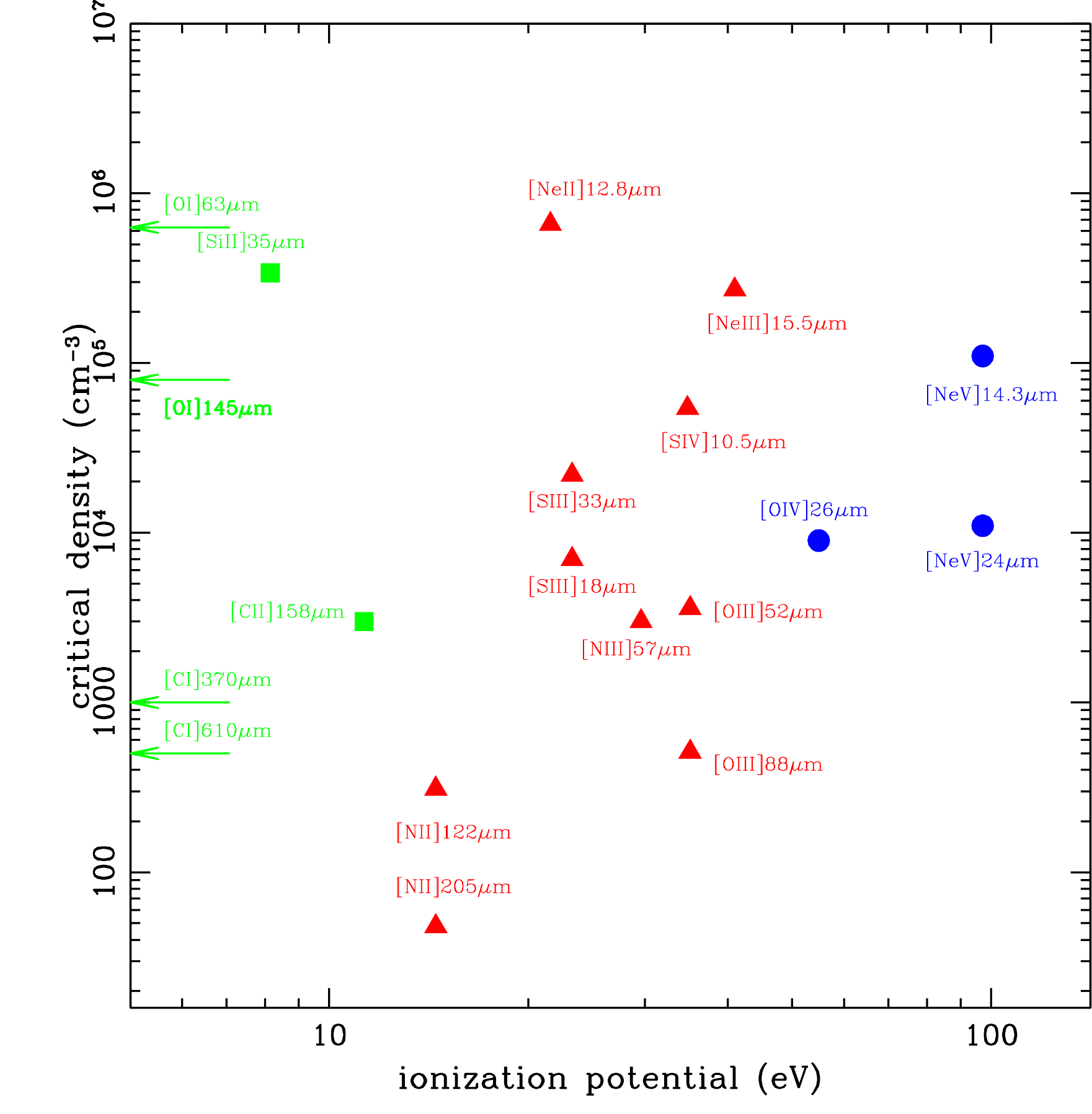}\includegraphics[width=8cm]{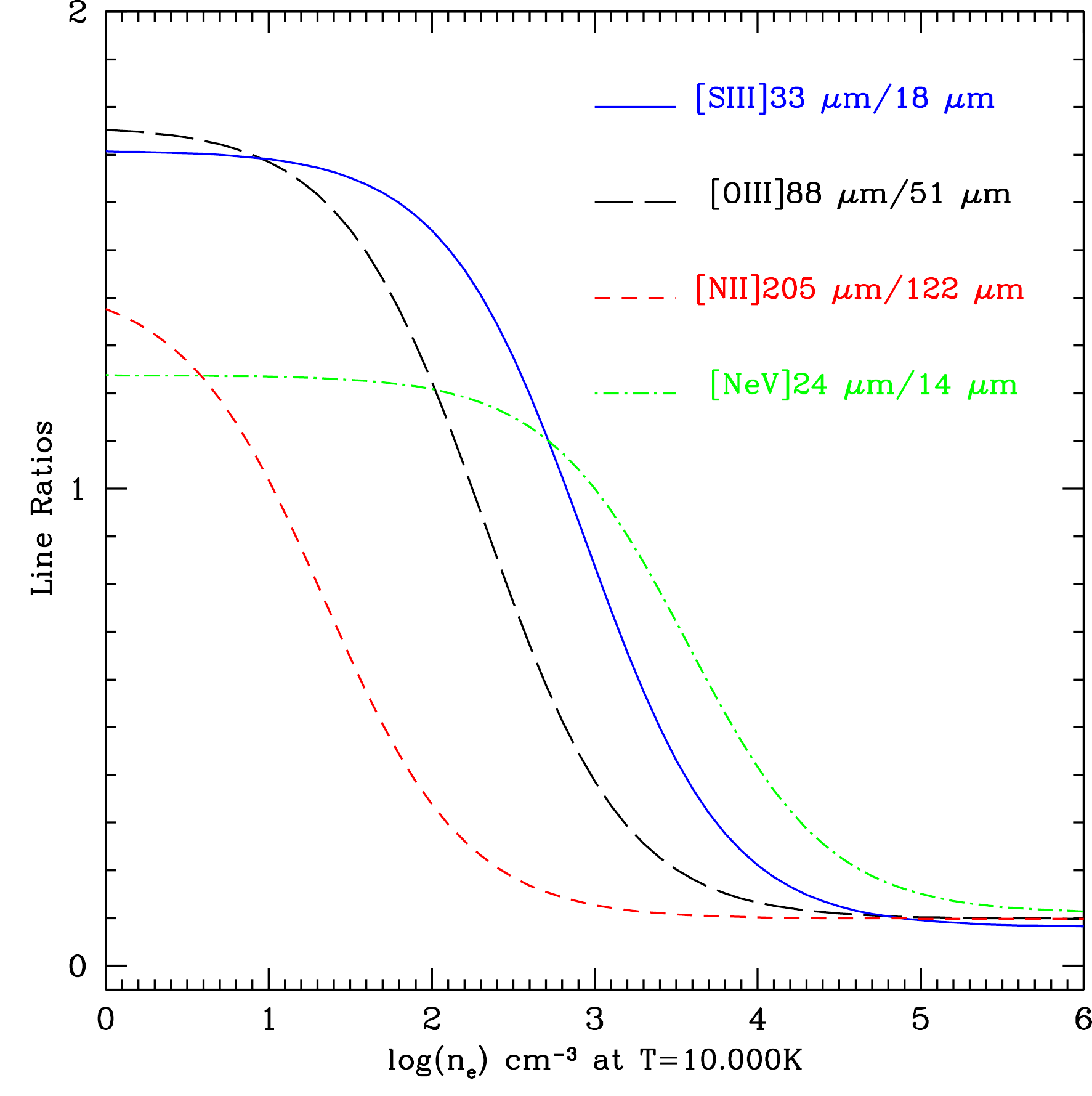}
\caption{\scriptsize {\bf Left: (a)} Critical density for collisional de-excitation vs ionisation potential of the IR fine-structure lines observed in the sample of Seyfert galaxies.  
{\bf Right: (b)} Ionic line ratios as density estimators. The density values are computed for a gas in a pure collisional regime at a temperature of T=10$^4$ K.}
\label{fig:line_ratios_density}
\end{figure*}

The ratio of two fine-structure lines from the same ion 
or neutral atom  
is frequently most sensitive to the electron densities of the gas \citep[e.g., ][]{rub94}.
Using the Herschel PACS, SPIRE and Spitzer-IRS line measurements (Tables \ref{tbl-3} and \ref{tbl-4}), we are able to use the [OIII], [NII], [SIII] and [NeV] lines to determine the electron densities
in the ionised line-emitting regions. Moreover, the [CI]  lines can be used to measure the densities in the neutral gas,
while the ratio of the [OI] lines, provides a measure of the neutral gas temperature \citep{th85}.  
We present in Fig.\ref{fig:line_ratios_density}(a), the critical densities of the fine structure lines considered in this work 
as a function of the ionisation potential of the ions \citep{sm92}.

The line ratios for thermalized gas at 10$^4$K that is purely collisionally excited are shown 
in Fig.\ref{fig:line_ratios_density}(b). As can be seen from the figure,  
the [NII] line ratio is sensitive at electron densities in the range of 1$<$ n$_{e}$ $<$ 10$^3$ cm$^{-3}$, 
the [OIII] line ratio at densities in the range of 10$<$ n$_{e}$ $<$ 10$^4$ cm$^{-3}$, the 
[SIII] line ratio at densities in the range of 10$^2$ $<$ n$_{e}$ $<$ 10$^4$ cm$^{-3}$, and, finally, the [NeV] 
line ratio at densities in the range of 10$^3$ $<$ n$_{e}$ $<$ 10$^5$ cm$^{-3}$.
In combination, these lines provide a sequence that can trace the density across a large range, from n$_{e}$=10 to n$_{e}$=10$^5$ cm$^{-3}$.

Using the level population equations of a constant density gas in a pure collisional regime \citep[see, e.g., ][]{of06}, we have determined the theoretical
line ratios, and compared them with the observations. 
We have considered gas temperatures of 1,000 and 10,000K for ratios of ionised lines and of
100K and 1,000K for the neutral lines line ratios, namely the [OI]145$\mu$m/63$\mu$m and the [CI]609$\mu$m/370$\mu$m line ratios. 
The atomic data used to calculate the  line ratios as a function of density and temperature have been taken from the IRON project for 
[SIII], [NII] and [OIII] \citep{hum93}, and from the LAMDA database for the other ions \citep{scho95}. 
Some caution has to be taken in comparing differences in the plots, because not all galaxies have been observed in all line ratios (see Tables \ref{tbl-3} and \ref{tbl-4}).
Moreover, because of the lack of knowledge of a precise calibration for the [OIII]52$\mu$m line flux (see Section \ref{sec:observations}), 
we warn the reader that the densities obtained from the [OIII] line ratio should be regarded as very rough approximations.

\begin{figure*}
\includegraphics[width=8cm]{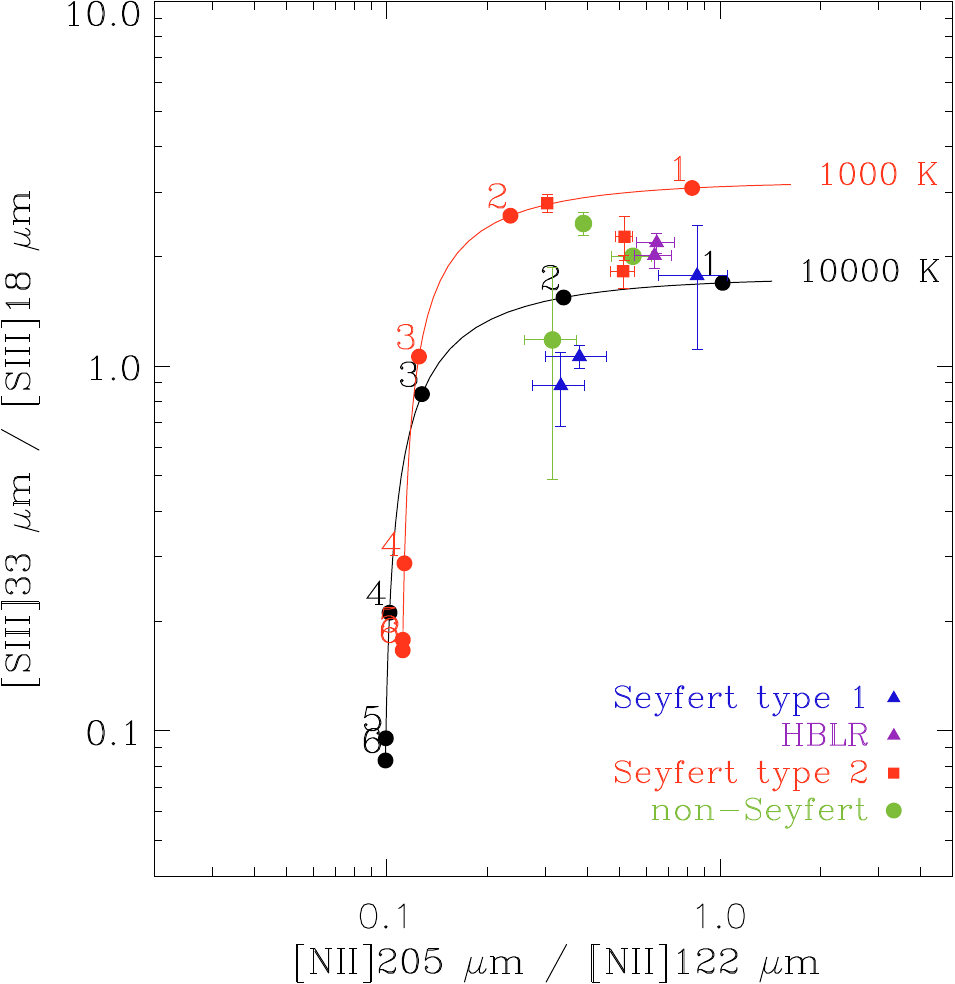}\includegraphics[width=8cm]{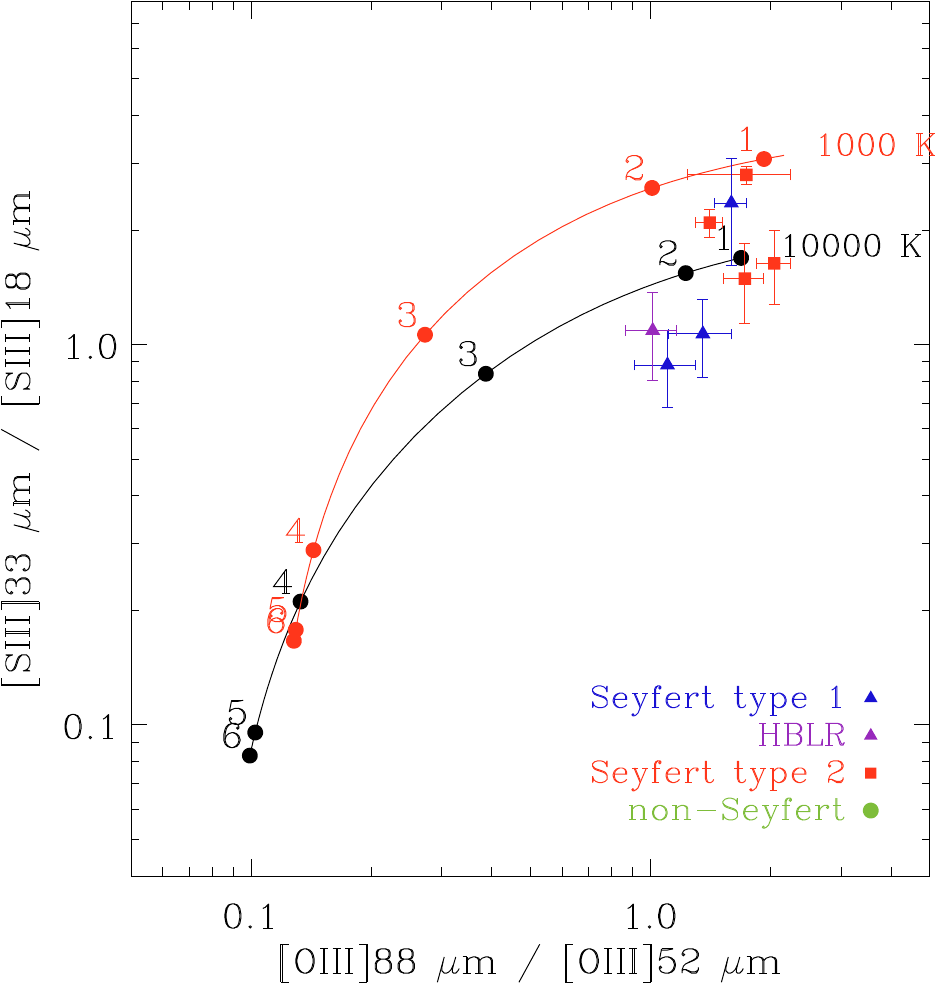}
\caption{\scriptsize Density diagnostics. The galaxies detected in each pair of line ratios are indicated with different symbols, depending on their classification in Table \ref{tbl-1}: Seyfert type 1's are shown as (blue) filled triangles; HBLR Seyfert 2's as (magenta) open triangles; "pure" Seyfert 2's as (red) filled squares and non-Seyfert nuclei (including LINERs, ULIRGs, etc.) are shown as (green) filled circles. The numbers indicate the logarithm of the densities for each of the two temperature values.
 {\bf Left: (a)} the [SIII]33$\mu$m/18$\mu$m line ratio vs the [NII]205$\mu$m/122$\mu$m line ratio. {\bf Right: (b)} the [SIII]33$\mu$m/18$\mu$m line ratio vs the [OIII]88$\mu$m/52$\mu$m line ratio.}
\label{fig:dens_1_2}
\end{figure*}

\begin{figure*}
\includegraphics[width=8cm]{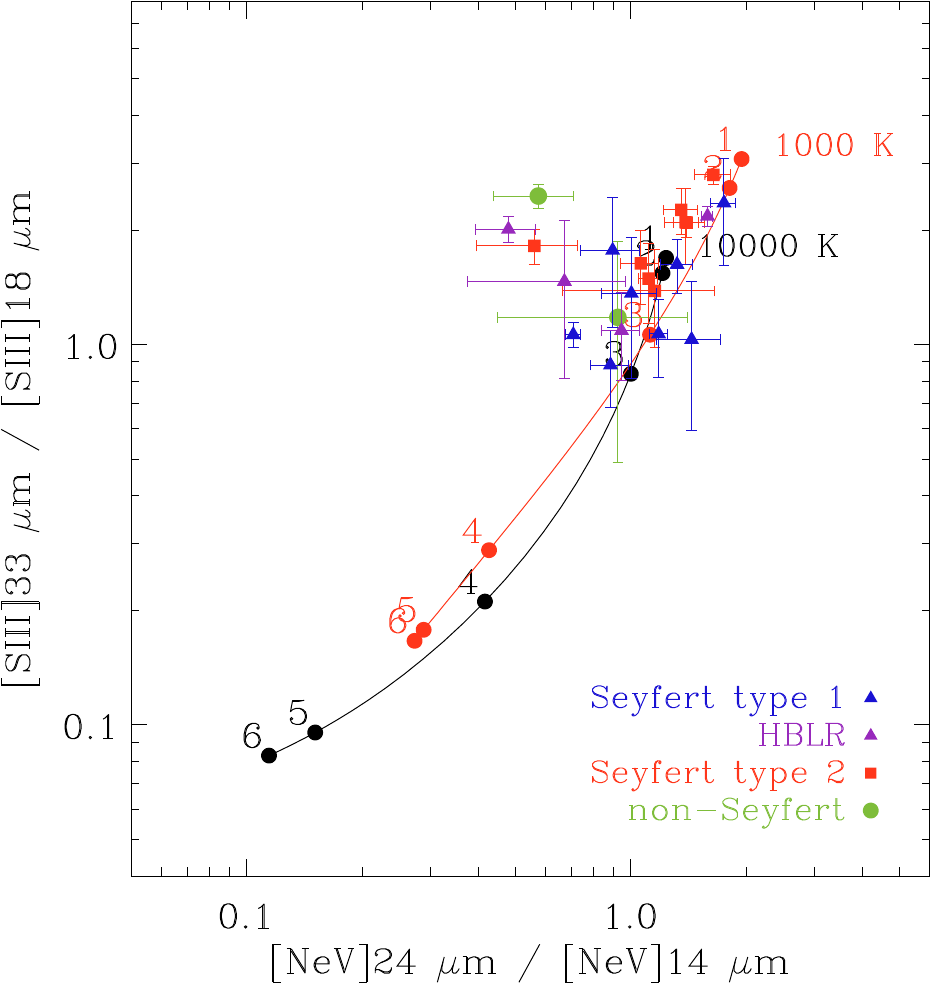}
\caption{\scriptsize Density diagnostics. Notations are the same as in Fig. \ref{fig:dens_1_2}. The [SIII]33$\mu$m/18$\mu$m line ratio vs the [NeV]24$\mu$m/14$\mu$m line ratio.}
\label{fig:dens_3}
\end{figure*}

Fig.\ref{fig:dens_1_2}(a) shows the [SIII]33$\mu$m/18$\mu$m line ratio, which traces the density in HII regions, versus the 
[NII]205$\mu$m/122$\mu$m line ratio, tracing the low-density diffuse ionised gas,  as well as the HII region gas. It appears that
the diffuse ionised gas of the observed galaxies has densities in the range of 10$\leq$n$\leq$10$^2$ cm$^{-3}$, while the HII regions 
gas has densities below 10$^3$ cm$^{-3}$, down to the low-density limit, where the [SIII] ratio no longer measures the density.  
Fig.\ref{fig:dens_1_2}(b) shows the [SIII]33$\mu$m/18$\mu$m line ratio vs the [OIII]88$\mu$m/52$\mu$m line ratio. 
These lines are primarily emitted in HII regions around young stars.
They indicate densities in the range 10$\leq$n$\leq$10$^3$ cm$^{-3}$.
Fig.\ref{fig:dens_3} plots the [SIII]33$\mu$m/18$\mu$m line ratio vs the [NeV]24$\mu$m/14$\mu$m line ratio. The observed galaxies 
here cluster around densities of 10$^2$-10$^3$ cm$^{-3}$. However there are some galaxies for which the density measured by the [SIII] ratio 
is below the low density limit (n$_e$ $\leq$ 10$^2$ cm$^{-3}$), while the [NeV] density is much higher (10$^3$ $\leq$ n$_e$ $\leq$ 10$^4$ cm$^{-3}$). 
This is because the [NeV] is emitted from the high excitation/ionisation NLR of the AGN, whereas the [SIII] lines are emitted mostly by stellar HII regions.
We refer to the discussion presented in \citet{tom10}, where it has been shown in detail that many observed line ratios indicate densities below the 
low density limit.

\begin{figure*}
\includegraphics[width=8cm]{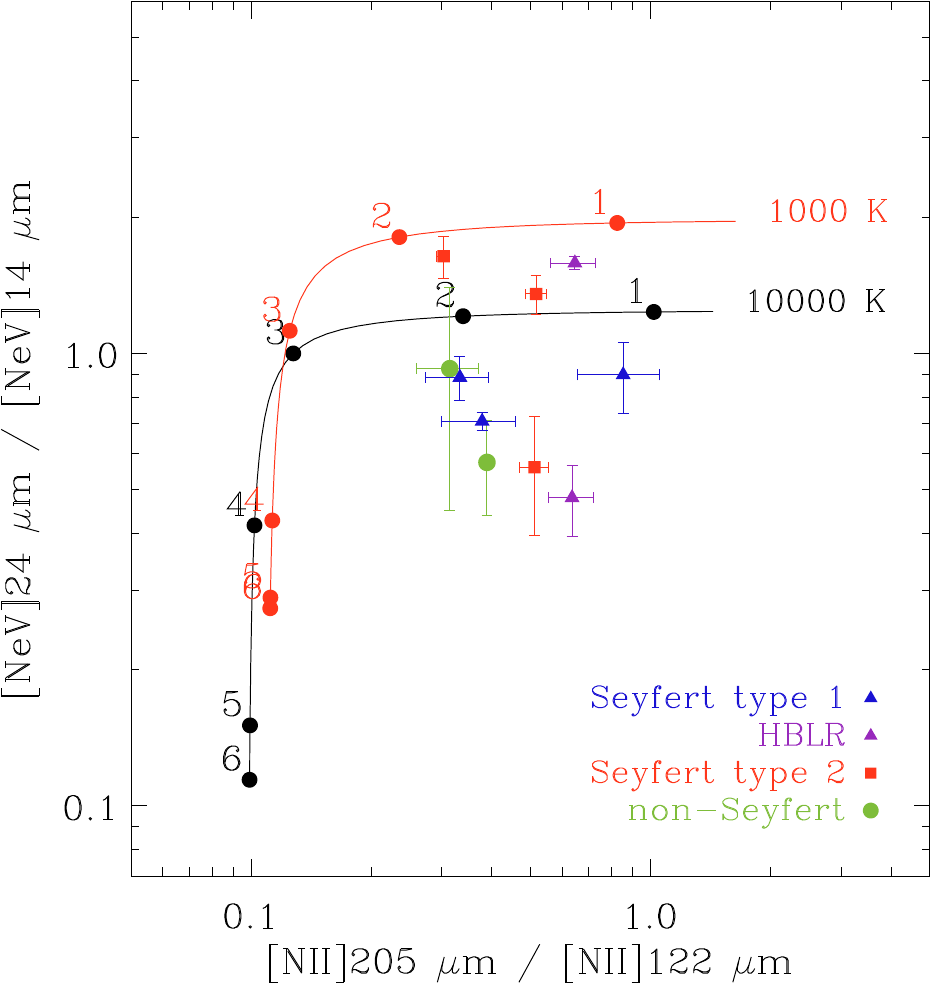}\includegraphics[width=8cm]{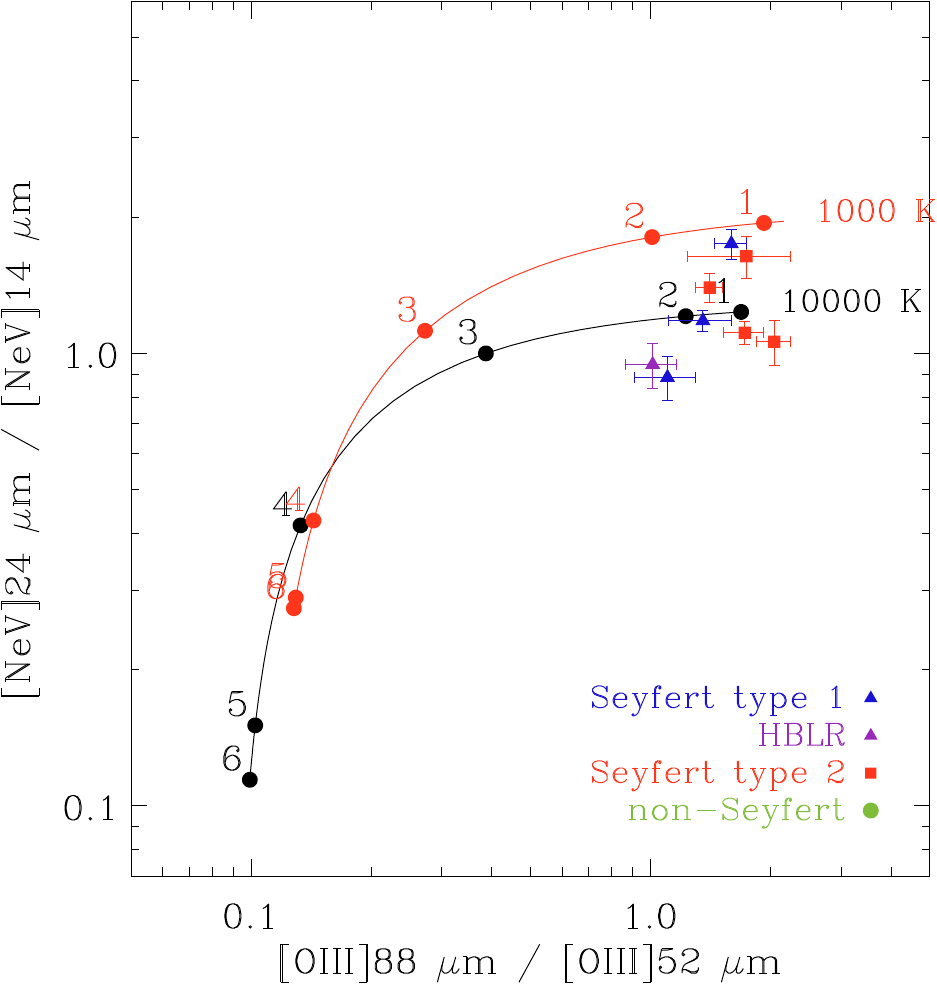}
\caption{\scriptsize Density diagnostics. Notations are the same as in Fig. \ref{fig:dens_1_2}. {\bf Left: (a)} the NeV]24$\mu$m/14$\mu$m line ratio vs the [NII]205$\mu$m/122$\mu$m line ratio. {\bf Right: (b)} the [NeV]24$\mu$m/14$\mu$m line ratio vs the  [OIII]88$\mu$m/52$\mu$m line ratio.}
\label{fig:dens_5_6}
\end{figure*}

Fig.\ref{fig:dens_5_6}(a) plots the [NeV]24$\mu$m/14$\mu$m line ratio vs the [NII]205$\mu$m/122$\mu$m line ratio.  
As in Fig.\ref{fig:dens_3}, the two line ratios are sampling different regions in the observed galaxies: the AGN NLR in the first case 
and the low density diffuse ionised ISM in the other ratio. 
Fig.\ref{fig:dens_5_6}(b) plots the [NeV]24$\mu$m/14$\mu$m line ratio vs the  [OIII]88$\mu$m/52$\mu$m line ratio, 
which indicates that the NLR detected in the few observed galaxies (for the 52$\mu$m line we only have 8 detections) has densities in the range  10 $\leq$ n$_e$ $\leq$ 10$^3$ cm$^{-3}$.

However, the [OIII] line ratio samples lower densities compared to the [NeV] line ratio, as can be seen from the difference in the critical densities of the two sets of transitions in Fig. \ref{fig:line_ratios_density}(a).  

\begin{figure*}
\includegraphics[width=7.9cm]{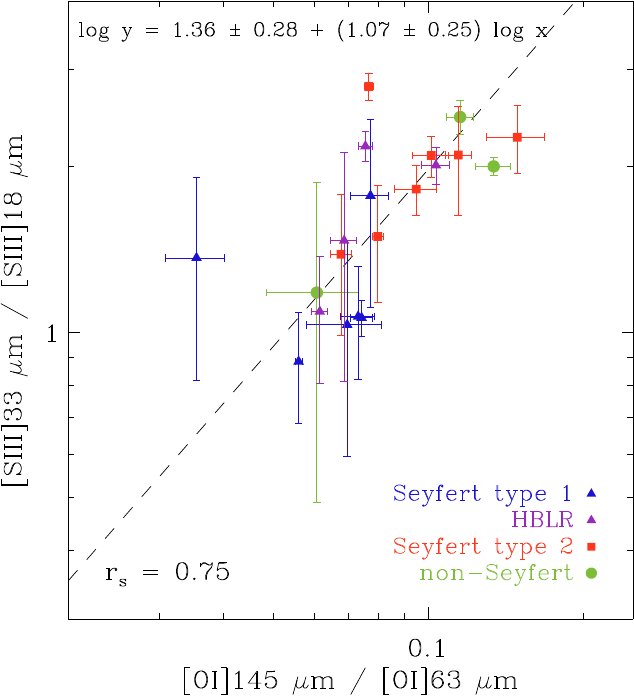} \includegraphics[width=8.15cm]{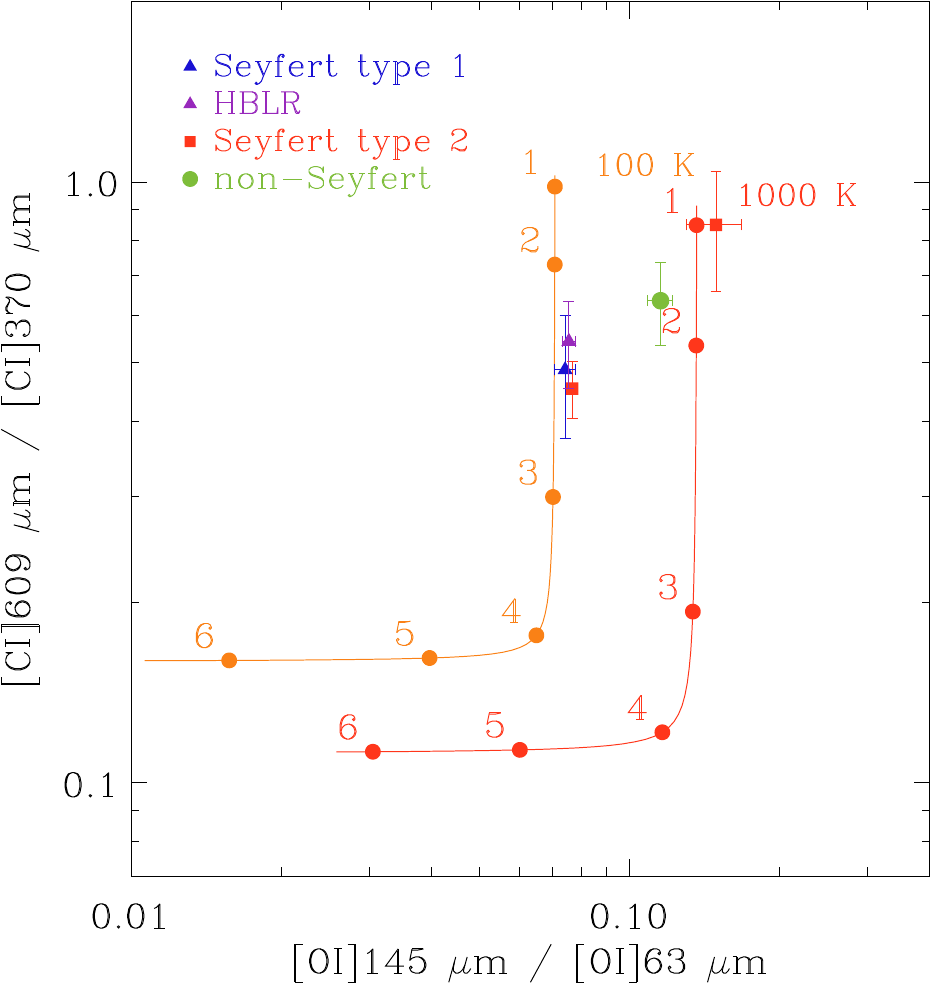} 
\caption{\scriptsize Density--temperature diagnostics. Notations are the same as in Fig. \ref{fig:dens_1_2}. {\bf Left: (a)} The [SIII]33$\mu$m/18$\mu$m line ratio vs the [OI]145$\mu$m/63$\mu$m line ratio.
{\bf Right: (b)} the  [CI]609$\mu$m/370$\mu$m line ratio vs the [OI]145$\mu$m/63$\mu$m line ratio.}
\label{fig:dens_7}
\end{figure*}

Fig.\ref{fig:dens_7}(a) plots the [SIII]33$\mu$m/18$\mu$m line ratio vs the [OI]145$\mu$m/63$\mu$m line ratio\footnote{No model has been included in 
this figure because the collision strengths for [SIII] are available only at 1,000K, and not at 100K while for the [OI] ratio the 100K data should be used.}.
A least-squares fit yields the following relation between the [SIII]33$\mu$m/18$\mu$m line ratio vs the [OI]145$\mu$m/63$\mu$m line ratio:
\begin{eqnarray*} 
{\rm log} F_{[SIII]33 \mu m}/F_{[SIII]18 \mu m}  = (1.03 \pm 0.22) \times  \\  \times {\rm log} F_{[OI]145 \mu m}/F_{[OI]63 \mu m}  + (1.33 \pm 0.24)~~~~~~(1)
\end{eqnarray*} 
 relation reflects an anti-correlation between the ionised gas density and the neutral ISM gas temperature.

Finally, Fig.\ref{fig:dens_7}(b) plots the [CI]609$\mu$m/370$\mu$m line ratio vs the [OI]145$\mu$m/63$\mu$m line ratio. 
The primarily molecular hydrogen gas traced by these lines spans densities from 10$^2$ $\leq$ n$_{H_2}$ $\leq$ 10$^3$ cm$^{-3}$ at the lower temperature 
of 100K to 10 $\leq$ n$_{H_2}$ $\leq$ 10$^2$ cm$^{-3}$ at the higher temperature of 1,000K. 

The [OI] ratio does not constrain the density, because it is below n$_H$ $<$ 10$^4$ cm$^{-3}$, but the gas temperature instead \citep{th85}. 
An anti-correlation between density and temperature seems present, which is expected at these low densities in the optically thin case, because 
the [CI] line ratio measures the molecular hydrogen density, while the [OI] line ratio is sensitive to the temperature. 
From Fig.\ref{fig:dens_7}(b), we can see that at a temperature of T=100K, the [CI] line ratios are consistent with densities of 10$^2$ $<$n$<$ 10$^3$  cm$^{-3}$. 
For the same Seyfert galaxies reported in this work, assuming that the [CI] emission is thermalized, \citet{p-s13} have derived a
temperature of  $<$ 30K and a density of n $>$ 10$^3$ cm$^{-3}$.  However, according to Fig. 9 of  \citet{p-s13}, the [CI] ratios 
presented in their work are also compatible with higher temperature and lower density (i.e. 10$^2$ $<$n$<$ 10$^3$  cm$^{-3}$ 
for 10$^2$ $<$T$<$ 10$^{2.5}$ K). In the present work, we have added the [OI] ratio in the diagram and the result might be different, 
also considering that the [CI] and [OI] emission might not be produced exactly in the same regions. 

\subsection{Density stratifications}

As already shown in the previous section, different lines in the same galaxy can originate in
different emission line regions, arising from physically distinct components in the galaxy. The nuclear region is characterised by emission 
of NLR gas directly illuminated by the AGN and excited by the hard primary ionising spectrum originating from black hole accretion. 
The NLR can have by itself a stratification of densities, that can be mapped by transitions with different collisional critical densities. 

Table \ref{tbl-5} gives, for each galaxy, the  [OIII]88$\mu m$/52$\mu m$, [SIII]33$\mu m$/18$\mu m$, 
[NII]205\-$\mu m$/122$\mu m$ and [NeV]24$\mu m$/14$\mu m$ line ratio values and their errors, 
and the average electron densities derived from each ratio, as well as the range in density due to the uncertainty of the line ratio,  
assuming a gas temperature of 10,000K.

\begin{figure*}
\includegraphics[width=8cm]{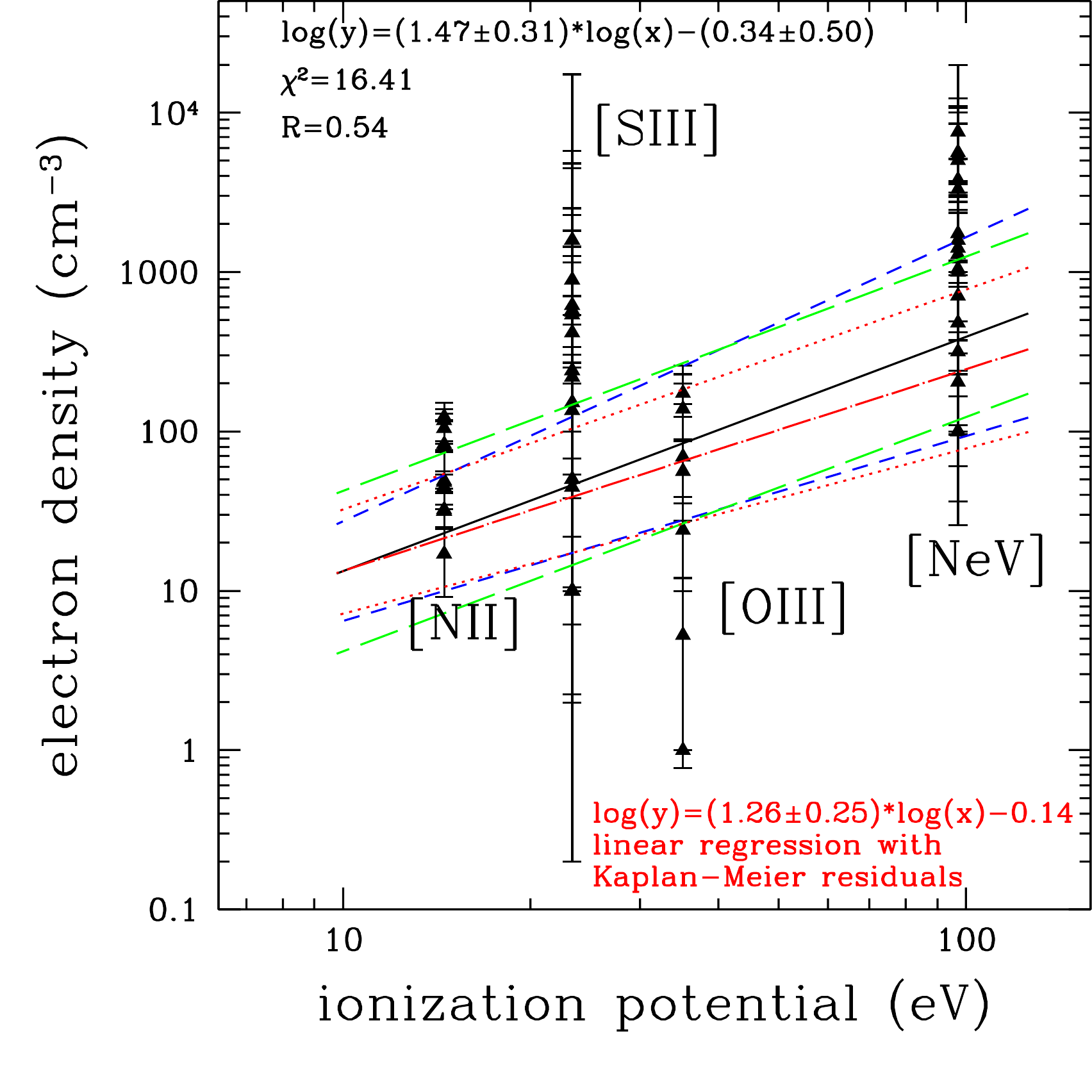}
\includegraphics[width=8cm]{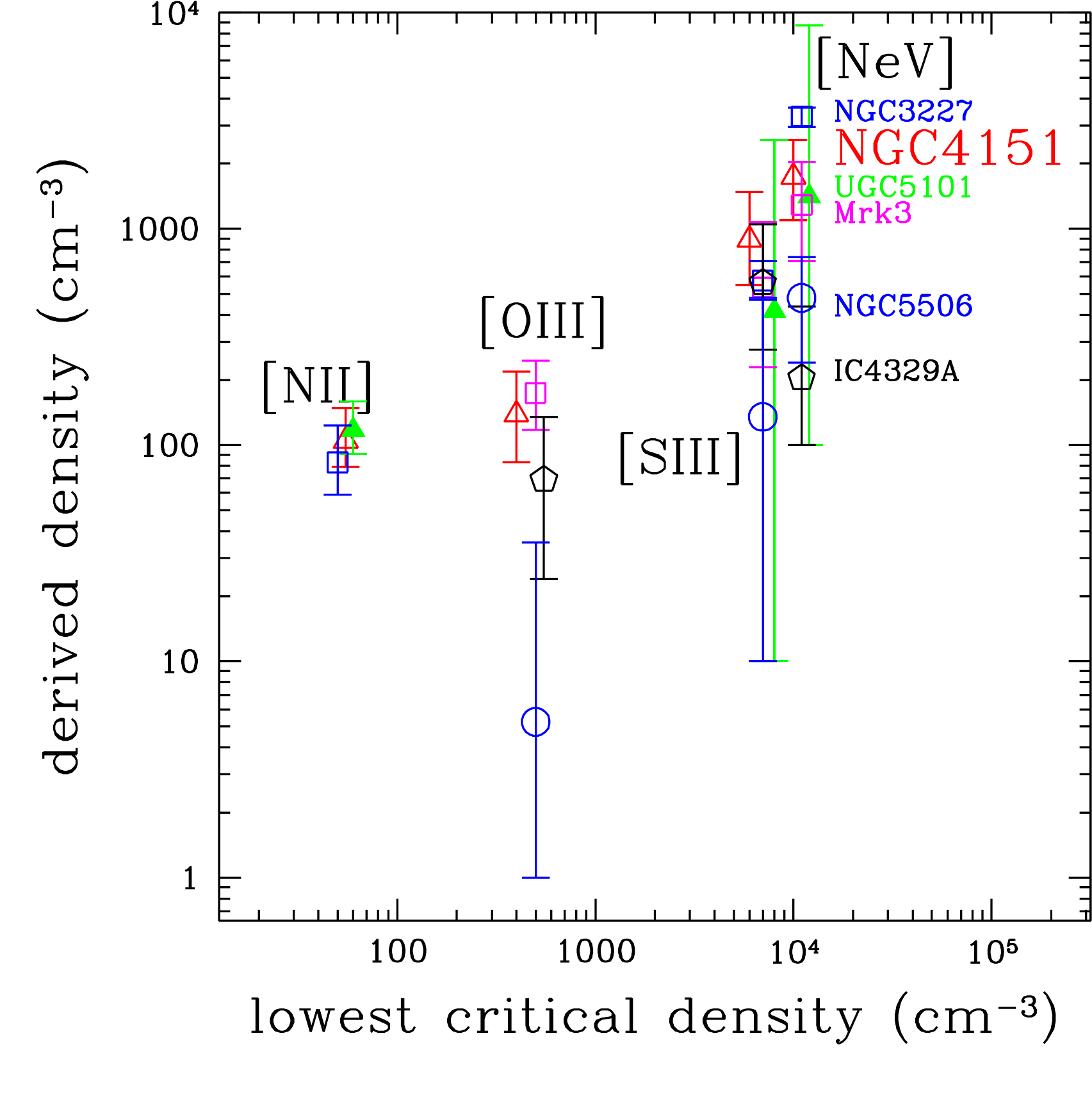}
\caption{\scriptsize Density stratifications. {\bf Left: (a)} Derived electron densities as a function of the ionisation potential for all galaxies in Table \ref{tbl-5}.
The solid line shows a weighted least squares fit to the data, indicating the presence of a correlation between density and ionisation, while the broken lines show the effect of the uncertainty in the slope and the intercept. The fit shown with the dot-dashed  line shows the result of the linear regression using the Kaplan-Meier residuals, while the dotted lines, the effects of the incertitude in the slope in this latter fit.
{\bf Right: (b)} Derived electron densities in 6 galaxies
for which at least three line ratios have been observed, as a function of the lowest critical 
density for collisional de-excitation of the pair of transitions from which the line ratio is formed. }
\label{fig:strat_1}
\end{figure*}

In Fig.\ref{fig:strat_1}(a) we present the derived electron densities as a function of the ionisation potential for all galaxies 
for which we determined the density in Table \ref{tbl-5}. 
A weighted least-squares fit to the whole set of data indicates a correlation between density and ionisation, with a 
correlation coefficient of R=0.54, a $\chi^2$ =16.4 and a regression line slope of $\alpha$ = 1.47$\pm$0.31. To compute this fit we have also included
the upper limits, by artificially assigning errors which are the semi-difference between the two values of the range given in Table \ref{tbl-5}, 
with the lower values of the range being set to one decade below the upper limit values. If no density range was assigned, an error of half a decade
was given. 
We also used another approach, that naturally includes the upper limits: a linear regression based on Kaplan-Meier residuals\footnote{The linear regression was 
performed using the BUCKLEYJAMES routine, available in the stsdas data analysis package. The description can be found at: http://stsdas.stsci.edu/cgi-bin/gethelp.cgi?buckleyjames.hlp}. 
A description of this method, which is part of the ASURV\footnote{ASTRONOMY SURVIVAL  ANALYSIS PACKAGE, whose description can be found at: http://stsdas.stsci.edu/cgi-bin/gethelp.cgi?survival} 
package, can be found in \citet{iso86}.  The results of this linear regression fit give a regression line slope of $\alpha$ = 1.26$\pm$0.25 
and are consistent with the weighted least squares fit, as shown in Fig. \ref{fig:strat_1}(a).

For the six galaxies for which at least three density-probing line ratios have been observed, we plotted the density estimate as a function of the
lowest collisional critical density of that particular pair of transitions from the same ion (i.e., the critical density of [NII]205$\mu$m, [OIII]88$\mu$m, 
[SIII]18$\mu$m and [NeV]24$\mu$m; Fig.\ref{fig:strat_1}(b)). From Fig.\ref{fig:strat_1}(b), it appears that the probed densities increase with the lowest critical density (as they should).
The trend between the derived density and the ionisation potential of the species emitting the lines is harder to deduce for this sample, with IC4329A not showing
a monotonic increase of the derived densities as a function of ionisation. In a previous study based on spectrally-resolved mid-IR lines, \citet{das11} 
reported a stratification of densities, with different fine-structure lines probing different (locations within the) clouds, with high ionisation potential ions being preferentially found nearer to the 
black hole.   

In conclusion we have found that there is a tendency for the higher ionisation lines to originate in gas with higher density. 
This can be seen as an average result from Fig.\ref{fig:strat_1}(a).
An explanation of this result is that the gas in the Narrow Line Region includes clouds of varying densities,
which are highest nearest to the ionising AGN, confirming the results of \citet{das11}. 

\subsection{[OI] diagnostics}

The [OI]145$\mu$m/63$\mu$m ratio is sensitive to the gas temperature 
at a given gas density below $\sim$ 10$^4$ cm$^{-3}$ \citep[see, e.g.,][]{th85}. 

In Table \ref{tbl-6} we present the temperature determinations computed from the [OI]145/$\mu$m/63$\mu$m
line ratio\footnote{When [OI]145/$\mu$m/63$\mu$m $>$ 0.14, no temperature value could be derived. }, 
using the analytic models presented in section \ref{dens_diag} for a density of n=10$^2$ cm$^{-3}$.

\begin{figure*}
\includegraphics[width=8.cm]{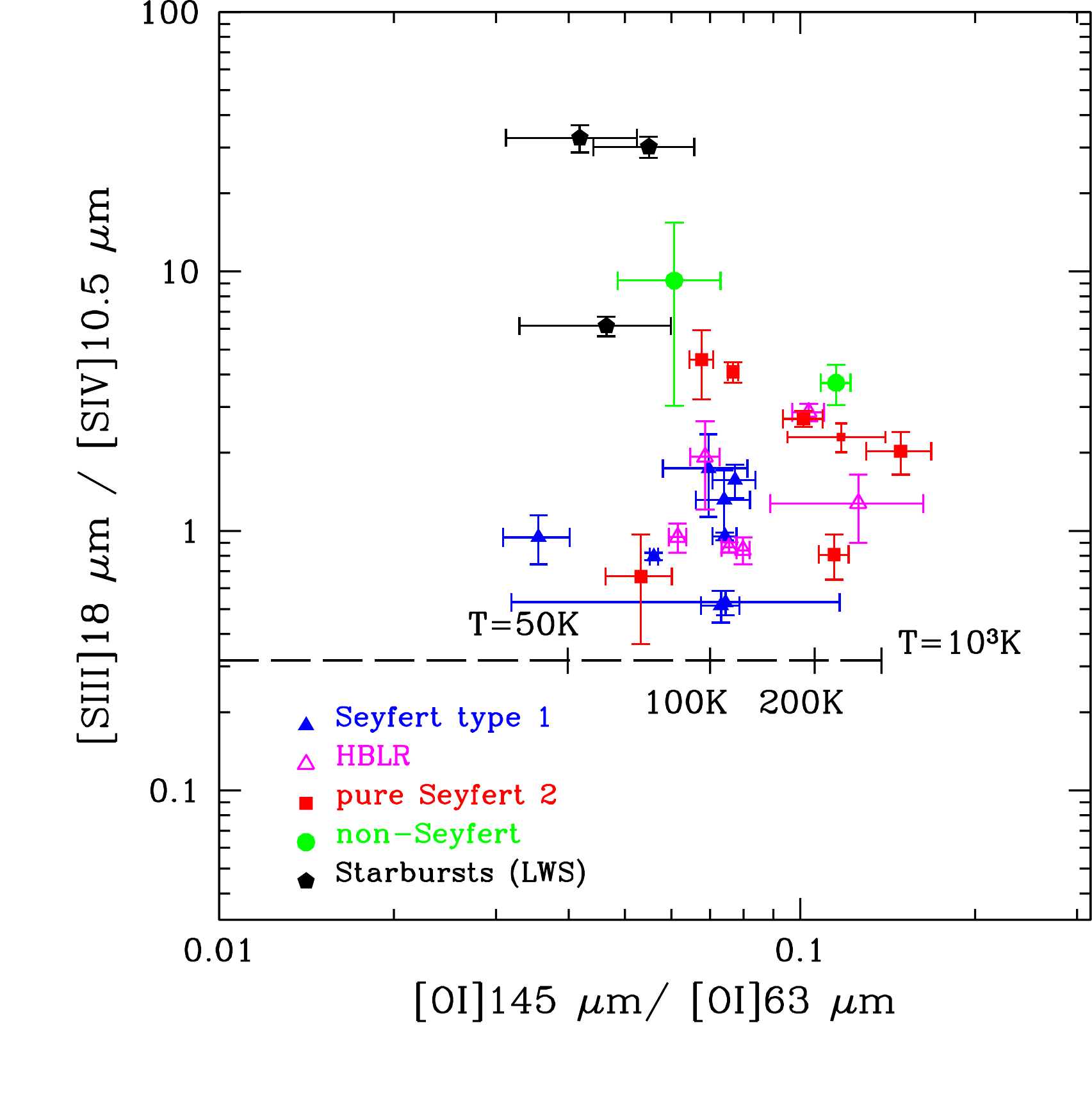}\includegraphics[width=8cm]{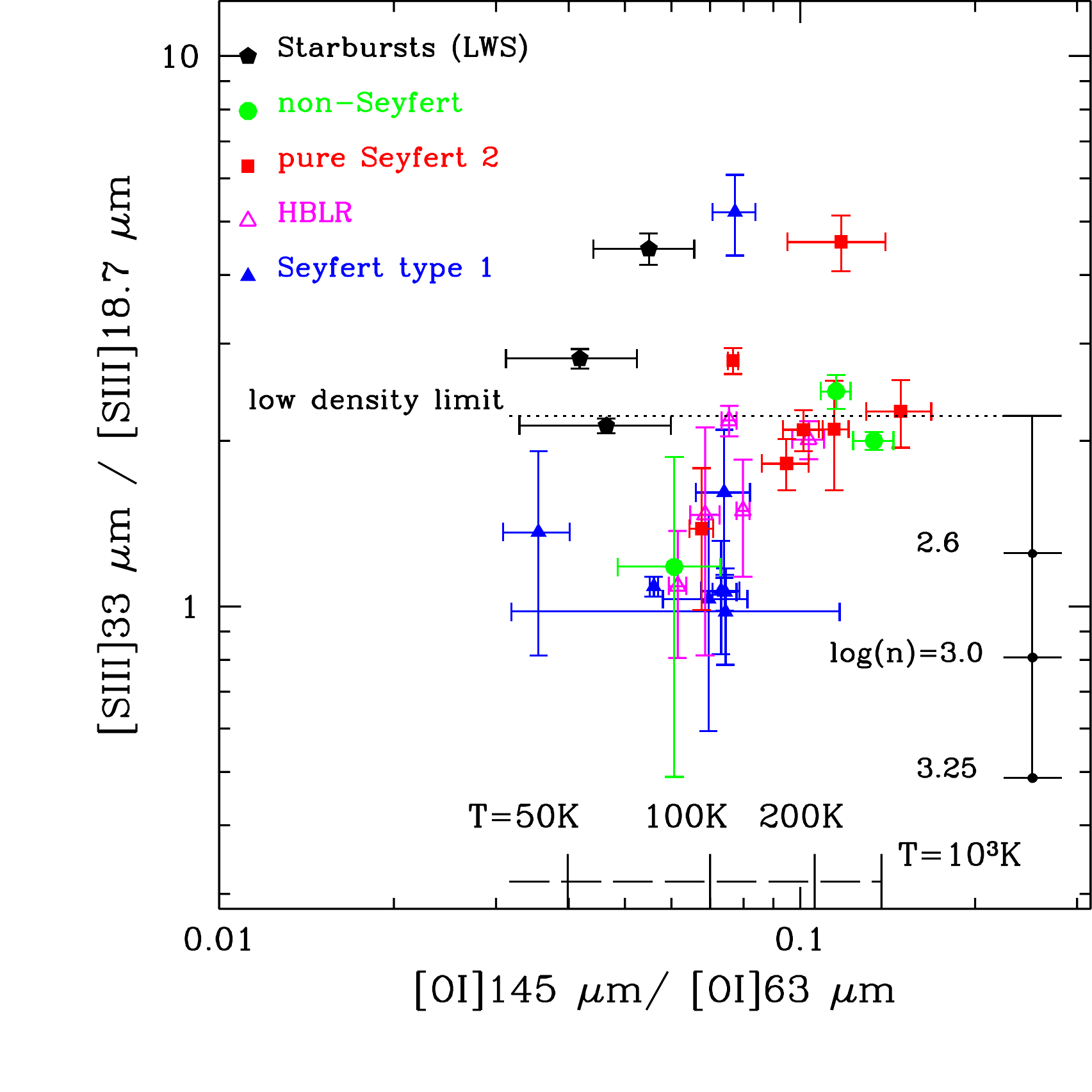}
\caption{\scriptsize Ionisation and density versus temperature. {\bf Left: (a)} the [OI]145$\mu$m/63$\mu$m vs the [SIII]/[SIV] line ratio. 
The [OI] line ratios corresponding to various gas temperatures and a density of n=10$^2$ cm$^{-3}$, as computed with the analytic models 
presented in section \ref{dens_diag}, are indicated in the bottom part of the diagram. 
{\bf Right: (b)} the [OI]145$\mu$m/63$\mu$m vs the [SIII]33/18$\mu$m, which, in turn, measures the electron density, whose values are marked on the right of the diagram.
Compared to Fig. \ref{fig:dens_7}(a), we have included here literature data for three starburst galaxies and two Seyfert 1's from ISO-LWS \citep{bra08}.} 
\label{fig:T_OI}
\end{figure*}

In order to test whether a harder radiation field corresponds to a warmer interstellar medium, we explored whether there is a correlation between
temperature and ionisation. For the latter, we used the ionisation-sensitive line ratio [SIV]10.5$\mu$m/[SIII]18.7\-$\mu$m.  According to Fig.\ref{fig:T_OI}(a), 
no such correlation holds. Even excluding the literature data, we could not find any correlation.
Fig.\ref{fig:T_OI}(b) shows the [OI]145$\mu$m/63$\mu$m vs the [SIII]33/18.7$\mu$m line ratio, where we can see a trend, 
which reflects into an anti-correlation between electron density and gas temperature, which is also shown for our data in Fig.\ref{fig:dens_7}(a).
The correlation seen in this latter figure disappears with the inclusion of the literature data (three starburst galaxies and two Seyfert 1's 
from observations of ISO-LWS reported in \citet{bra08}) because both the different types of the objects included and the different instrument used 
(see Appendix \ref{app.b}) could introduce a bias. 

\subsection{AGN vs starburst diagnostics}

We have used the Cloudy\footnote{Available at: http://www.nublado.org} photoionisation code (version 10.00), 
last described by \citet{fer98},  to make constant density, constant ionisation parameter models for both AGN NLR 
and starburst emission regions. For the AGN models, we chose a slope of the ionising 
continuum of $\alpha$ = -1.4, with a grid of hydrogen densities 
n$_H$ = 10$^2$, 10$^3$, 10$^4$, 10$^5$ and 10$^6$ cm$^{-3}$ and ionisation parameters of 
U = 10$^{-1.5}$, 10$^{-2.0}$ and 10$^{-2.5}$. 
To allow the models to also reproduce the photodissociation regions (PDR) located outside the 
ionised regions, the integration was allowed to continue until the low temperature of 50K was reached. 

The starburst models adopted the ionising spectrum from the Starburst'99 code \citep{lei99} for continuous 
star formation with an age of 20 $\times$ 10$^6$ yr. The starburst grid spans values of densities of 
n$_H$ = 10$^2$, 10$^3$, 10$^4$, 
10$^5$ and 10$^6$   
cm$^{-3}$ and ionisation parameters of U = 
10$^{-2.0}$,  
10$^{-2.5}$, 
10$^{-3.0}$, 10$^{-3.5}$, 10$^{-4.0}$ 
and 10$^{-4.5}$. 
To be able to directly compare the AGN with the starburst models, 
we have used almost the same parameter grid for the two sets of models, the difference being in the ionisation parameters that extend up to
U=10$^{-1.5}$ for AGN, while the starburst models extend down to U=10$^{-4.5}$. We note that in the following figures (from
\ref{fig:c2_o1a} to \ref{fig:c2_n2vsne3_ne2}) we were not able to plot all models for all values of the two parameters, because some
lines are not produced in some models. 
Also for the starburst models, the integration was continued till the low temperature of 50K was reached
to model the photodissociation regions (PDR) located outside the HII regions. 
However, we have also computed "pure" starburst models, with the stopping temperature set to the default of 1000K,
that will be used to separate the PDR emission from that due to photoionisation. 

We show in the following the predicted
line ratio diagrams, compared with observations of the galaxies 
observed in this work,
the PACS data for MRK463 and HII region galaxies and LINERs \citep{far13},
as well as starburst and Seyfert galaxies from the
ISO-LWS literature \citep{bra08}.

The first sequence of diagrams, Fig. \ref{fig:c2_o1a}(a,b)  and Fig. \ref{fig:c2_o1b}(a,b), show the 
[CII]157$\mu$m/[OI]63$\mu$m line ratio as a function of various ionisation-sensitive line ratios: [OIII]88$\mu$m/[OI]63$\mu$m, 
[OIII]88$\mu$m\-/[NII]122$\mu$m,  [NeII]12.8$\mu$m\-/[NeIII]15.5$\mu$m and [OIII]88$\mu$m/[OIV]26$\mu$m. 
It can be seen that the [CII]157$\mu$m/[OI]63$\mu$m line ratio is mainly sensitive to the gas density. 
Only in Fig. \ref{fig:c2_o1a}(a), because it includes [OI]63$\mu$m also in the x-axis, some ionisation dependence of the 
[CII]157$\mu$m/\-[OI]63$\mu$m line ratio is seen. In general, the combination of these two line ratios is only able to weakly 
separate the AGN and starburst emission.  
AGN and starburst models overlap when we consider a possible high-ionization
starburst model with U=10$^{-2.5}$, which is considered to be the extreme value for energetic starburst galaxies. 

In Fig. \ref{fig:c2_o1a}(b) the [CII]157$\mu$m/\-[OI]63$\mu$m versus [OIII]88$\mu$m/[NII]122$\mu$m line ratio diagram, is
only able to weakly separate, through photoionisation models, some of the Seyfert galaxies
from others which are photoionised purely by starbursts.
The greatest ambiguities occur when starburst models are included with $U$ as large as 10$^{-2.5}$. 
This {\it only partially} confirms the results of the ISO-LWS spectrometer \citep{spi00}. 
It appears therefore that in galaxies where the nucleus dominates over the galactic emission, 
the [OI]63$\mu$m/ [CII]157$\mu$m ratio increases. This latter line ratio is indeed higher in X-ray Dominated Regions 
(XDR), characterised by the presence of an AGN, compared to classical photodissociation regions, at
a given density, column and radiation field strength. This is because in PDRs, the fine-structure lines are only produced 
at the edge of the cloud, whereas the deeply penetrating AGN X-rays in XDRs generate emission  
throughout the entire volume. While in PDRs the [OI]63$\mu$m/ [CII]157$\mu$m ratio 
depends only on density and radiation field, in XDR it depends also on column density. Moreover [OI] emission
dominates over [CII] emission in XDR because carbon does not become fully ionised
\citep{mei07}. 

Fig. \ref{fig:c2_o1b}(a)  shows that the combination of the [CII]157$\mu$m/[OI]63$\mu$m line ratio with the mid-IR line ratios from the neon lines, 
observed with the IRS onboard of {\it Spitzer}, 
is not only able to separate AGN from starburst galaxies, but it also defines a sequence of excitation among the Seyfert nuclei.
This diagram shows most Seyfert 1's in the region of pure AGN models, HBLR galaxies at intermediate locations between AGN and starburst models 
and ''pure" Seyfert 2's towards the starburst models. In these plots most starburst and even LINER galaxies 
(from the observations of \citet{far13}) are well explained by starburst photoionisation models. Fig. \ref{fig:c2_o1b}(b), 
showing the [CII]157$\mu$m/[OI]63$\mu$m  versus the [OIII]88$\mu$m/[OIV]26$\mu$m line ratio, is able to separate, 
by two orders of magnitude on 
its x-axis, AGN from starbursts: in this diagram all detected Seyfert 1's, HBLR and 
Seyfert 2's are close to the AGN grid models, while starburst galaxies and what we considered non-Seyfert nuclei, 
including a few LINERs and ULIRGs, are located at intermediate positions between AGN and starburst models.

Fig. \ref{fig:n2_o1a}(a) shows the [NII]122$\mu$m/[OI]63$\mu$m line ratio, which is mostly sensitive to density, mainly 
because the lines of [NII] and [OI] are characterised by very different critical densities for collisional de-excitation 
(namely n$_{crit}$ $\sim$ 4 $\times$10$^2$ cm$^{-3}$ for [NII]122$\mu$m)  as a function of the [OIII]88\-$\mu$m/[OIV]26$\mu$m ratio. 
The diagram of Fig. \ref{fig:n2_o1b}(a) shows that most Seyfert~2 galaxies have line ratios consistent with a combination of AGN 
and starburst emission, while Seyfert 1's and HBLR's ratios are well reproduced by pure AGN models. 
Fig. \ref{fig:n2_o1a}(b)  
shows the [OI]145$\mu$m/[OIII]88$\mu$m line ratio, which is mostly sensitive to density, again as a function of the 
ionisation-sensitive ratio of [SIII]18$\mu$m/[SIV]10.5$\mu$m. The [OI]145$\mu$m/[OIII]\-88$\mu$m line ratio does not seem to be able 
to separate AGN and starburst emission, while, as seen already, the ionisation sensitive ratio in the x-axis 
can do this.

Fig. \ref{fig:n2_o1b} shows the [NII]122$\mu$m/[NIII]57$\mu$m line ratio {\it vs.} the [OIII]88$\mu$m/[OIV]26$\mu$m ratio. 
As noted above, the [OIII]88$\mu$m/[OIV]26$\mu$m ratio is able to separate AGN versus starburst galaxies, while the 
observed [NII]122$\mu$m/[NIII]57$\mu$m line ratios in most galaxies show an excess compared to photoionisation models, which is higher
in the lower excitation active galaxies (the pure Seyfert 2's and the non-Seyferts).  All observed ratios are offset with respect to any line connecting
the two grids of models, indicating that they cannot be reconciled with any combination of AGN and starburst models. 
However, no conclusion can be drawn from this diagram because of the few data points and also because of the lack of predictions of the 
[OIV]26$\mu$m line for low ionisation starburst models, which prevents their inclusion in the diagram. 
However, the low ionisation parameter (logU=-3.5 and -4) starburst models predict [NII]122$\mu$m/[NIII]57$\mu$m line ratios of $\sim$2.5 and of
$\sim$1.2 for logU=-3.5 and $\sim$20 and $\sim$10 for logU=-4, at densities of $n=10^2$ and $n=10^{3}$ cm$^{-3}$, respectively,
indicating that the high [NII]/[NIII] ratios observed in a few galaxies in Fig. \ref{fig:n2_o1b} are consistent with starburst models. 

Fig. \ref{fig:c2_n2vsne3_ne2}(a) and (b) show the [NeII]12.8$\mu$m\-/[NeIII]15.5$\mu$m ionisation-sensitive line ratio versus 
[CII]157$\mu$m/[NII]122$\mu$m and [CII]157$\mu$m/[NII]205$\mu$m, which are able to quantitatively separate the photoionisation 
and photodissociation regions contributions \citep[e.g.,][]{obe06}. 
In particular, Fig. \ref{fig:c2_n2vsne3_ne2}(a) 
shows that most of the Seyfert galaxies of this study can well be represented in this line ratio diagram by starburst photoionisation models 
which include photodissociation region emission, which is accounted for by Cloudy models, when the integration is continued down to the 
low gas temperature of 50K. In the figure we have included for comparison the ``bare" starburst models which have only
the photoionisation component, but not the photodissociated one, by using the default stopping criterion in Cloudy for HII regions: T=1000K.
As can be seen from the figure, no one object can be reproduced by the photoionised-only models, but the PDR component
is always needed to reproduce the observations. 

Fig. \ref{fig:c2_n2vsne3_ne2}(b) shows the [NeII]12.8$\mu$m\-/[NeIII]15.5$\mu$m ratio as a function of the [CII]157$\mu$m/[NII]\-205$\mu$m ratio. 
Although here we have fewer objects observed in the [NII]205$\mu$m line, we can use this 
figure to measure how much [CII] is due to PDR emission. 
The comparison of the observed  [CII]/[NII] ratios with 
starburst photoionisation models that included PDR emission, which adequately represent the data, 
shows that the [CII] emission due to the PDR component is at least $\sim 4$ times brighter that the photoionised component for most detected galaxies. 

\section {Conclusions and summary}

We have reported here the results of a spectroscopic survey of 26 local Seyfert galaxies in the far-IR ionic 
and neutral  
fine-structure lines 
with the PACS Spectrometer, onboard {\it Herschel}. PACS data have been combined with {\it Herschel}-SPIRE and {\it Spitzer}-IRS
spectroscopy of fine-structure lines at both longer and shorter wavelengths. 
This allowed us to use the line ratios of the ions [SIII], [NeV], [OIII], [NII], 
and the neutral atoms of  
[OI] and [CI] as density estimators of the gas, 
assumed to be in a pure collisional regime. The [OIII]88$\mu$m/52$\mu$m and the [NII]205$\mu$m/122$\mu$m 
and the [CI]609$\mu$m/370$\mu$m line ratios
are confirmed to provide good density estimates 
of ionised and neutral gas, respectively, at densities of n$_{H}$ = 10$^1$-10$^3$ cm$^{-3}$.
The mid-IR line ratios, measured by {\it Spitzer}, [SIII]33$\mu$m/18$\mu$m and the [NeV]24$\mu$m/14$\mu$m, 
are density-sensitive for n$_{H}$ $\leq$ 10$^4$ cm$^{-3}$. 
Using the different line ratios of ionised lines, we have determined the gas densities in the sample galaxies. Because each line ratio
is sensitive to a particular density range, we have also identified in this way density stratifications which are present in 
the same galaxy. We have found  that higher ionisation lines are, on average, emitted from higher density gas.

The [OI]145$\mu$m/63$\mu$m line ratio provides a measure of the gas temperature of the neutral ISM gas, for n$_{H}$ $\leq$ 10$^4$ cm$^{-3}$ 
and it anti-correlates with the [SIII]33$\mu$m/18$\mu$m line ratio. This relation reflects into an anti-correlation between the ionised gas density 
and the neutral ISM gas temperature.

The use of the observed [CII]157$\mu$m/[OI]63$\mu$m line ratio with other far-IR line ratios  
(e.g. [OIII]88$\mu$m/[OI]63$\mu$m, or  [OIII]88$\mu$m/[NII]122$\mu$m), in combination with photoionisation models, is only moderately able 
to discriminate the excitation mechanism 
at the origin of the line emission, between young stars, or black hole accretion.
This separation is better traced by mid-IR line ratios (e.g.  [NeII]12.8$\mu$m/[NeIII]15.5$\mu$m or [SIII]18$\mu$m/[SIV]10.5$\mu$m) 
or mixed mid- and far-IR ratios (e.g. [OIII]88$\mu$m/[OIV]26$\mu$m). 

The [CII]157$\mu$m/[NII]122$\mu$m  and [CII]157$\mu$m/[NII]205$\mu$m line ratios are able to quantify the amount of PDR emission
in a galaxy compared to what is expected by photoionisation. 
We find that most of the [CII] emission arises from photodissociation regions,
while only about one fifth of it is on average produced in pure HII regions.   

\acknowledgments

We thank the anonymous referee for the constructive and exhaustive
report, that helped improving this paper.
We also thank the Herschel Science Center for the help in the reduction and 
the PACS ICC at MPE, Garching, in particular Helmut Feuchtgruber and Eckhard Sturm, for providing the 
calibration for the 52$\mu$m spectroscopic data.
Finally, we thank Nicola Marchili for useful discussions on the statistical analysis and Scige John Liu
for support in the software. 
This work has been funded in Italy from ASI (Italian Space Agency) under contract I/005/11/0.
KMD acknowledges support by a CNES fellowship, and by the European Research Council in the frame of the Advanced 
Grant Program Number 267399-Momentum (PI Combes).
GB is supported by the Spanish MICINN grant AYA2011-30228-C03-01 (co-funded with FEDER funds).
PACS has been developed by a consortium of institutes led
by MPE (Germany) and including UVIE (Austria); KU
Leuven, CSL, IMEC (Belgium); CEA, LAM (France);
MPIA (Germany); INAF-IFSI/OAA/OAP/OAT, LENS,
SISSA (Italy); IAC (Spain). This development has been
supported by the funding agencies BMVIT (Austria),
ESA-PRODEX (Belgium), CEA/CNES (France), DLR
(Germany), ASI/INAF (Italy), and CICYT/MCYT
(Spain). Data presented in this paper were analysed
using The Herschel Interactive Processing Environment
(HIPE), a joint development by the Herschel Science
Ground Segment Consortium, consisting of ESA, the
NASA Herschel Science Center, and the HIFI, PACS
and SPIRE consortia.

\clearpage

\begin{deluxetable}{lcccccc}
\tabletypesize{\footnotesize}
\tablecaption{  Sample of bright Seyfert galaxies observed with the PACS spectrometer \label{tbl-1}} 
\tablehead{ \colhead{Target}& \colhead{R.A.(2000)}  & \colhead{Dec.(2000)} & \colhead{$cz$}  & \colhead{type} &  \colhead{12$\mu$m}  & \colhead{Ref.} \\ 
                      \colhead{}           &   \colhead{(h:m:s)}           & \colhead{(d:m:s)}     &\colhead{km~s$^{-1}$}     &  \colhead{}  &   \colhead{sample} &   \colhead{}   }
\startdata
NGC1056 	& 	02:42:48.3  &   +28:34:27       & 1545  & non-Sy	& \checkmark	& (1) 	\\   
MRK\,1066  		& 02:59:58.6   & +36:49:14     & 3605  &  Sy2/nHBLR &           &  \\  
UGC\,2608  		& 03:15:01.4   & +42:02:09     & 6998  &  Sy2/nHBLR &          &   \\ 
3C120  	               &  04:33:11.1   & +05:21:16     & 9896  & Sy1             &\checkmark	& (1)  \\  
MRK\,3                    &  06:15:36.3  & +71:02:15     &  4050  &  Sy2/HBLR    &           &  \\  
UGC5101 	& 	09:35:51.6  &   +61:21:11      &11802   &	non-Sy	& \checkmark	&              \\ 
NGC3227 	& 	10:23:30.6  &	 +19:51:54    & 1157    & Sy1            & \checkmark	& 	 (2) 	\\ 
NGC\,3516             & 11:06:47.5	& +72:34:07  &	2649  &   Sy1                &  \checkmark	&  (3)   \\  
NGC3982 	& 	11:56:28.1  &	 +55:07:31     & 1109   & Sy2/nHBLR  & \checkmark	& 	 (1)	\\  
NGC4051 	& 	12:03:09.6  &	 +44:31:53     &  700    & Sy1           & \checkmark	& 	 (1)	\\  
NGC4151 	& 	 12:10:32.6  &	 +39:24:21     &   995   & Sy1           & \checkmark	& 	 (2) 	\\  
NGC4388 	& 	12:25:46.7  &	 +12:39:44     & 2524   & Sy2/HBLR & \checkmark	& 	 (1) 	\\   
NGC\,4507  	&       12:35:36.6   & --39:54:33       & 3538    &  Sy2/HBLR &           &    \\   
TOL1238-364 (IC3639) & 	12:40:52.8 & --36:45:21  & 3275   & Sy2/HBLR & \checkmark	& 	 (3) 	\\   
NGC\,5256  (Mrk266) & 13:38:17.5 & +48:16:37    & 8353    &  Sy2/nHBLR & \checkmark	&     (1)     \\ 
IRAS13451+1232 	  & 13:47:33.3  &   +12:17:24 & 36497  &  Sy2/HBLR &              &  \\ 
IC\,4329A  	          & 13:49:19.2  & --30:18:34    & 4813    & Sy1             &  \checkmark	&         (1)   \\
NGC\,5506                  & 14:13:14.9  & --03:12:27      &1853   & Sy2/nHBLR & \checkmark	& (1)   \\ 
NGC\,5728 		& 14:42:23.9   & --17:15:11      & 2804   & Sy2/nHBLR &            &  \\ 
IRAS18216+6418       & 18:21:57.3  &  +64:20:36  & 89038  & Sy1             &            & \\
ESO103-G035          & 18:38:20.3   & --65:25:39       & 3983 & Sy2/nHBLR &             &    \\
ESO140-G043          & 18:44:54.0    & --62:21:53      & 4250 & Sy1             &              &   \\ 
MRK\,509            	  &  20:44:09.7	& --10:43:25 &  10312  & Sy1                &\checkmark	& (1)     \\   
NGC7130 	& 	21:48:19.5  &	  --34:57:05    & 4842  &  non-Sy      & \checkmark	& 	(1) 	\\   
NGC7172 	& 	22:02:01.9  &	  --31:52:11    & 2603  & Sy2/nHBLR & \checkmark	& 	(1) 	\\  
NGC7582 	& 	23:18:23.5  &	  --42:22:14    & 1575  & Sy2/nHBLR & \checkmark	& 	(1)	\\  
\enddata
\tablecomments{References for type: (1): \citet{tom10};  (2): unpublished data; (3): \citet{tom08}; }
\end{deluxetable}

\begin{table}
\caption{Fine-structure lines observed with the PACS spectrometer}
\begin{center}
\label{tbl-2}
\tabletypesize{\footnotesize}
\begin{tabular}{lcccc}
\hline
\hline
Line &$\lambda$ &$\nu$   &Spec. Resolution     &Angular Resolution   \\
     &(\mum)    &(GHz)   &(\kms)               &(arcsec) \\  
\hline
[OIII]$^3$P$_2$--$^3$P$_1$ &  \phn51.81 & 5787.57 &$\sim$105		    &\phn9.7$^{\rm{1}}$ \\  

[NIII]$^2$P$_{3/2}$--$^2$P$_{1/2}$  &\phn57.32           &5230.43    &$\sim$105		    &\phn9.7 \\ 

[OI]$^3$P$_1$--$^3$P$_2$   &\phn63.18 		&4744.77    &$\sim\phn86$   &\phn9.7 \\

[OIII]$^3$P$_1$--$^3$P$_0$ 		  &\phn88.36 	&3393.01    &$\sim124$      &\phn9.7 \\

[NII]$^3$P$_2$--$^3$P$_1$   &121.90    		&2459.38    &$\sim290$      &\phn9.7 \\

[OI]$^3$P$_0$--$^3$P$_1$   &145.52    		&2060.07    &$\sim256$      &10.3 \\

[CII]$^2$P$_{3/2}$--$^2$P$_{1/2}$ 	& 157.74	&1900.54    &$\sim238$      &11.2 \\

\hline
\end{tabular}

{Notes: $^{\rm{1}}$: beam size is dominated by the spaxel size (9.7") below  $\sim$120$\mu$m} 
\end{center}
\end{table}

\clearpage
\LongTables
\begin{landscape}
\begin{deluxetable}{lcccccccccc}
\tabletypesize{\scriptsize} 
\tablecaption{Far-infrared Fine Structure Lines in our Sample from {\it Herschel} PACS  and SPIRE observations \label{tbl-3}}\tablewidth{0pt} 
\tablehead{      & \multicolumn{8}{c}{Line fluxes (10$^{-17}~W~m^{-2}$)}\\
\colhead{NAME} & \colhead{[OIII]} & \colhead{[NIII]} & \colhead{[OI] } & \colhead{[OIII]}& \colhead{[NII]} & \colhead{[OI]} & \colhead{[CII]} & \colhead{[NII]$^{\rm{1}}$}  & \colhead{[CI]$^{\rm{1}}$} & \colhead{[CI]$^{\rm{1}}$}\\
                             &(51.81$\mu m$) &(57.32$\mu m$) &(63.18$\mu m$)&(88.36$\mu m$)&(121.90$\mu m$)&(145.52$\mu m$)&(157.74$\mu m$)& (205.18$\mu m$) &(370.52$\mu m$)& (609.31$\mu m$) }
\startdata
NGC1056 &     \nodata &  8.32$\pm$0.64  &   55.52$\pm$1.56   &  41.47$\pm$0.96   & 9.22$\pm$0.63   &  7.44$\pm$0.38   &  213.1$\pm$0.86   &  5.05$\pm$0.34   & 0.77$\pm$0.15 & $<$1.22  \\
MRK1066 &     \nodata &      \nodata         &   136.82$\pm$2.35 &  30.42$\pm$0.84    & 11.20$\pm$0.36 &  9.26$\pm$0.28   & 109.72$\pm$1.18  &    \nodata              & \nodata          & \nodata     \\
UGC2608 & 31.66$\pm$2.72 & \nodata &    77.65$\pm$1.50 &   64.59$\pm$0.75  &	\nodata &    \nodata	 & 165.3$\pm$0.85    & 4.66$\pm$0.32$^{\rm{2}}$ & 1.36$\pm$0.02$^{\rm{2}}$& $<$2.48$^{\rm{2}}$\\
3C120   & 46.02$\pm$3.33 &      \nodata &    21.41$\pm$1.39 &   73.29$\pm$1.37  & 1.56$\pm$0.45  &  \nodata	 &  39.51$\pm$0.54  &  \nodata  & \nodata          & \nodata \\
MRK3    & 54.4$\pm$7.16 & 15.12$\pm$2.68&   124.08$\pm$2.41 &  55.05$\pm$0.76   & 6.50$\pm$0.38  &  7.64$\pm$0.12   &  45.42$\pm$0.34 & \nodata  & \nodata          & \nodata \\
UGC5101 &     \nodata &     5.89$\pm$0.41 &   19.09$\pm$1.22  &  14.91$\pm$0.96   & 12.07$\pm$0.73 &  1.16$\pm$0.16   &  78.78$\pm$0.85  & 3.79.$\pm$0.44             & 1.58$\pm$0.21 & $<$1.6  \\
NGC3227 &     \nodata &  $<$3.00     &    133.1$\pm$1.34 &    22.08$\pm$1.06 & 10.2$\pm$0.99  &  9.88$\pm$0.37   &  130.81$\pm$0.79 & 3.86$\pm$0.43              &  4.66$\pm$0.19 &  2.27$\pm$0.43 \\
NGC3516 &     \nodata &      \nodata &    23.10$\pm$1.93 &    22.13$\pm$1.18 & 4.14$\pm$0.53  & \nodata	 &  20.72$\pm$0.53  &  \nodata & \nodata          & \nodata  \\
NGC3982 &     \nodata &   $<$4.33   &    52.12$\pm$1.22 &   15.94$\pm$0.77  & 17.21$\pm$0.68 &  4.94$\pm$0.34	 & 239.88$\pm$0.91 & 8.81$\pm$0.37                          & 1.0$\pm$0.15   & $<$1.32    \\
NGC4051 &     \nodata &    $<$1.50   &    26.80$\pm$0.72 &    6.22$\pm$0.60  & 2.93$\pm$0.18  &  2.07$\pm$0.12   &  50.09$\pm$1.42  & 2.50$\pm$0.43       &   0.78$\pm$0.13        & $<$1.13     \\
NGC4151   &	40.85$\pm$6.30  &  12.78$\pm$0.85 &   325.83$\pm$1.68 &   45.0$\pm$0.92  &  6.37$\pm$0.22  &  18.25$\pm$0.19 &   66.53$\pm$0.52 & 2.12$\pm$0.31       &   0.84$\pm$0.10        & $<$0.69    \\
NGC4388 &     \nodata & 26.01$\pm$1.66 &    199.69$\pm$1.50 &     127.96$\pm$1.30 & 7.34$\pm$0.65 & 15.08$\pm$0.34   & 245.51$\pm$0.86 & 4.74$\pm$0.19 &  2.67$\pm$0.07  & 1.45$\pm$0.2 \\
NGC4507 &     \nodata &      \nodata &   128.31$\pm$2.03 &  22.51$\pm$0.78   & 4.64$\pm$0.32  &  8.81$\pm$0.37   &  64.36$\pm$0.45   &  \nodata   & \nodata          & \nodata  \\
IC3639&     \nodata &   $<$2.35   &    82.80$\pm$1.43 &   40.52$\pm$0.67  & 7.16$\pm$0.59  &  8.57$\pm$0.40   & 183.66$\pm$1.26  & 4.55$\pm$0.21 &   0.58$\pm$0.07  &  $<$1.17 \\
NGC5256 & 48.37$\pm$2.89&      \nodata &   91.26$\pm$1.69 &   68.01$\pm$1.27 &	 11.56$\pm$0.44 &  9.25$\pm$0.56  & 158.72$\pm$0.68 &  \nodata     & \nodata          & \nodata\\
IRAS 13451+1232 &  \nodata &      \nodata &   10.39$\pm$0.84 & $<$3.05   &	 	\nodata &  0.42$\pm$0.06  &  5.61$\pm$0.24   &  \nodata   & \nodata          & \nodata\\
IC4329A &  19.07$\pm$2.81 & 4.66 $\pm$0.51 &  	39.96$\pm$1.38   &   25.77$\pm$0.84   &  2.25$\pm$0.19  &  2.92$\pm$0.12  & 17.0$\pm$0.41    &    \nodata   &   \nodata  &   \nodata    \\
NGC5506 & 55.05$\pm$5.85& 26.21$\pm$0.84 &   150.16$\pm$1.41 &  94.75$\pm$0.88   & 10.54$\pm$0.21 & 11.98$\pm$0.19   & 110.82$\pm$0.58 &  \nodata  & \nodata          & \nodata \\
NGC5728 &     \nodata &      \nodata &   112.82$\pm$1.95 &   68.5$\pm$1.26  & 15.49$\pm$0.47 &  12.91$\pm$0.53   & 136.9$\pm$0.48  & \nodata    & \nodata          & \nodata   \\ 
IRAS18216+6418    &     \nodata &      \nodata &    13.14$\pm$1.01 &    11.64$\pm$0.38 &	$<$0.75 & 	$<$2.10 &   $<$2.05 	  & \nodata      & \nodata          & \nodata   \\   
ESO103-G035  &     \nodata &      \nodata &    48.3$\pm$2.19 &   7.84$\pm$0.59   & 0.42$\pm$0.06  &  2.57$\pm$0.22   &   6.76$\pm$0.46 &  \nodata  & \nodata          & \nodata \\
ESO140-G043  &     \nodata &      \nodata &    27.06$\pm$1.30 &   11.27$\pm$0.78  & 1.96$\pm$0.24  &  0.96$\pm$0.08   &  19.26$\pm$0.34 &  \nodata  & \nodata          & \nodata \\
MRK509  &     \nodata &      \nodata &    26.44$\pm$1.52 &    15.94$\pm$0.58 & \nodata  &  1.84$\pm$0.20   &  33.72$\pm$0.63  & \nodata      & \nodata          & \nodata  \\
NGC7130 &     \nodata &  19.20$\pm$2.35 &   95.46$\pm$1.21  &  21.98$\pm$0.87   & 31.62$\pm$0.70 &  11.01$\pm$0.51   & 258.77$\pm$1.31 & 12.3$\pm$0.18     & 2.82$\pm$0.10  &  1.79$\pm$0.22 \\
NGC7172 &     \nodata & 6.30$\pm$0.70 &    32.42$\pm$1.46 &   27.43$\pm$1.31  & 23.24$\pm$0.61 &  4.83$\pm$0.40   & 173.40$\pm$0.78  & 12.0$\pm$0.41   & 4.5$\pm$0.4   &   3.82$\pm$0.52 \\
NGC7582 &  108.1$\pm$30.40     & 90.97$\pm$2.57 &   294.14$\pm$1.93 &  187.68$\pm$1.48 & 63.37$\pm$0.59 & 22.55$\pm$0.29   & 486.6$\pm$1.27 & 19.2$\pm$0.54 & 7.59$\pm$0.20  & 3.44$\pm$0.28 \\
\enddata
\tablenotetext{*}{Notes: $^{\rm{1}}$: SPIRE flux from \citet{p-s13}; 
$^{\rm{2}}$: SPIRE fluxes from archive observations reduced in this work.} 
\end{deluxetable}
\clearpage
\end{landscape}

\clearpage
\LongTables
\begin{landscape}
\begin{deluxetable}{lcccccccccc}
\tabletypesize{\scriptsize}
\tablecaption{Mid-infrared Fine Structure Lines in our Sample from {\it Spitzer} IRS High resolution observations \label{tbl-4}}\tablewidth{0pt} 
\tablehead{   
&  \multicolumn{5}{c}{Line fluxes (10$^{-17}~W~m^{-2}$) in SH (4.7\arcsec $\times$ 11.3\arcsec)} & \multicolumn{4}{c}{Line fluxes (10$^{-17}~W~m^{-2}$) in LH (11.1\arcsec  $\times$ 22.3\arcsec )} & Ref.\\
&  \cline{1-5}  \cline{7-10} & \\
\colhead{NAME} & \colhead{[SIV]} & \colhead{[NeII]}  & \colhead{[NeV] } & \colhead{[NeIII]} & \colhead{[SIII]} & \colhead{[NeV]} & \colhead{[OIV]} & \colhead{[SIII]}   & \colhead{[SiII]}  & \colhead{}\\
                               &(10.51$\mu m$) &(12.81$\mu m$) &(14.32$\mu m$)  &(15.56$\mu m$)  &(18.71$\mu m$)&(24.32$\mu m$)  &(25.89$\mu$m)& (33.48$\mu$m)&(34.82$\mu$m) &                 \\}
\startdata
NGC1056         & $<$1.31               &  33.6$\pm$1.04   &   $<$1.80            &  10.4$\pm$0.34 & 18.3$\pm$0.35 &    $<$1.23            &  1.40$\pm$0.30  &  36.6$\pm$0.69  &  49.1$\pm$0.97 & (1) \\ 
MRK1066        & 9.91$\pm$2.02  & 101.1$\pm$8.93   &  7.08$\pm$1.32  &   46.4$\pm$3.18  & 45.3$\pm$4.23 &   8.18$\pm$1.95 &   38.1$\pm$2.54  &   62.7$\pm$12.2  &   93.8$\pm$14.4 & (2) \\
UGC2608        & 28.74$\pm$2.07 &  58.27$\pm$4.91 & 30.33$\pm$2.13 & 69.46$\pm$4.52 & 28.50$\pm$2.62 &  32.18$\pm$1.42 &  134.37$\pm$8.39 &  46.61$\pm$5.89 & 83.19$\pm$13.26 &(2)\\
3C120	       & 24.1$\pm$0.86    & 7.84 $\pm$0.63  & 16.6$\pm$0.89  & 27.6$\pm$0.93  & 7.34$\pm$1.48 &   29$\pm$0.62    &123$\pm$0.76    & 17.3$\pm$1.93   &  36.3$\pm$3.02  & (1) \\
MRK3                & 55.9$\pm$2.92   &  95.30$\pm$7.47   &  63.8$\pm$3.22  & 174.$\pm$11.7  &  52.9$\pm$4.09  &  60.3$\pm$3.87 &   178.$\pm$15.7  &   57.6 $\pm$10.5  &   85.0$\pm$6.7 & (2) \\
UGC5101        &1.15$\pm$0.450  &   34.74$\pm$2.61 &    2.94$\pm$0.41  &   13.4$\pm$1.49  &   10.6$\pm$2.97 &   2.72$\pm$1.02 &   7.00$\pm$1.43   &  12.5$\pm$3.80    & \nodata  & (2) \\
NGC3227        & 25.6$\pm$0.33  & 74.3$\pm$0.86   &  22.6$\pm$0.38  & 74.4$\pm$0.43   & 24.4$\pm$0.57 & 16.0$\pm$0.45    & 61.1$\pm$0.38 & 25.9$\pm$1.33    & 55.9$\pm$1.60   & (3) \\
NGC3516       & 13.33$\pm$0.38 & 8.07$\pm$0.25 &   7.88$\pm$0.50  & 17.72$\pm$0.33 & 5.86$\pm$0.35  & 10.39$\pm$0.33  &  46.92$\pm$0.35  &  9.52$\pm$0.96  & 22.14$\pm$0.54 & (4) \\ 
NGC3982         & 1.48$\pm$0.33 &  11.4$\pm$0.34   &   2.89$\pm$0.31  &  6.79$\pm$0.30 & 3.16$\pm$0.39  &  1.62$\pm$0.30  &  5.11$\pm$0.40 &  15.4$\pm$0.61 & 32.8$\pm$1.07 & (1) \\
NGC4051         & 4.75$\pm$0.38 &   21.2$\pm$0.31   &  10.7$\pm$0.35  &  17.1$\pm$0.36 &  7.45$\pm$0.49 &  32.2$\pm$0.21  &  33.7$\pm$0.52  &  38.8$\pm$3.93 & 39.6$\pm$1.01 & (1,5) \\
NGC4151         & 84.64$\pm$4.37 &	132.9$\pm$7.78  & 76.96$\pm$	4.10	& 205.3$\pm$9.72  & 69.48$\pm$5.55 & 68.1$\pm$3.97 & 243.6$\pm$13.26 & 61.40$\pm$9.17 & 132.9$\pm$4.12 & (2) \\
NGC4388         & 45.3$\pm$0.76  &  76.6$\pm$0.97    & 46.1$\pm$0.86   & 106$\pm$1.11  &  39.1$\pm$1.19 & 73.0$\pm$1.05   &  340$\pm$2.15  & 85.1$\pm$2.78  & 135$\pm$2.76  & (1) \\
NGC4507        &8.35$\pm$1.63 &    30.55$\pm$1.84  &   12.5$\pm$2.93  &   28.5$\pm$2.21  &   16.1$\pm$2.85  &  8.39$\pm$1.73  &  33.8$\pm$4.53  &   23.6$\pm$6.31  &   45.6$\pm$8.21 & (2) \\
IC3639             & 5.70$\pm$0.28 &  45.15$\pm$0.54 & 11.15$\pm$0.61 & 27.00$\pm$0.44 & 16.32$\pm$0.48  &  5.35$\pm$0.67 &  21.21$\pm$0.58  & 32.80$\pm$1.62  & 44.99$\pm$1.62 & (4)\\ 
NGC5256       & 2.57$\pm$0.25   & 19.8$\pm$0.20   &  2.31$\pm$0.15   &  10.6$\pm$0.23  &  8.38$\pm$0.39   &  11.9$\pm$0.31 &  56.8$\pm$0.76  &  48.2$\pm$1.09  & 92.3$\pm$1.46 & (1) \\
IRAS13451+1232   &     $<$0.26 &   4.34$\pm$0.57  &  $<$0.81              &  3.10$\pm$0.65   &      $<$1.89          &    $<$1.31             &      $<$0.60          &            \nodata      &          \nodata        & (2)  \\
IC4329A         & 29.1$\pm$1.32   &  27.6$\pm$0.73   &  29.3$\pm$0.88  &  57.0$\pm$0.97  & 15.0$\pm$1.44    &  34.6$\pm$0.85  &  117$\pm$1.42   &  16.0$\pm$2.19  &  32.5$\pm$3.06 & (1) \\   
NGC5506      & 25.4$\pm$0.32    &  26.4$\pm$0.40   & 18.5$\pm$0.28   &  45.6$\pm$0.24  & 19.1$\pm$0.50    &  56.5$\pm$0.50  & 239$\pm$0.87 &  90.1$\pm$1.71  &137$\pm$2.40 & (1) \\   
NGC5728     &  28.9$\pm$3.63  &   28.33$\pm$1.61 &    21.2$\pm$1.28  &   51.7$\pm$1.44 &    23.4$\pm$1.71 &   29.4$\pm$1.76  &  113.$\pm$13.2  &   49.1$\pm$7.44  &   67.3$\pm$9.94 & (2) \\
IRAS18216+6418 & 5.67$\pm$0.32 & 3.34$\pm$0.32 &  4.63$\pm$0.58 &  10.03$\pm$1.75 & 3.35$\pm$0.51 &  4.58$\pm$1.13  &  23.04$\pm$2.58  &            \nodata      &          \nodata        &(2)  \\
ESO103-G035  &  12.0$\pm$2.05  &   29.23$\pm$2.53  &   16.7$\pm$2.41  &   39.0$\pm$1.60  &   8.01$\pm$2.24  &  9.91$\pm$1.10  &  33.3$\pm$2.29   &  $<$4.52  &     $<$4.18     & (2) \\
ESO140-G043  &  8.39$\pm$0.78  &   9.42$\pm$0.54   &  7.05$\pm$0.65  &   13.8$\pm$0.58 &    7.92$\pm$0.95  &  7.07$\pm$0.51 &   23.5$\pm$1.17  &   10.8$\pm$3.05  &   10.9$\pm$1.80 & (2) \\
MRK509		& 4.13$\pm$0.76    &  14.0$\pm$0.79   &   4.74$\pm$0.59  &  14.5$\pm$0.60 & 7.19$\pm$1.20  &  6.82$\pm$0.46 &  27.5$\pm$0.57  &  7.41$\pm$1.91  & 14.5$\pm$3.41  & (1) \\
NGC7130		& 5.27$\pm$0.84    &  79.3$\pm$0.93   &   9.09$\pm$0.64  &  29.4$\pm$0.77 & 19.6$\pm$0.33  &  5.22$\pm$0.86  &  19.7$\pm$0.84 &  48.2$\pm$2.59  & 93.9$\pm$4.90   & (1) \\
NGC7172         & 5.87$\pm$0.61    &  33.0$\pm$1.01    & 10.2$\pm$0.67   & 17.1$\pm$0.68  &  11.9$\pm$1.00  & 13.8$\pm$0.49  & 45.4$\pm$0.48   &  26.9$\pm$1.51  &    59.3$\pm$2.42 & (1) \\
NGC7582         & 21.3$\pm$1.43    &  322$\pm$6.41     & 38.8$\pm$1.54   & 105$\pm$2.05   &  87.3$\pm$1.99  & 63.6$\pm$4.29  &  262$\pm$5.54   &  244$\pm$7.85   &        \nodata          & (1)\\    
\enddata
\tablenotetext{*}{References: (1): \citet{tom10}; (2): this work, data reduced from the {\it Spitzer} archive; (3): unpublished data; (4): \citet{tom08}; (5): \citet{das08} for the [OIV]26$\mu$m line flux.}
\end{deluxetable}
\clearpage

\end{landscape}

\LongTables
\begin{landscape}
\begin{deluxetable}{lccccccccc}
\tabletypesize{\scriptsize} 
\tablecaption{Density determinations from Fine Structure Lines from {\it Herschel} PACS  and SPIRE and {\it Spitzer} IRS observations \label{tbl-5}}\tablewidth{0pt} 
\tablehead{ \colhead{NAME} & \colhead{class} & \colhead{[OIII]} & \colhead{log(n$_{e}$) (range)$^{\rm{1}}$}&  \colhead{[SIII]}& \colhead{log(n$_{e}$)(range)} & \colhead{[NII]} & \colhead{log(n$_{e}$)(range)} & \colhead{[NeV]}  & \colhead{log(n$_{e}$)(range)}  \\
                           &                           & 88.36/51.81 &   (cm$^{-3}$) & 33.48/18.71 & (cm$^{-3}$) & 205.18/121.90 &  (cm$^{-3}$) & 24.32/14.32 &   (cm$^{-3}$) }
\startdata
NGC1056		& no-S	& \nodata       & \nodata   \nodata    & 2.00$\pm$0.08 & $\leq$1.$^{\rm{2}}$\nodata   & 0.54$\pm$0.07& 1.65     (1.55-1.76) &  \nodata	     & \nodata   \nodata  \\  
MRK1066		& Sy2	& \nodata       & \nodata   \nodata    & 1.38$\pm$0.40 &  2.34 (1.-2.85)   &  \nodata     & \nodata   \nodata    &  1.15$\pm$0.49  & 2.50    (2.-3.59) \\ 
UGC2608		& Sy2	& 2.04$\pm$0.20 & $\leq$0.\nodata   & 1.63$\pm$0.36 &  1.65 (1.-2.51)   &  \nodata     & \nodata   \nodata    &  1.06$\pm$0.12  & 2.85    (2.31-3.13) \\ 
3C120		& Sy1	& 1.59$\pm$0.14 & 1.38     (0.59-1.68) & 2.36$\pm$0.74 & $\leq$1.(0.-1.70) &  \nodata     & \nodata   \nodata    &  1.75$\pm$0.13  & $\leq$2.	 \nodata  \\ 
MRK3		& HBLR  & 1.01$\pm$0.15 & 2.24     (2.07-2.39) & 1.09$\pm$0.28 &  2.73 (2.36-3.03) &  \nodata     & \nodata.  \nodata    &  0.95$\pm$0.11  & 3.11    (2.85-3.31) \\ 
UGC5101		& no-S	& \nodata       & \nodata   \nodata    & 1.18$\pm$0.69 &  2.62 (1.-3.41)   & 0.31$\pm$0.05&  2.07    (1.96-2.20) &  0.93$\pm$0.48  & 3.15    (2.-3.94) \\ 
NGC3227		& Sy1	& \nodata       & \nodata   \nodata    & 1.06$\pm$0.08 &  2.76 (2.67-2.85) & 0.38$\pm$0.08&  1.92    (1.77-2.09) &  0.71$\pm$0.03  & 3.52    (3.47-3.56) \\ 
NGC3516		& Sy1	& \nodata       & \nodata   \nodata    & 1.62$\pm$0.26 &  1.70 (1.-2.38)   &  \nodata     & \nodata   \nodata    &  1.32$\pm$0.13  & $\leq$2.(1.-2.22) \\ 
NGC3982		& Sy2	& \nodata       & \nodata   \nodata    & 1.82$\pm$0.19 & $\leq$1.(0.-1.65) & 0.51$\pm$0.04&  1.69    (1.63-1.76) &  0.56$\pm$0.16  & 3.75    (3.50-4.03) \\ 
NGC4051		& Sy1	& \nodata       & \nodata   \nodata    & 1.77$\pm$0.66 & $\leq$1.(0.-2.70) & 0.85$\pm$0.20&  1.23    (0.95-1.50) &  0.90$\pm$0.16  & 3.20    (2.85-3.47) \\ 
NGC4151		& Sy1	& 1.10$\pm$0.19 & 2.14     (1.92-2.34) & 0.88$\pm$0.20 &  2.95 (2.74-3.17) & 0.33$\pm$0.06&  2.02    (1.90-2.17) &  0.88$\pm$0.10  & 3.24    (3.04-3.41) \\ 
NGC4388		& HBLR  & \nodata       & \nodata   \nodata    & 2.18$\pm$0.14 &  $\leq$1.\nodata  & 0.65$\pm$0.08&  1.50    (1.39-1.61) &  1.58$\pm$0.05  & $\leq$2.	 \nodata  \\ 
NGC4507		& HBLR  & \nodata       & \nodata   \nodata    & 1.47$\pm$0.65 &  2.18 (1.-3.02)   &  \nodata     & \nodata   \nodata    &  0.67$\pm$0.30  & 3.58    (3.07-4.10) \\   
IC3639		& HBLR  & \nodata       & \nodata   \nodata    & 2.01$\pm$0.16 &  $\leq$1.\nodata  & 0.64$\pm$0.08&  1.51    (1.40-1.62) &  0.48$\pm$0.09  & 3.88    (3.73-4.05) \\ 
NGC5256		& Sy2	& 1.41$\pm$0.11 & 1.75     (1.55-1.91) & 2.09$\pm$0.18 &  $\leq$1.\nodata  &  \nodata     & \nodata   \nodata    &  1.40$\pm$0.10  & $\leq$2.	 \nodata  \\ 
IC4329A		& Sy1	& 1.35$\pm$0.24 & 1.84     (1.38-2.13) & 1.07$\pm$0.25 &  2.75 (2.44-3.02) &  \nodata     & \nodata   \nodata    &  1.18$\pm$0.06  & 2.31    (2.-2.64) \\ 
NGC5506		& Sy2	& 1.72$\pm$0.20 & 0.72     (0.-1.55)   & 1.49$\pm$0.36 &  2.13 (1.-2.68)   &  \nodata     & \nodata   \nodata    &  1.11$\pm$0.06  & 2.68    (2.38-2.87) \\ 
NGC5728		& Sy2	& \nodata       & \nodata   \nodata    & 2.10$\pm$0.47 & $\leq$1.(0.-1.65) &  \nodata     & \nodata   \nodata    &  1.39$\pm$0.17  & $\leq$2.(1.-1.78) \\ 
IRAS18216+6418	& Sy1	& \nodata       & \nodata   \nodata    & \nodata       & \nodata  \nodata  &  \nodata     & \nodata   \nodata    &  0.99$\pm$0.37  & 3.02    (2.-3.66) \\ 
ESO103-G035	& Sy2	& \nodata       & \nodata   \nodata    & 0.66$\pm$0.41 &  3.20 (2.75-3.87) &  \nodata     & \nodata   \nodata    &  0.59$\pm$0.15  & 3.70    (3.47-3.96) \\ 
ESO140-G043	& Sy1	& \nodata       & \nodata   \nodata    & 1.36$\pm$0.55 &  2.38 (1.-3.03)   &  \nodata     & \nodata   \nodata    &  1.00$\pm$0.16  & 3.00    (2.44-3.31) \\ 
MRK509		& Sy1	& \nodata       & \nodata   \nodata    & 1.03$\pm$0.44 &  2.79 (2.18-3.28) &  \nodata     & \nodata   \nodata    &  1.44$\pm$0.28  & $\leq$2.(1.-2.44) \\ 
NGC7130		& no-S	& \nodata       & \nodata   \nodata    & 2.46$\pm$0.17 &  $\leq$1. \nodata & 0.39$\pm$0.01&  1.90    (1.88-1.92) &  0.57$\pm$0.13  & 3.73    (3.53-3.96) \\ 
NGC7172		& Sy2	& \nodata       & \nodata   \nodata    & 2.26$\pm$0.32 &  $\leq$1. \nodata & 0.52$\pm$0.03&  1.68    (1.63-1.72) &  1.35$\pm$0.14  & $\leq$2.(1.-1.98) \\ 
NGC7582		& Sy2	& \nodata       & \nodata   \nodata    & 2.79$\pm$0.15 &  $\leq$1. \nodata & 0.30$\pm$0.01&  2.09    (2.07-2.12) &  1.64$\pm$0.18  & $\leq$2.\nodata  \\ \enddata
\tablenotetext{*}{$^{\rm{1}}$: in brackets are given the ranges in density due to the uncertainties of the line ratio; where the density value is an upper limit, then the lower value in the range is set to one decade below the upper limit value;
$^{\rm{2}}$: upper limits indicate that the average density is at or below the low density limit for that line ratio, in these cases we assumed the logarithmic value of the density to be: log(n$_{e}$)=0., 1. 
and 2. for the [OIII], the [SIII] and [NeV] ratios, respectively.
} 
\end{deluxetable}
\clearpage
\end{landscape}

\clearpage

\begin{table}
\caption{Temperature determinations from the [OI]145$\mu$m/63$\mu$m ratio observed from {\it Herschel} PACS  observations \label{tbl-6}}
\begin{center}
\tabletypesize{\footnotesize}
\begin{tabular}{lcc}
\hline
\hline
NAME & class & Derived Temperature \\
           &          &  (K)                              \\
\hline
NGC1056		& no-S	&   \nodata \\  
MRK1066		& Sy2	&   93.7$\pm$6.5  \\ 
UGC2608		& Sy2	&   \nodata \\  
3C120		& Sy1	&   \nodata \\  
MRK3		& HBLR   &   82.8$\pm$3.6  \\ 
UGC5101		& no-S	&  81.5$\pm$22.8  \\ 
NGC3227		& Sy1	& 106.5$\pm$6.6  \\ 
NGC3516		& Sy1	&  \nodata    \\ 
NGC3982		& Sy2	& 154.3$\pm$32.1  \\ 
NGC4051		& Sy1	& 112.1$\pm$3.4  \\ 
NGC4151		& Sy1	& 74.6$\pm$1.2  \\ 
NGC4388		& HBLR    & 108.9$\pm$4.2   \\ 
NGC4507		& HBLR    & 95.7$\pm$7.7  \\   
IC3639		& HBLR    & 186.3$\pm$33.7  \\ 
NGC5256		& Sy2	& 177.9$\pm$37.2   \\ 
IRAS13451+1232 & HBLR  &  56.6$\pm$9.7    \\    
IC4329A		& Sy1	& 104.3$\pm$10.3  \\ 
NGC5506		& Sy2	& 116.9$\pm$3.9  \\ 
NGC5728		& Sy2	& 247.8$\pm$68.1  \\ 
IRAS18216+6418 & Sy1   & \nodata    \\ 
ESO103-G035	& Sy2	& 70.7$\pm$9.7  \\ 
ESO140-G043	& Sy1	& 51.8$\pm$4.5  \\ 
MRK509		& Sy1	& 97.6$\pm$21.8  \\ 
NGC7130		& no-S	& 255.3$\pm$76.1  \\ 
NGC7172		& Sy2	& \nodata \\ 
NGC7582		& Sy2	&111.0$\pm$2.8 \\ 
\hline
\end{tabular}
\end{center}
\end{table}



\begin{figure*}
\includegraphics[width=8cm]{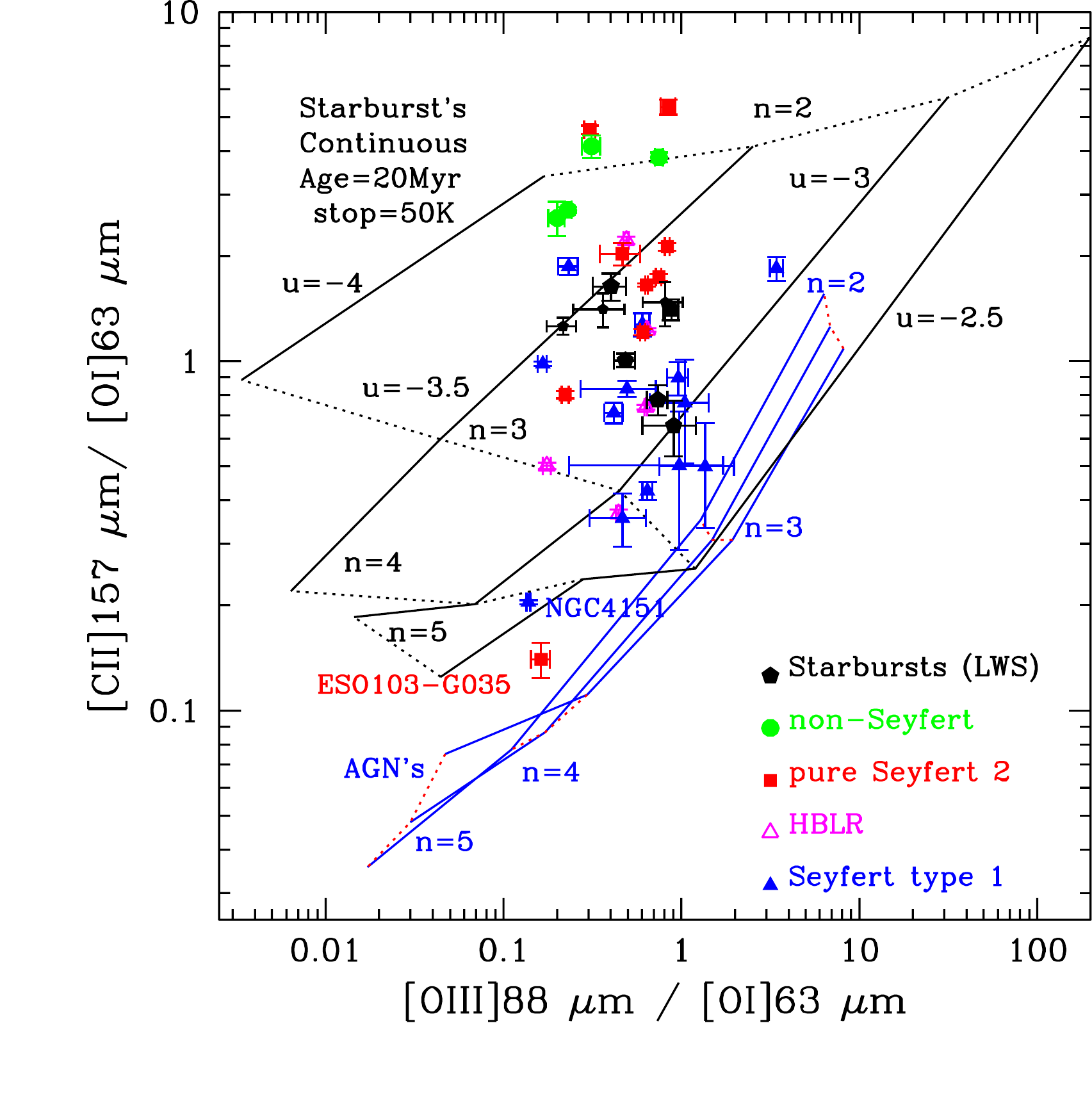}\includegraphics[width=8cm]{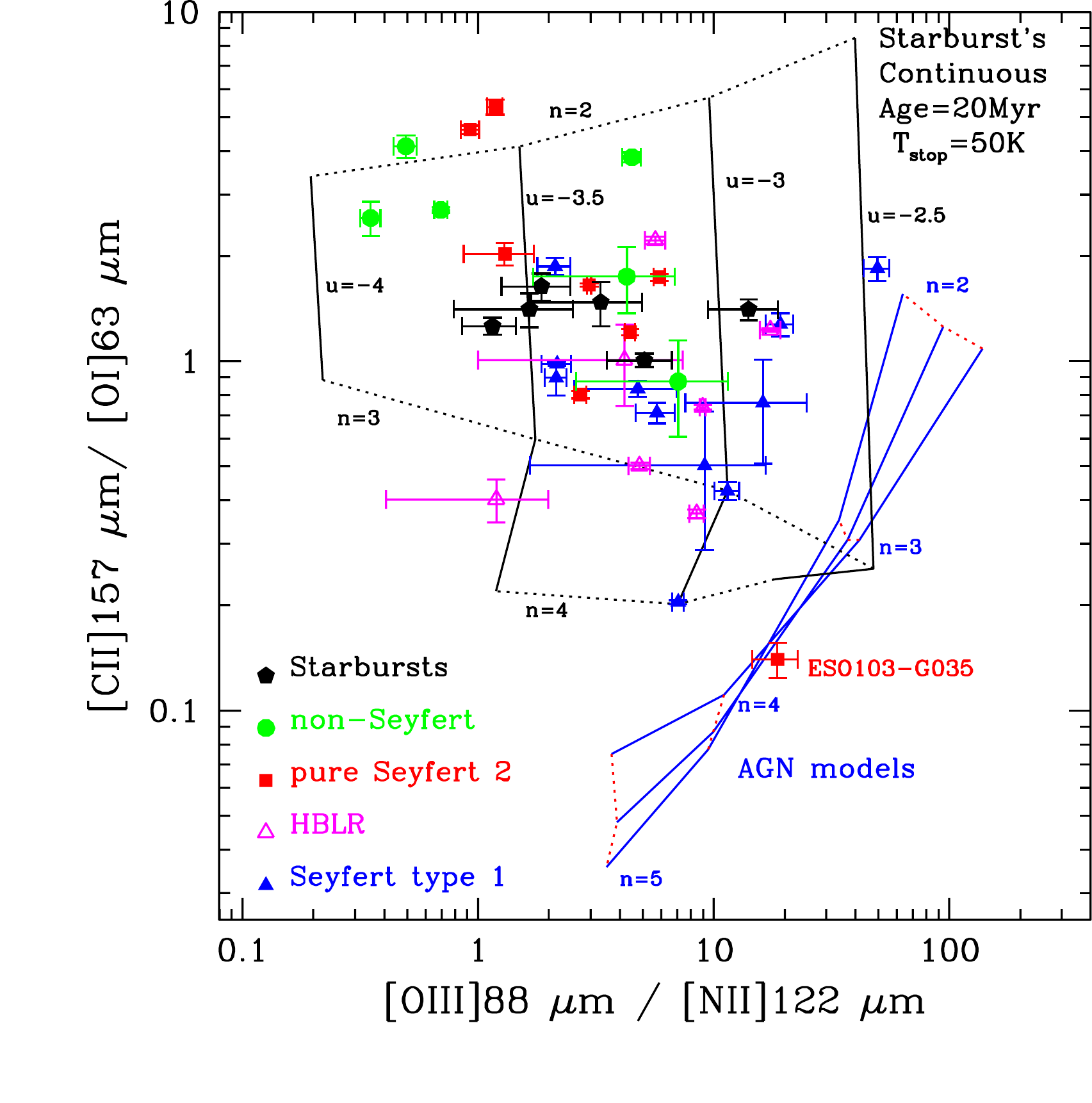}
\caption{\scriptsize Line ratio diagrams: 
{\bf Left: (a) } The [CII]157$\mu$m/[OI]63$\mu$m line ratio {\it vs.} the [OIII]88\-$\mu$m/[OI]63$\mu$m ratio: observations are shown as different types of points (Starburst galaxies: filled pentagons,
"non-Seyfert"  active galaxies: filled circles; "pure" Seyfert 2's: filled squares, HBLRs: open triangles, Seyfert 1's: filled triangles), photoionisation models of AGNs and Starburst galaxies 
are shown as blue and black grids, respectively. In the plot are indicated the logarithmic values of the density and ionisation potential of the photoionisation models, with $n$ and $U$, respectively.
 {\bf Right: (b)} The [CII]157$\mu$m/[OI]63$\mu$m line ratio {\it vs.} the [OIII]88\-$\mu$m/[NII]122$\mu$m ratio. }
\label{fig:c2_o1a}
\end{figure*}

\begin{figure*}
\includegraphics[width=8cm]{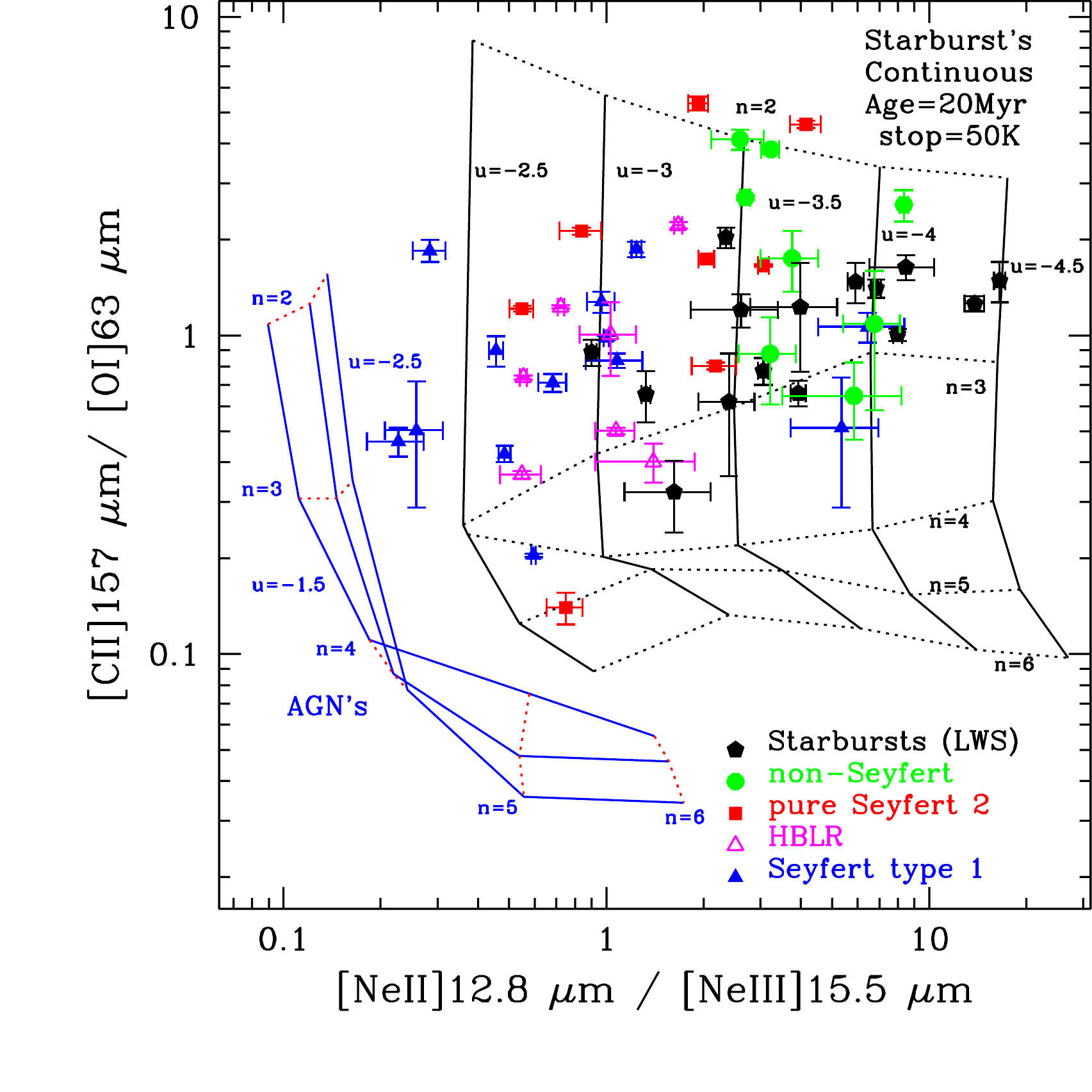}\includegraphics[width=8cm]{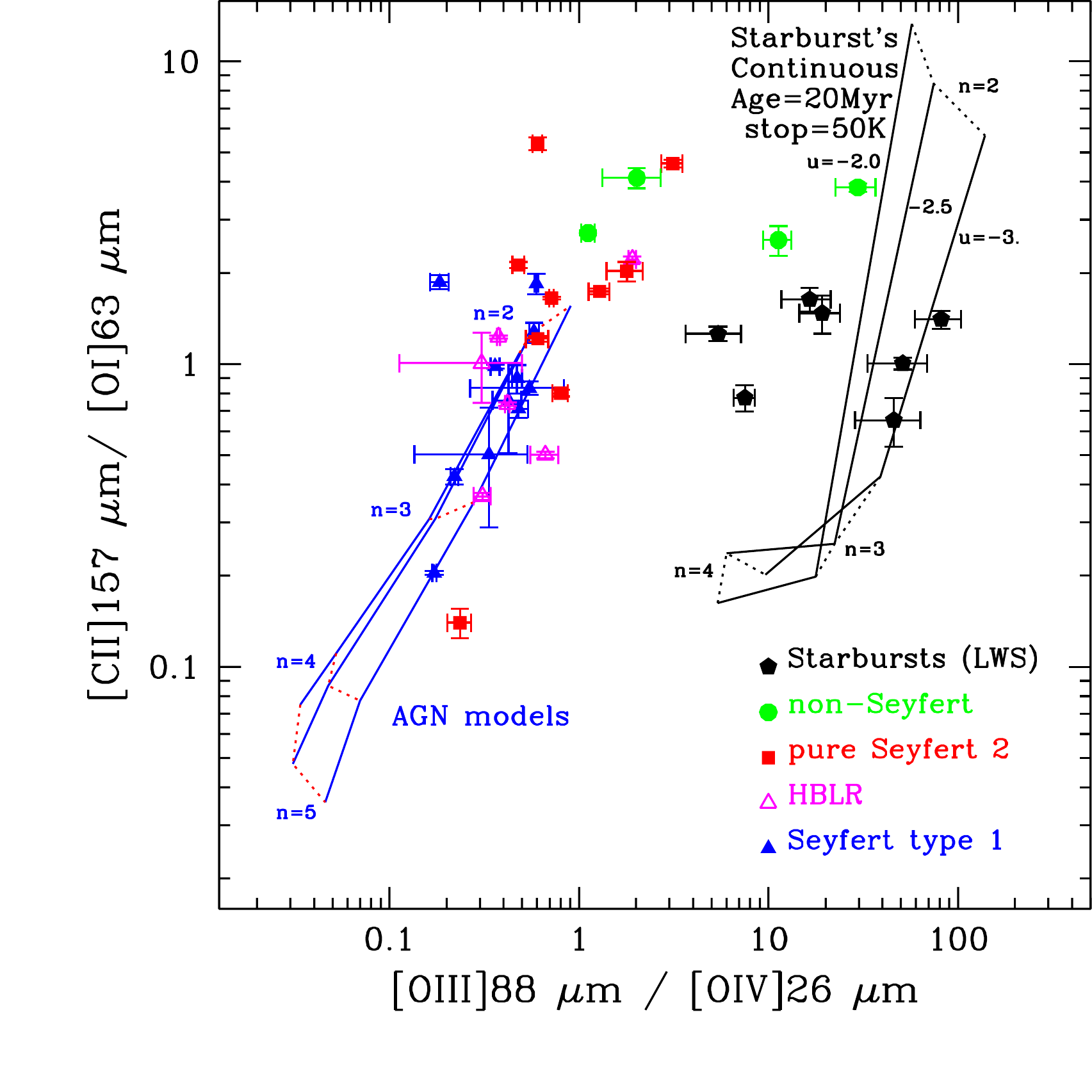}
\caption{\scriptsize Line ratio diagram (same notations as in Fig \ref{fig:c2_o1a}):
{\bf Left: (a) } The [CII]157$\mu$m/[OI]63$\mu$m line ratio {\it vs.} the [NeII]12.8\-$\mu$m/[NeIII]15.5$\mu$m ratio.  
 {\bf Right: (b)} The [CII]157$\mu$m/[OI]63$\mu$m line ratio {\it vs.} the [OIII]88\-$\mu$m/[OIV]26$\mu$m ratio. }
\label{fig:c2_o1b}
\end{figure*}

\begin{figure*}
\includegraphics[width=8cm]{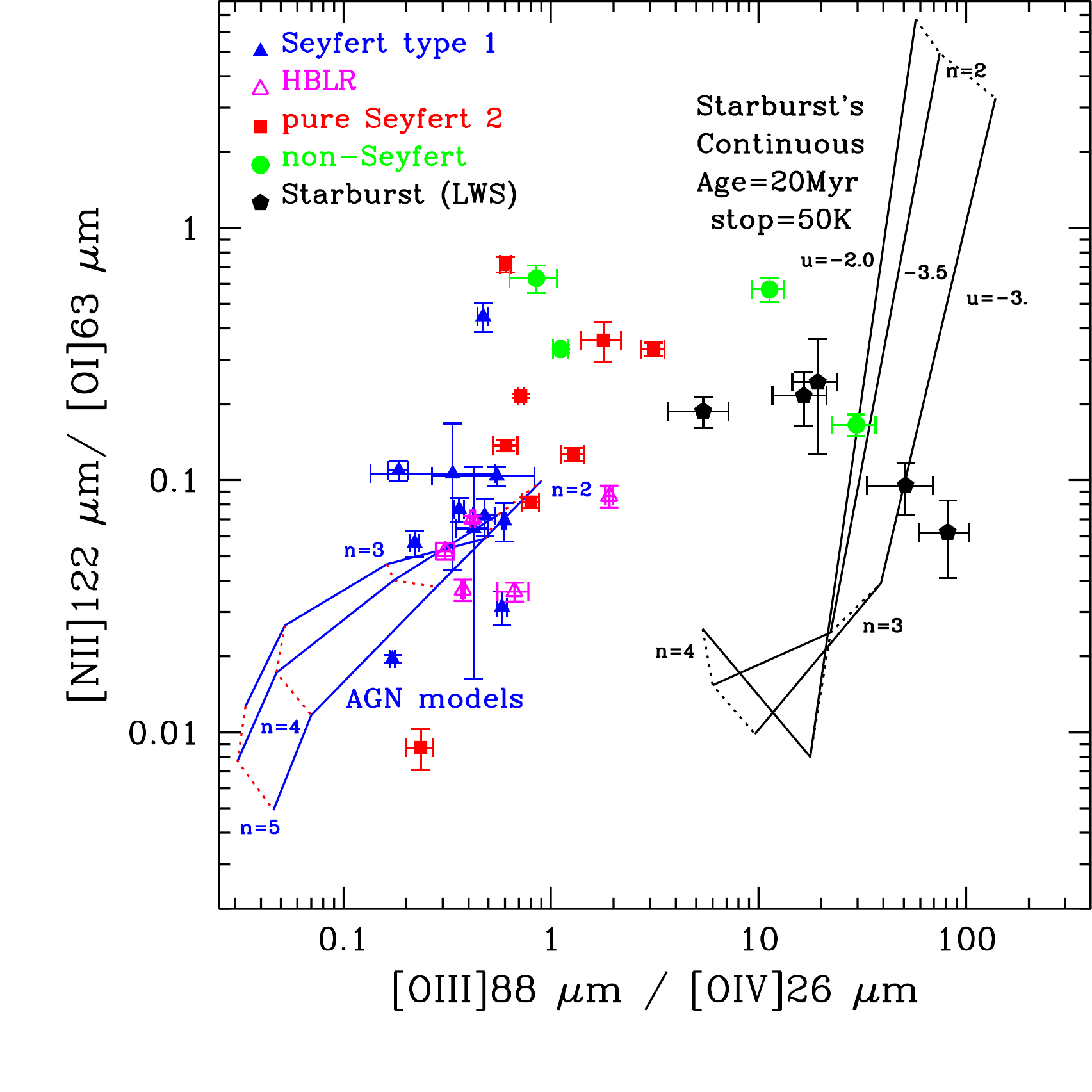}\includegraphics[width=8cm]{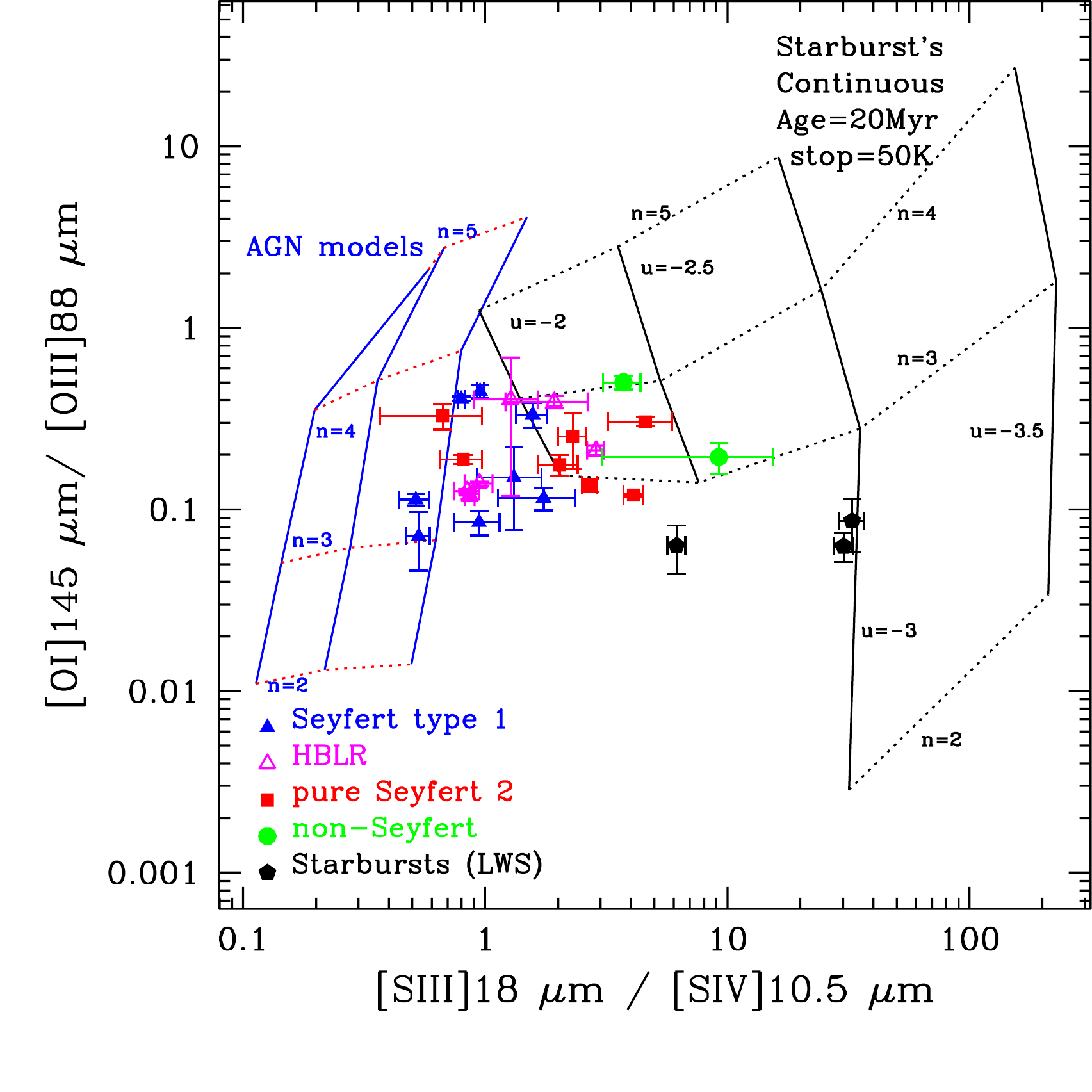}\caption{\scriptsize Line ratio diagram (same notations as in Fig \ref{fig:c2_o1a}):
{\bf Left: (a) } The [NII]122$\mu$m/[OI]63$\mu$m line ratio {\it vs.} the [OIII]88$\mu$m/[OIV]26$\mu$m ratio.  
 {\bf Right: (b)} The [OI]145$\mu$m/[OIII]88$\mu$m line ratio {\it vs.} the [SIII]18$\mu$m/[SIV]10.5$\mu$m ratio. }
\label{fig:n2_o1a}
\end{figure*}

\begin{figure*}
\includegraphics[width=8cm]{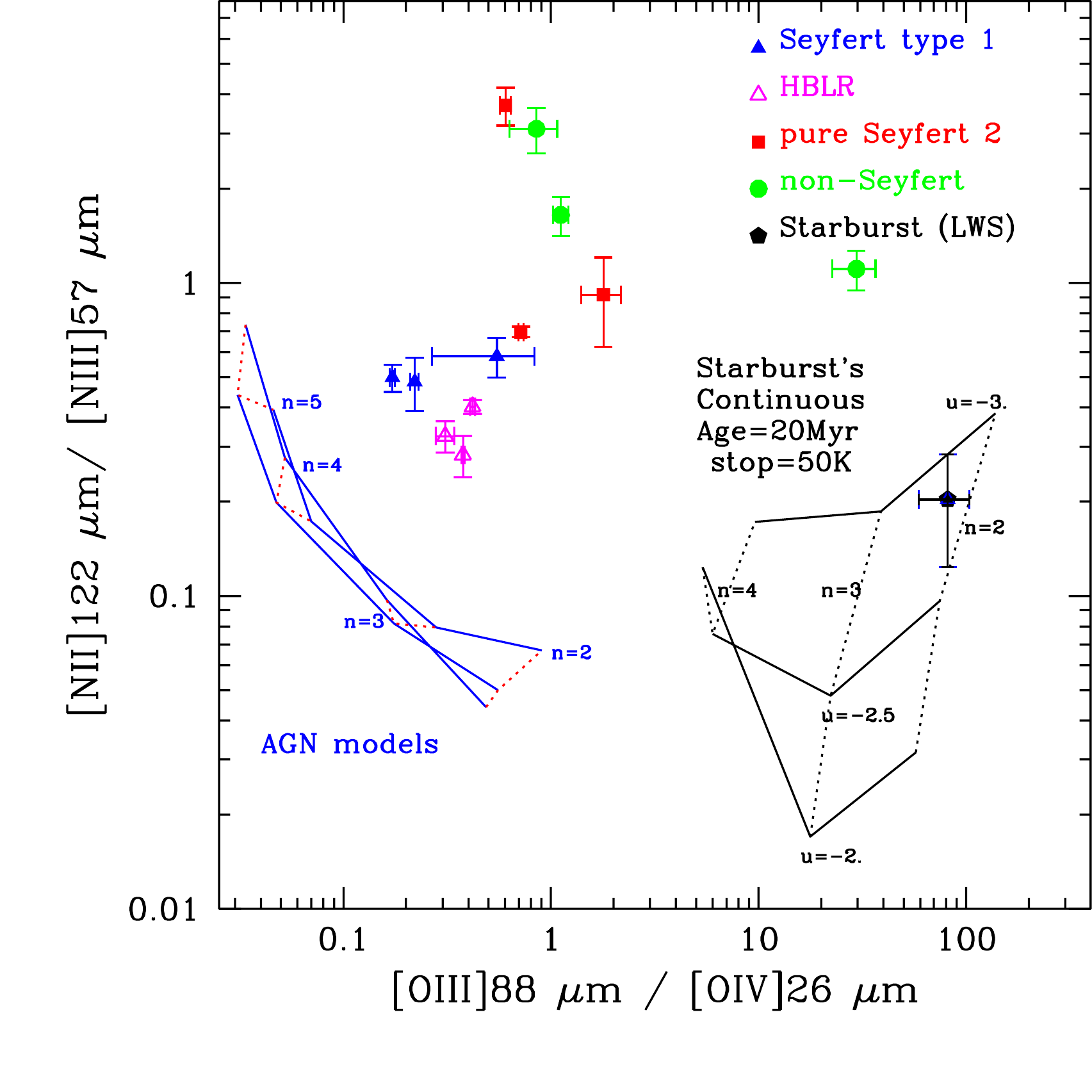}
\caption{\scriptsize Line ratio diagram (same notations as in Fig \ref{fig:c2_o1a}):
The [NII]122$\mu$m/[NIII]57$\mu$m line ratio {\it vs.} the [OIII]88$\mu$m/[OIV]26$\mu$m ratio.  
 }
\label{fig:n2_o1b}
\end{figure*}

\begin{figure*}
\includegraphics[width=8cm]{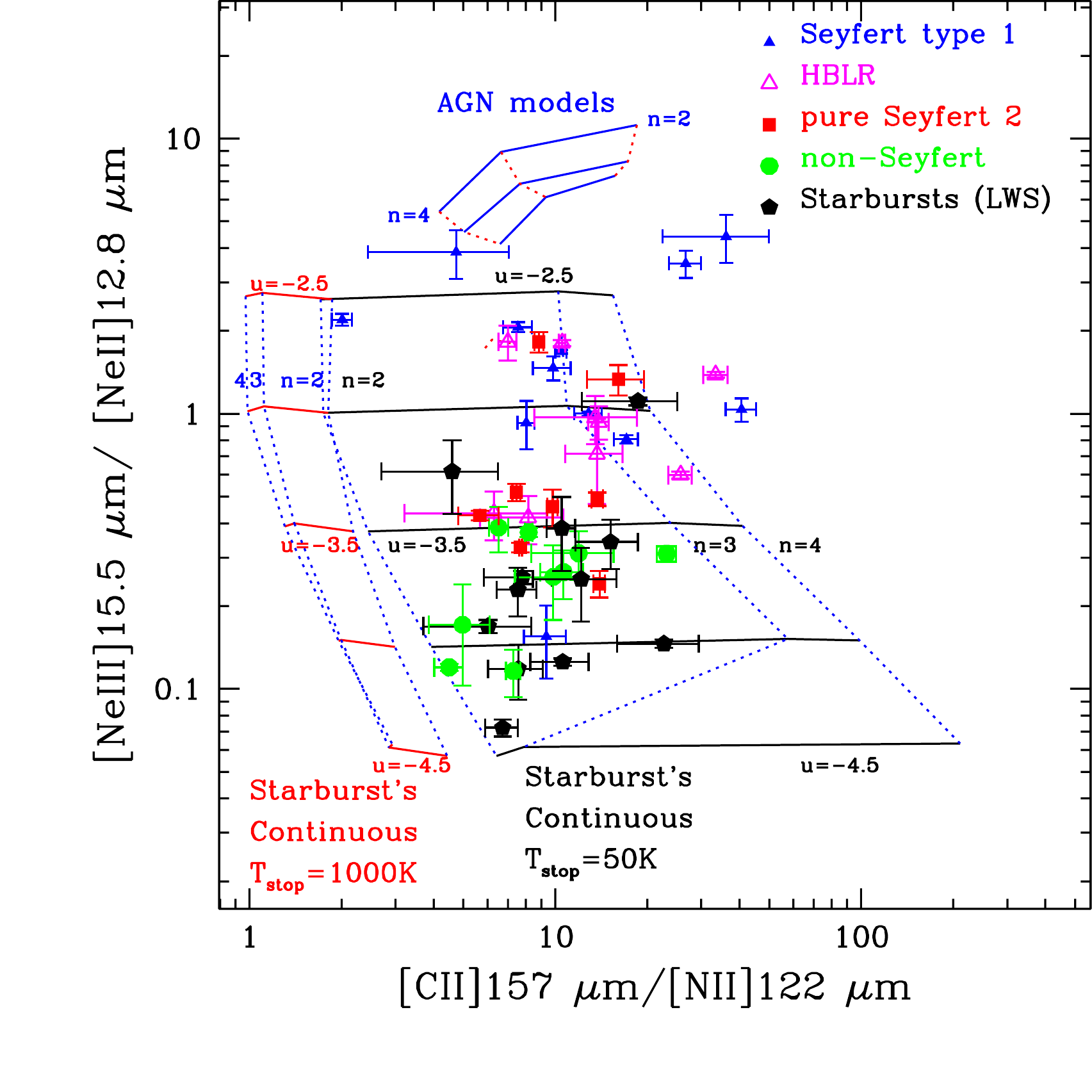}\includegraphics[width=8cm]{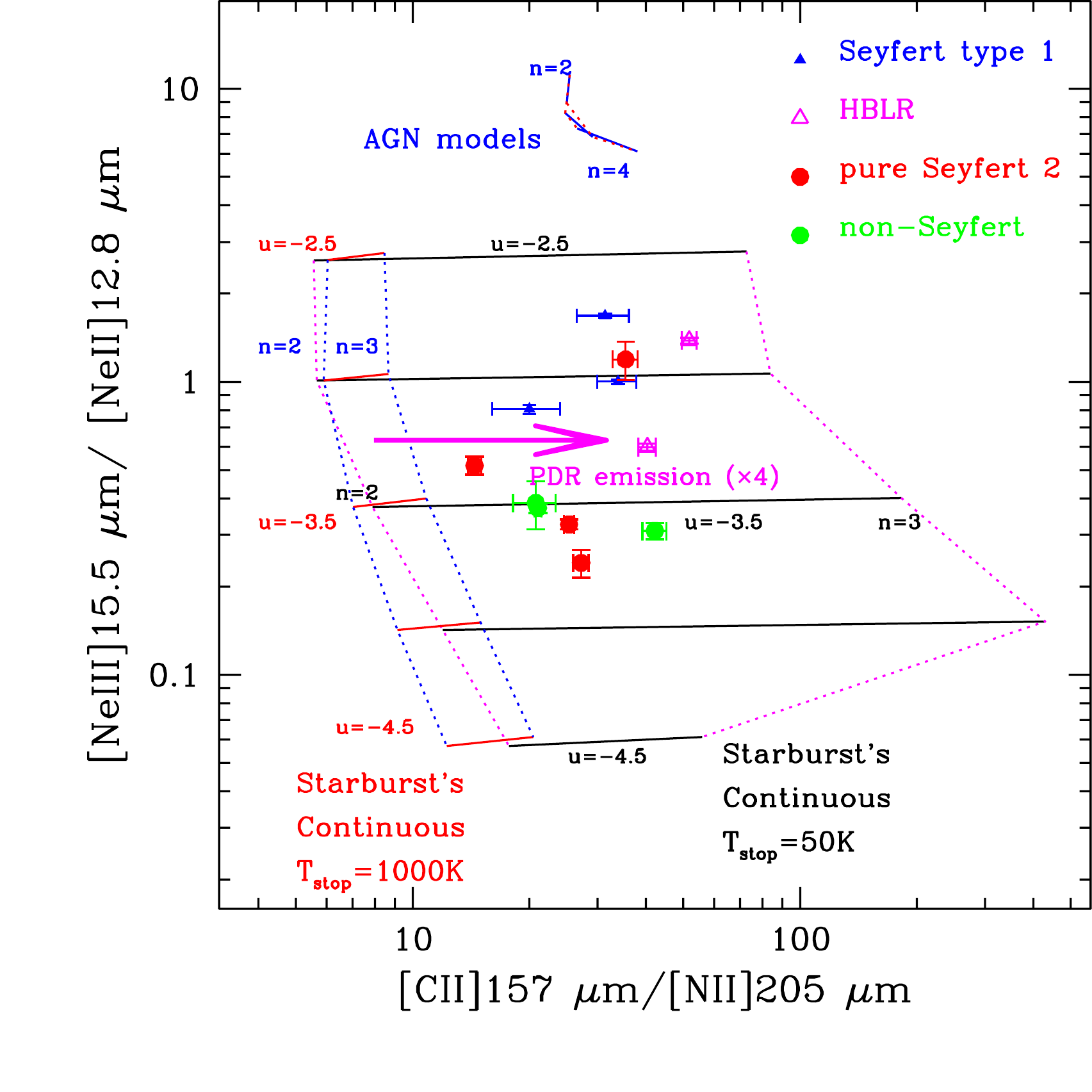}
\caption{\scriptsize Line ratio diagrams (same notations as in Fig \ref{fig:c2_o1a}):
{\bf Left: (a) } The [NeIII]15.5$\mu$m/[NeII]12.8$\mu$m line ratio {\it vs.} the [CII]157$\mu$m/[NII]122$\mu$m ratio.  The grid of models in the centre shows 
the usual starburst+PDR models, with the integration going down to the temperature of 50K, while the grid at the left of the diagram shows the pure photoionised models
(T$_{stop}$ = 1000K).  {\bf Right: (b)} The [[NeIII]15.5$\mu$m/[NeII]12.8$\mu$m line ratio {\it vs.} the [CII]157$\mu$m/[NII]205$\mu$m ratio. Same notations as in {\bf a)}. The 
arrow at the center shows a factor 4 increase in this ratio, as due to photodissociation regions (see the text).}
\label{fig:c2_n2vsne3_ne2}
\end{figure*}

\appendix

\section{A. Line profiles}\label{app.a}

In this section we present the line profiles in Fig.\ref{fig:spectra1}. 

\section{B. Comparision of line fluxes from ISO-LWS and Herschel-PACS}\label{app.b}

We present in Fig. \ref{lws_vs_pacs} the line fluxes observed with ISO-LWS \citep{bra08,spi97} and PACS (this work).
Besides the [OIII]52$\mu$m line, reported in the text, most discrepancies happens to be for the [OIII]88$\mu$m line, for which a significantly higher flux for all sources
has been measured from ISO-LWS, with respect to PACS. In two cases, for 3C120 and IC4329A, also the [OI]63$\mu$m 
flux is higher. We argue that this result is  unlikely to be due to aperture effects, because the [OIII] lines are produced in high-ionisation environments, which are
located in the central regions of galaxies. We do not have an explanation for this behaviour, other from instrumental systematic effects. 

\section{C. Spitzer images of the observed galaxies}\label{app.c}

In this section we present the 3.6$\mu$m IRAC {\it Spitzer} images of the sample galaxies, overplotted to the frame of the PACS spectrometer of 47$\arcsec$ $\times$ 47$\arcsec$
and the slit of 22.3$\arcsec$ $\times$ 11.1$\arcsec$ of the long-wavelength high-resolution mode (LH) of the IRS spectrometer of \spi~ 
and the short- and long-wavelengths beams of the SPIRE FTS spectrometer, for the galaxies for which these observations are available,  
in Fig.\ref{fig:spitzima_1} and \ref{fig:spitzima_2 }

\clearpage

\begin{figure*}
\centering
\includegraphics[width=16cm]{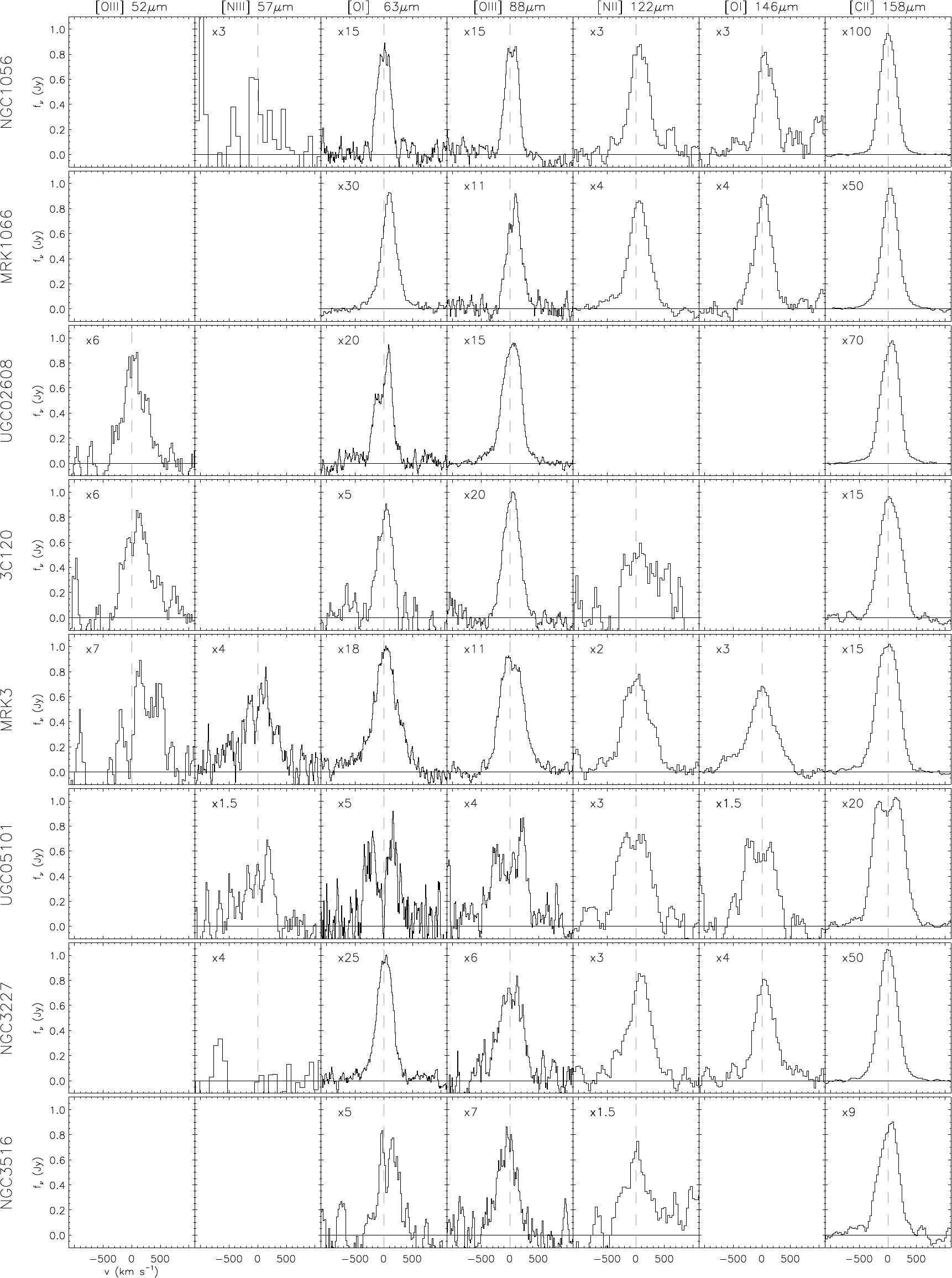}
\caption{Spectra of the detected lines. The spectra were extracted from the central 3 $\times$ 3 spaxels and the point-source aperture correction has been applied.
Line detections with low S/N (see Table \ref{tbl-3}) were rebinned for more clearly displaying the line profiles. }
\label{fig:spectra1}
\end{figure*}

\begin{figure*}
\centering
\addtocounter{figure}{-1}
\includegraphics[width=16cm]{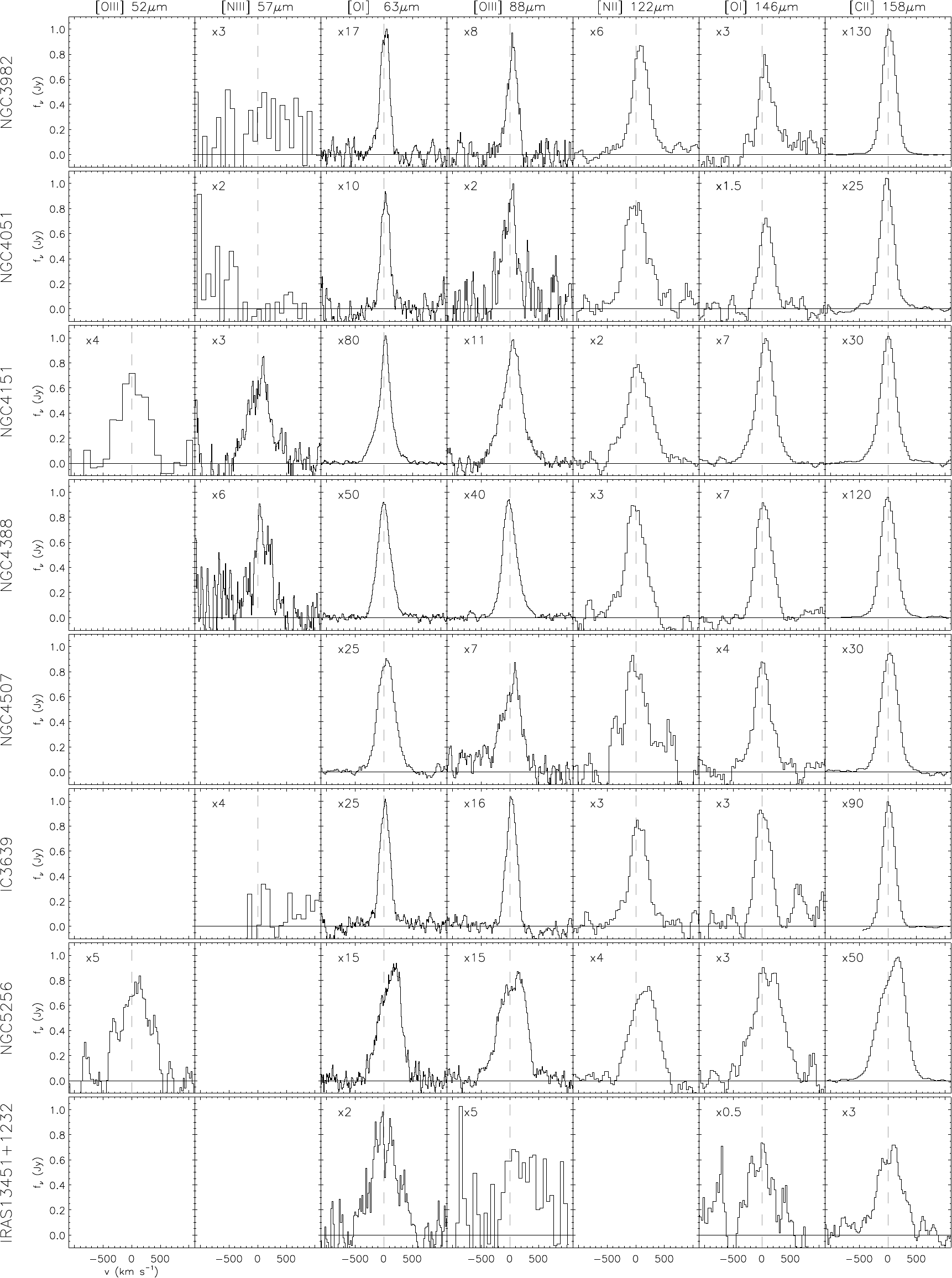}
\caption{(Continued)}
\label{fig:spectra2}
\end{figure*}

\begin{figure*}
\centering
\addtocounter{figure}{-1}
\includegraphics[width=16cm]{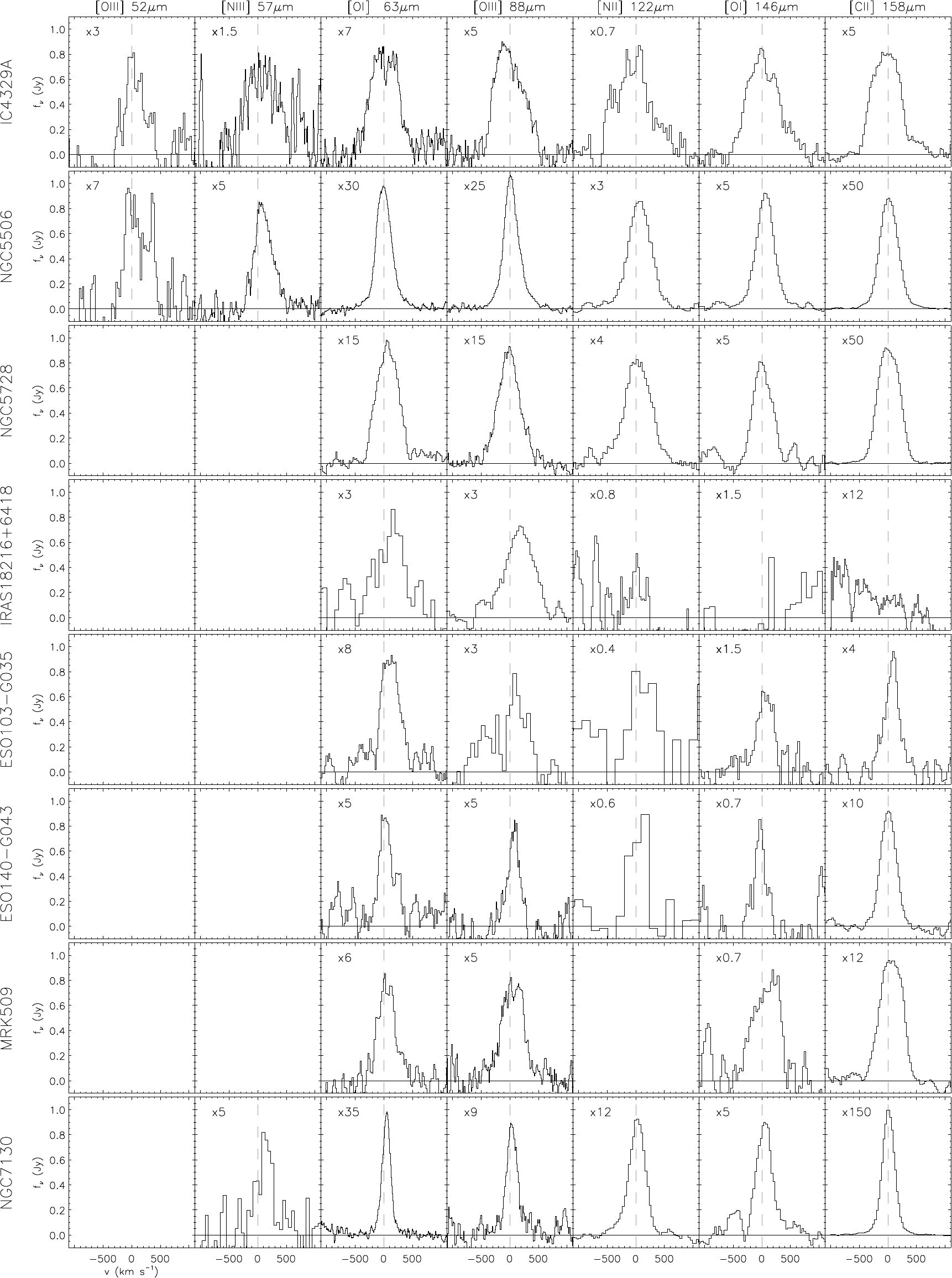}
\caption{(Continued)}
\label{fig:spectra3}
\end{figure*}

\begin{figure*}
\centering
\addtocounter{figure}{-1}
\includegraphics[width=16cm]{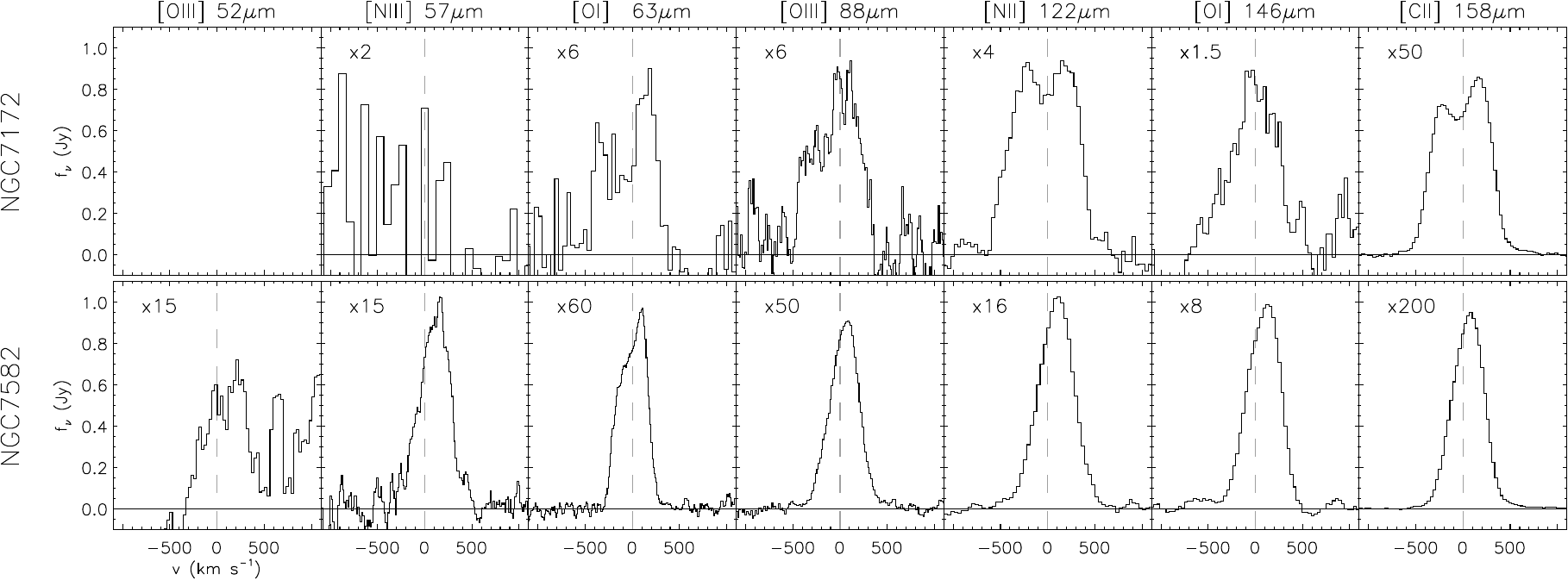}
\caption{(Continued)}
\label{fig:spectra}
\end{figure*}

\begin{figure*}
\includegraphics[width=8cm]{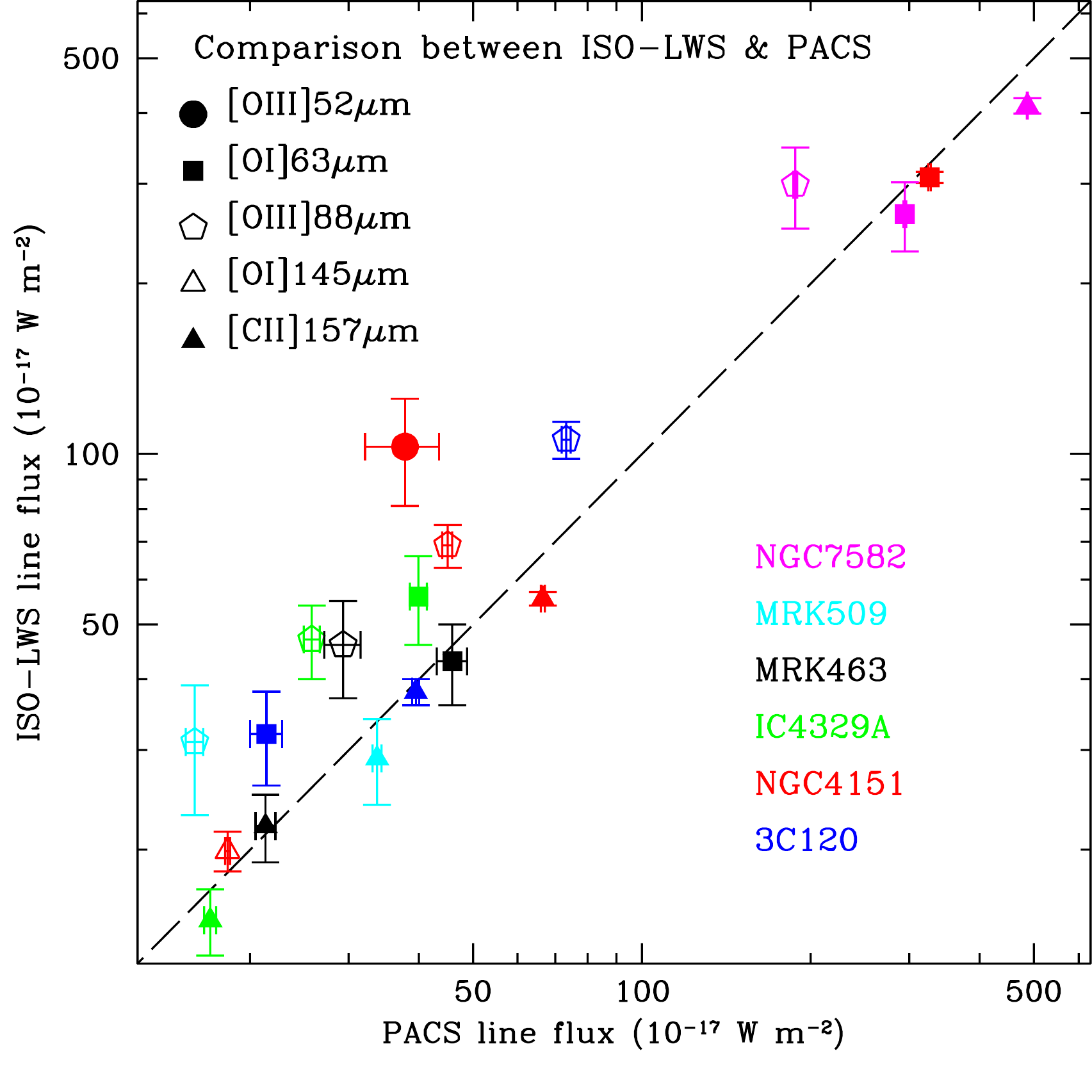}
\caption{\scriptsize ISO-LWS fluxes as a function of the Herschel-PACS fluxes for the galaxies observed by both spectrometers.}
\label{lws_vs_pacs}
\end{figure*}

\begin{figure*}
\includegraphics[width=4.cm]{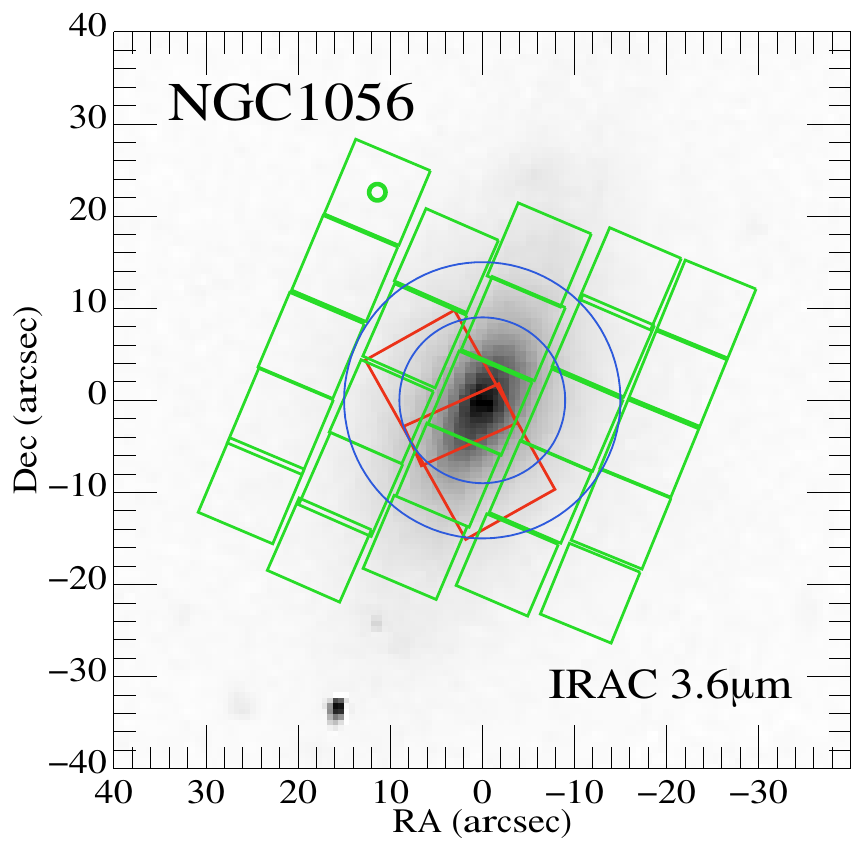}\includegraphics[width=4.2cm]{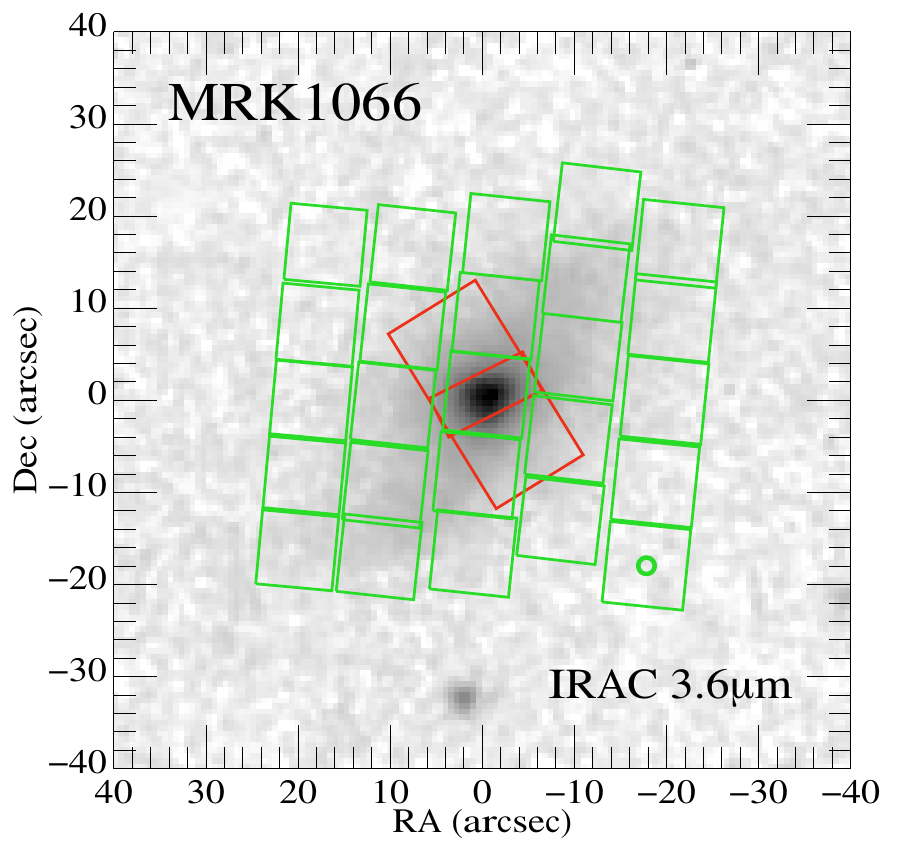}\includegraphics[width=4.2cm]{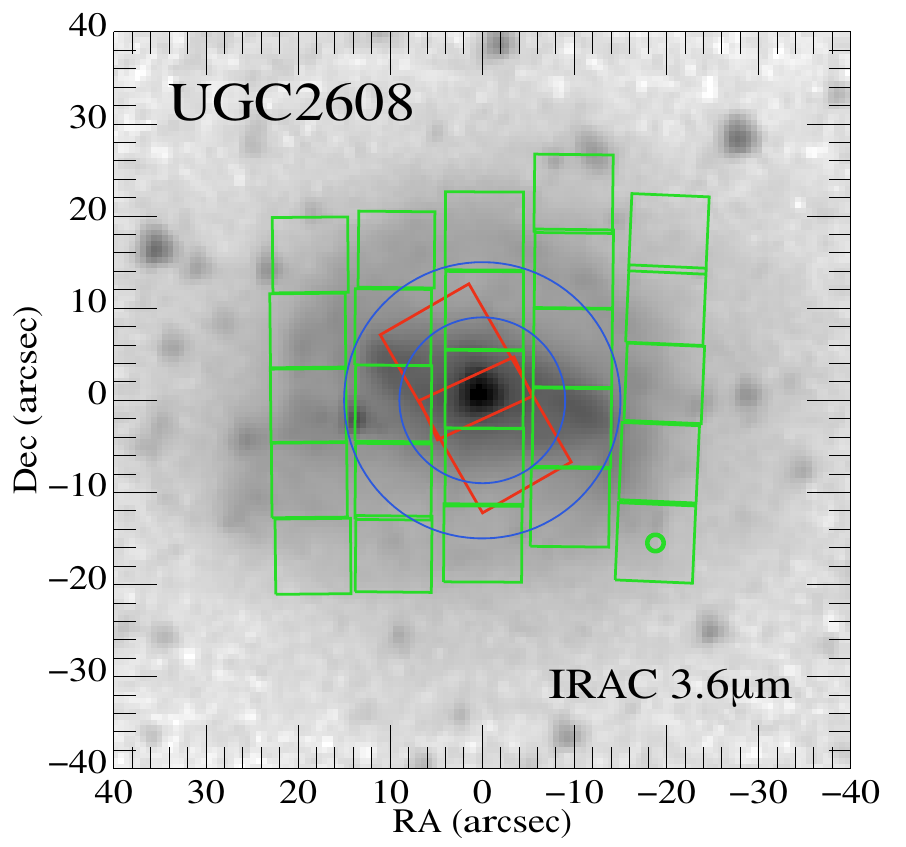}\includegraphics[width=4.2cm]{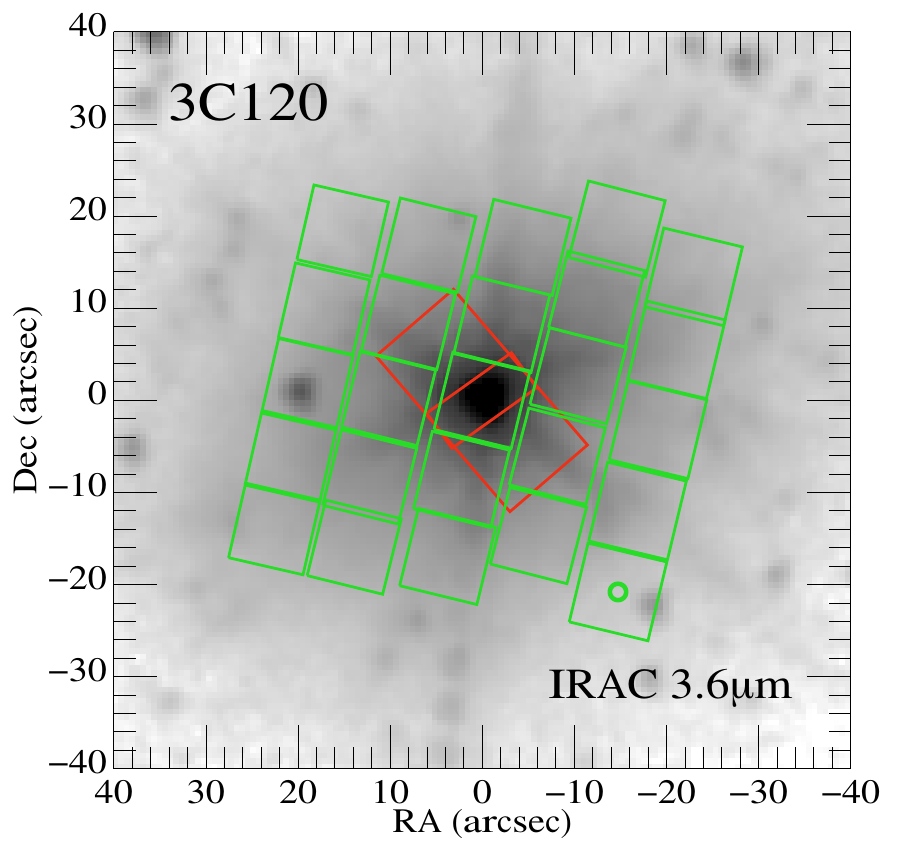}
\includegraphics[width=4.2cm]{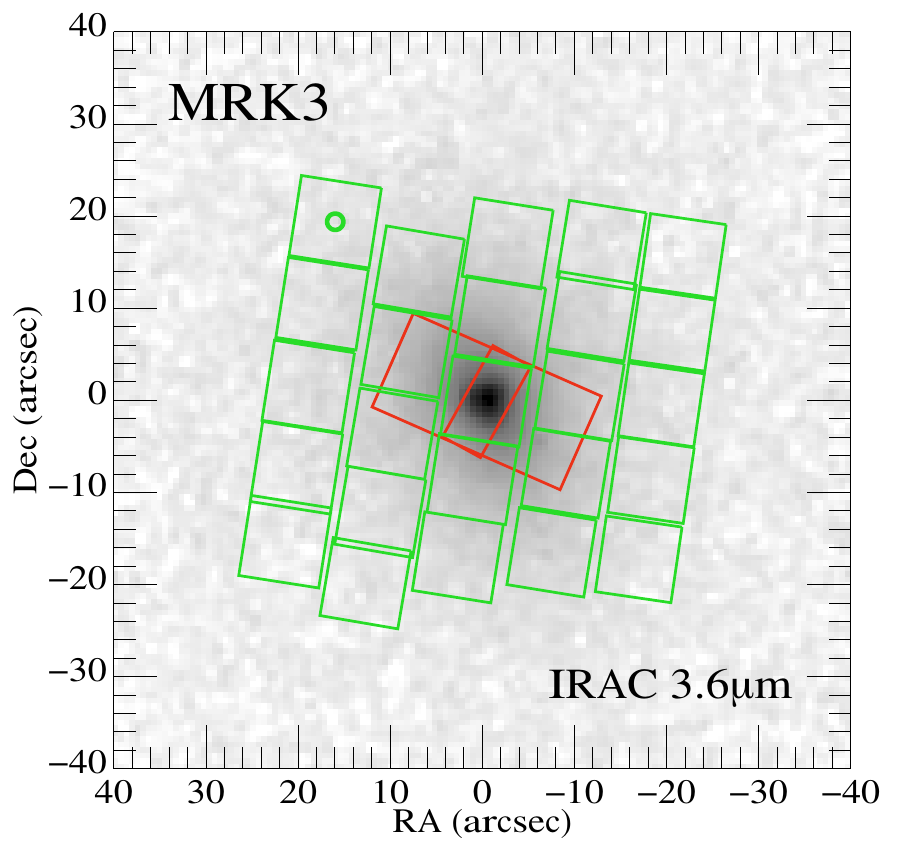}\includegraphics[width=4cm]{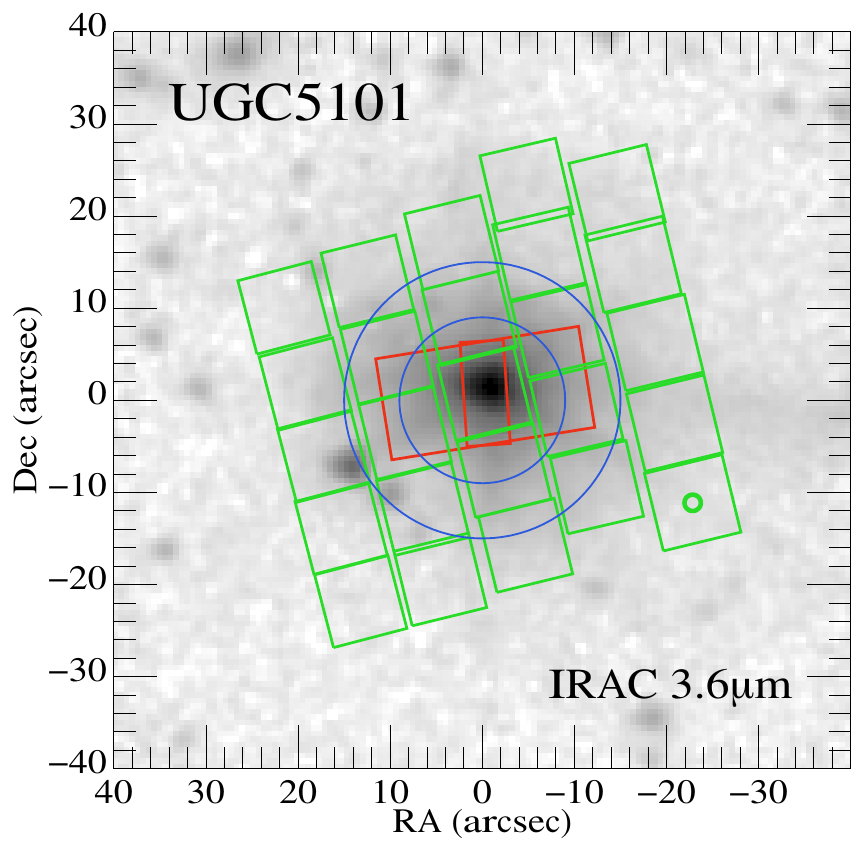}\includegraphics[width=4.2cm]{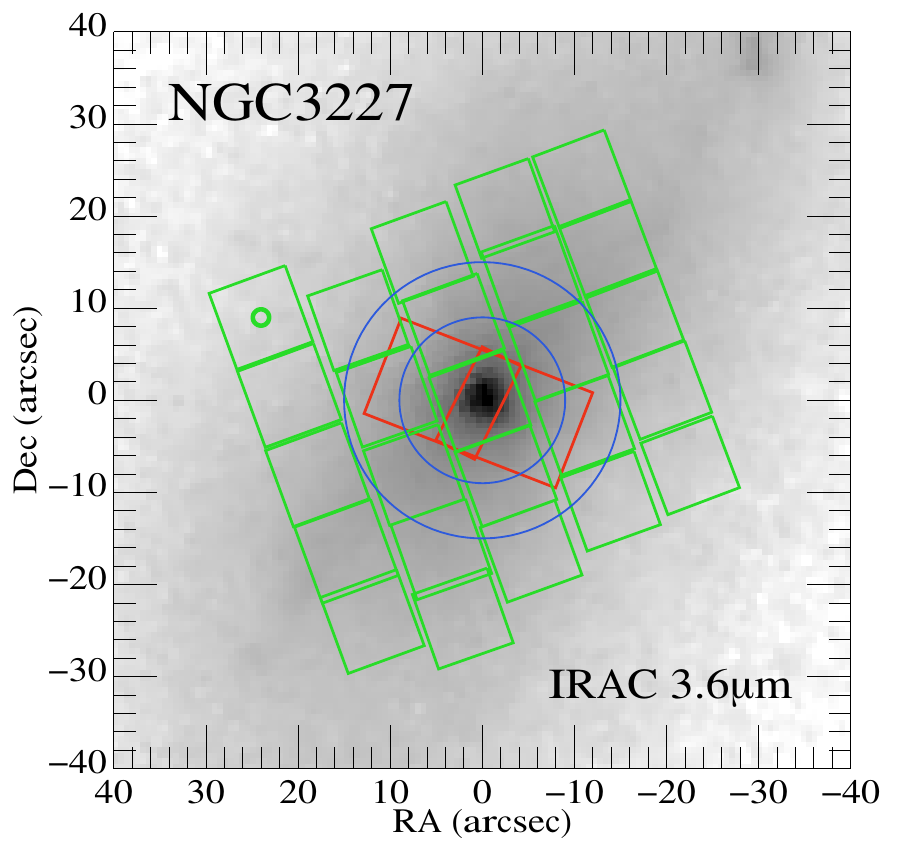}\includegraphics[width=4.2cm]{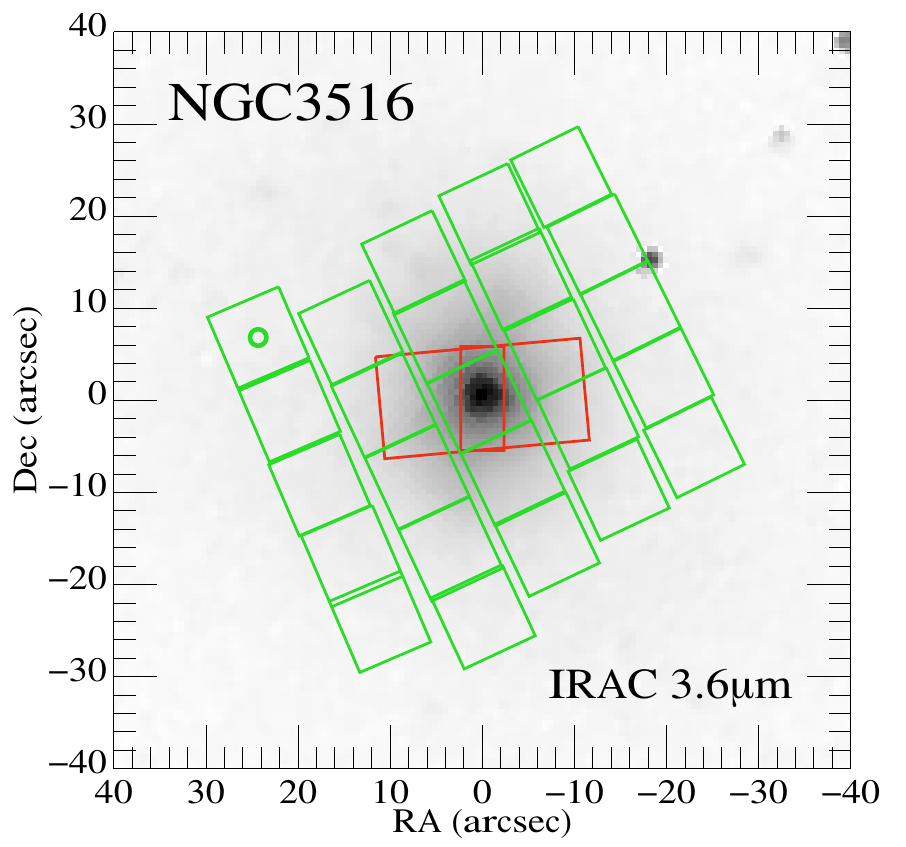}
\includegraphics[width=4.2cm]{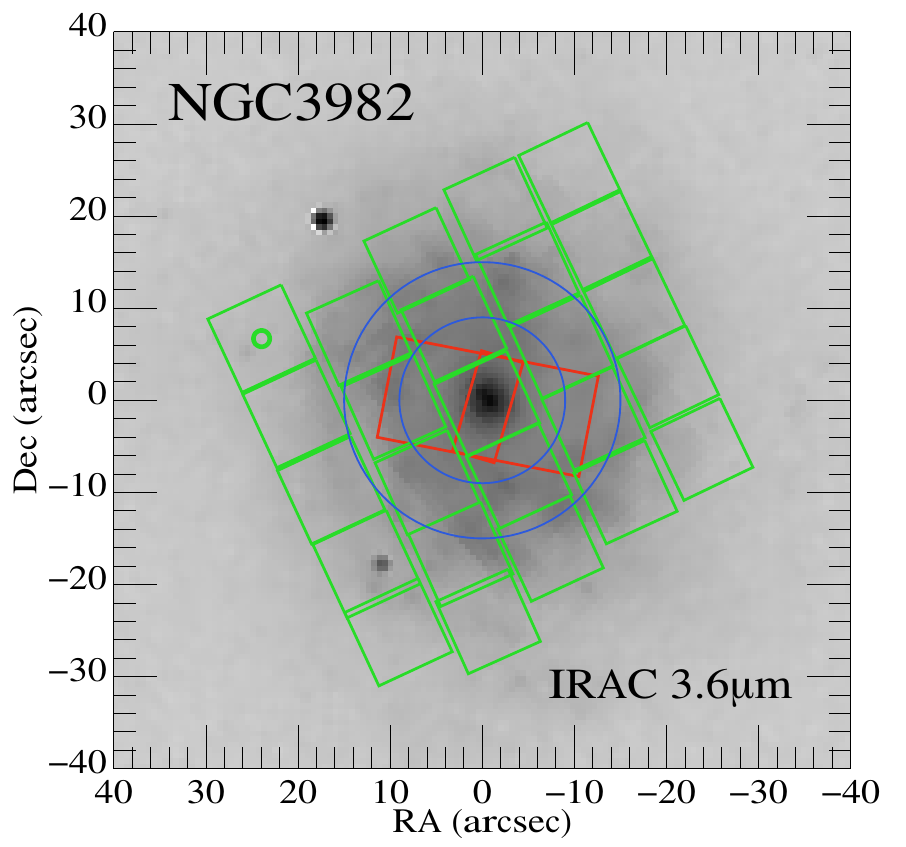}\includegraphics[width=4cm]{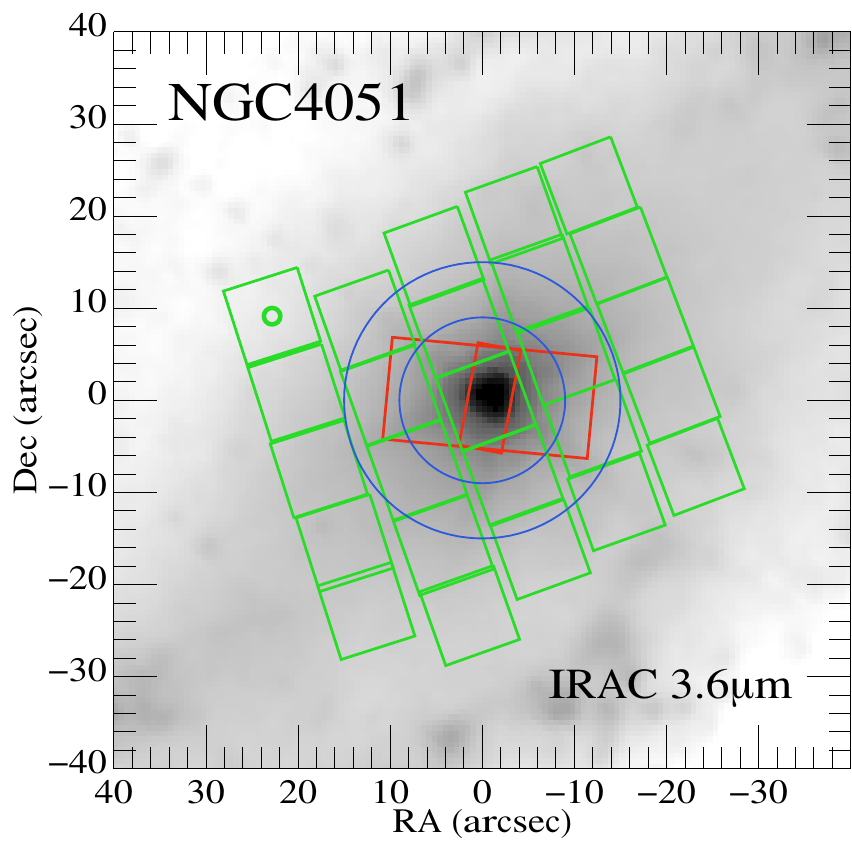}\includegraphics[width=4cm]{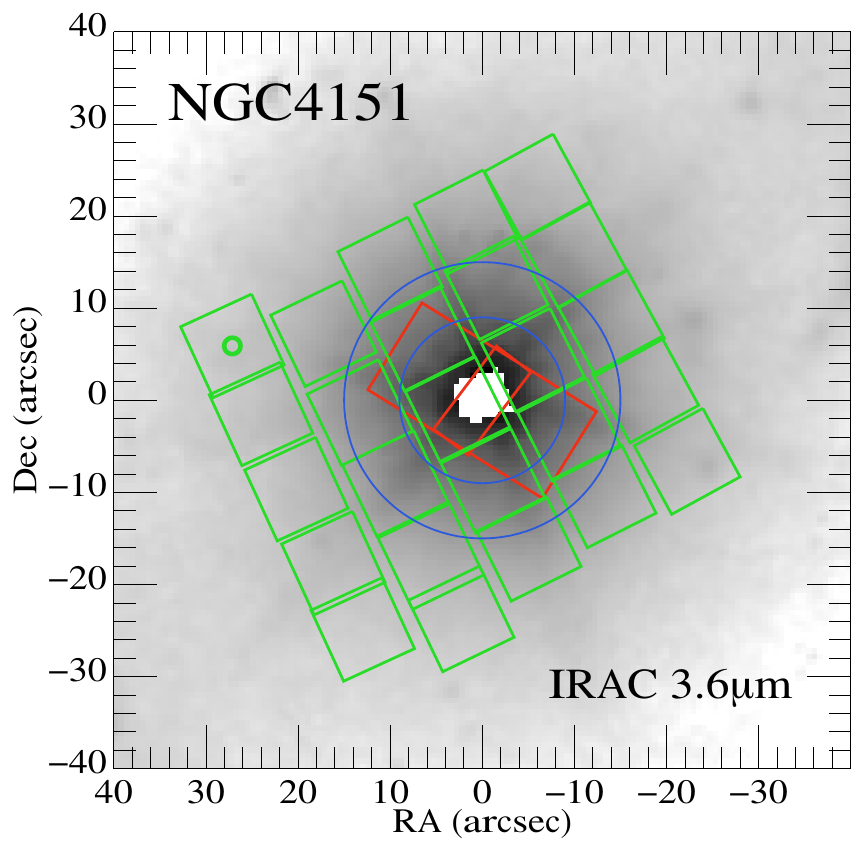}\includegraphics[width=4.3cm]{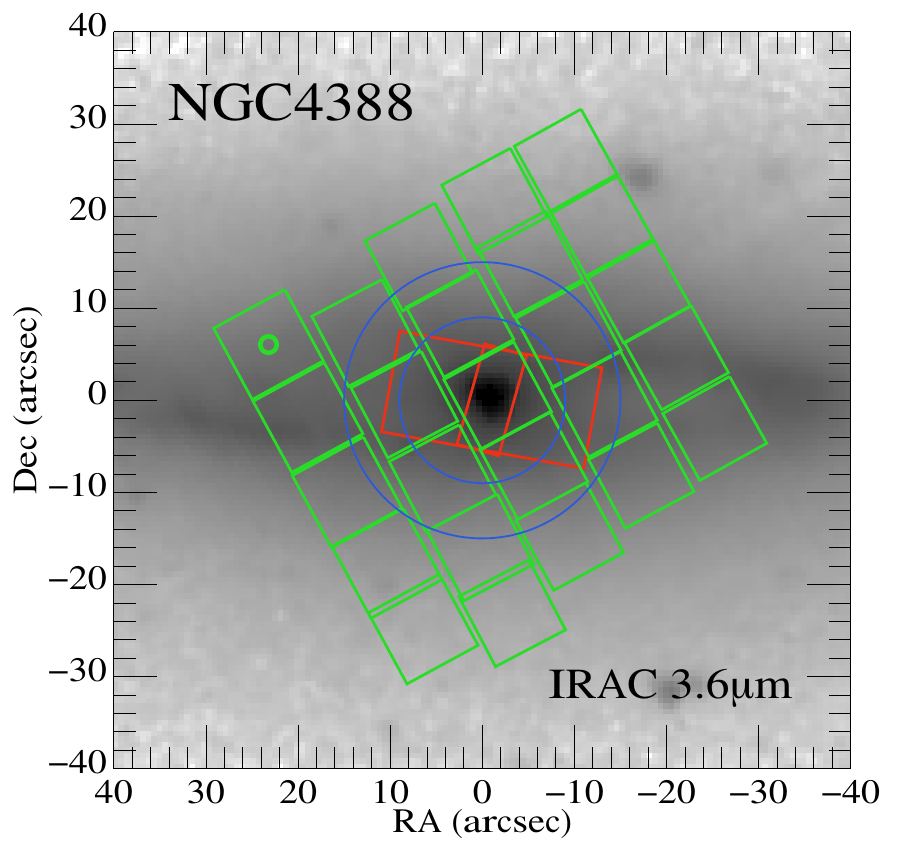}
\includegraphics[width=4.2cm]{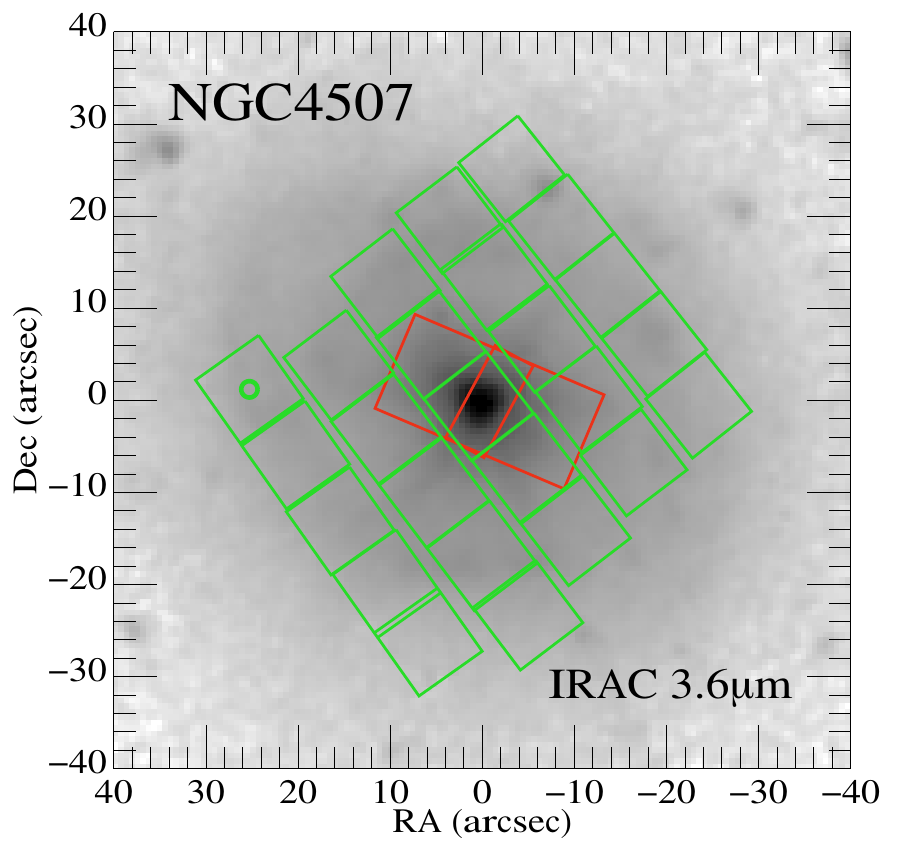}\includegraphics[width=4.2cm]{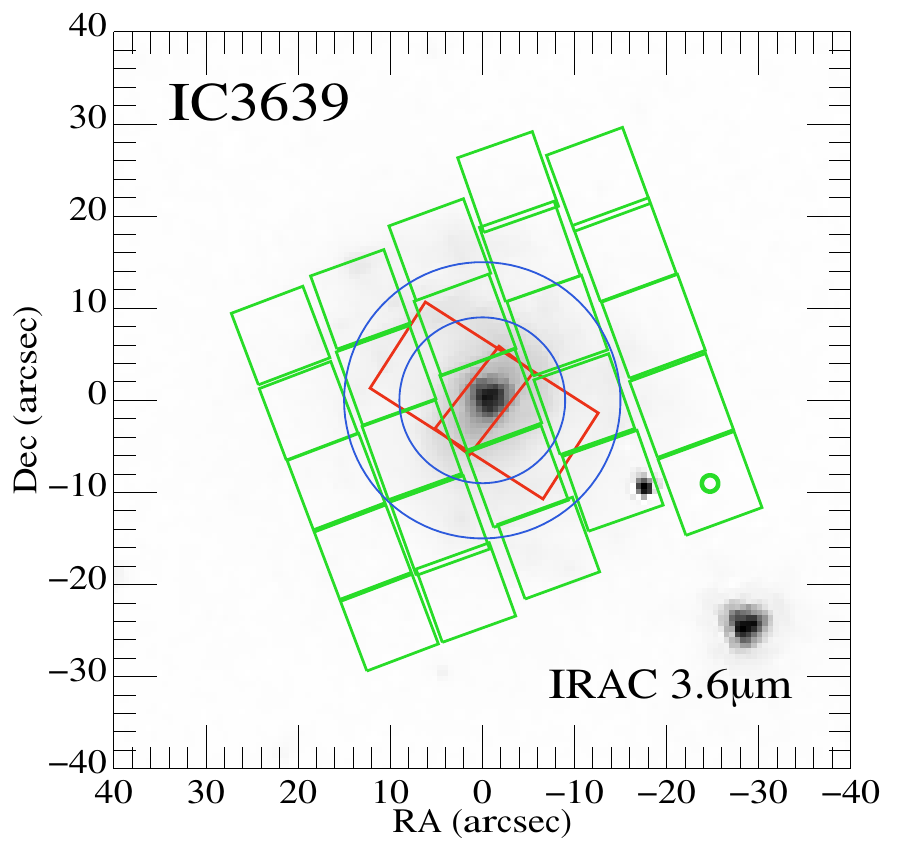}\includegraphics[width=3.9cm]{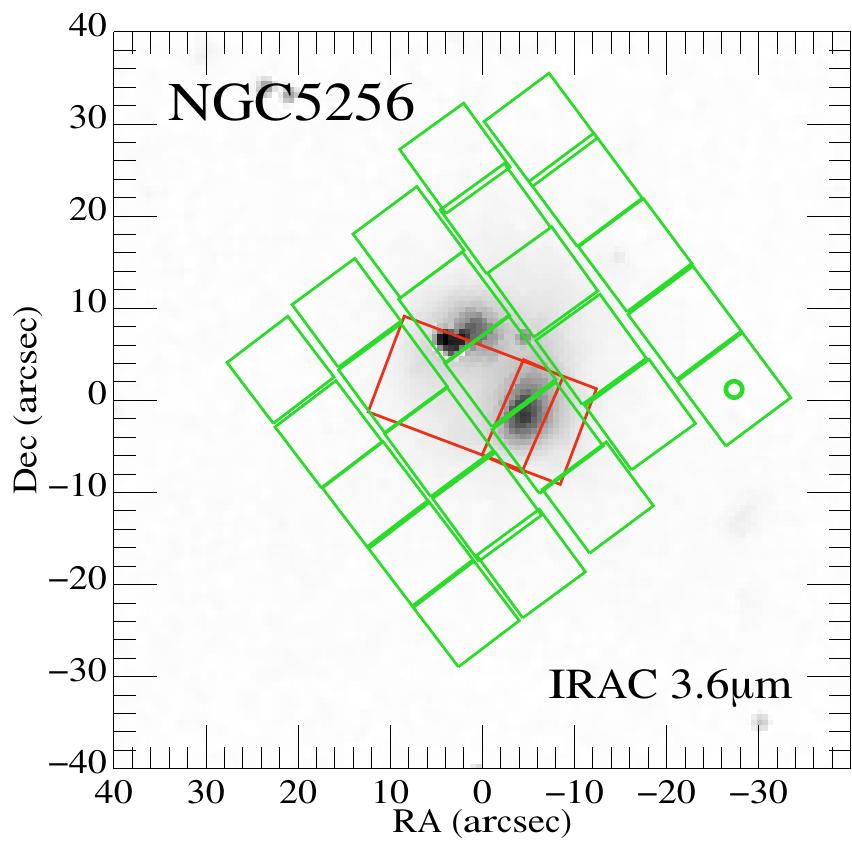}
\includegraphics[width=4cm]{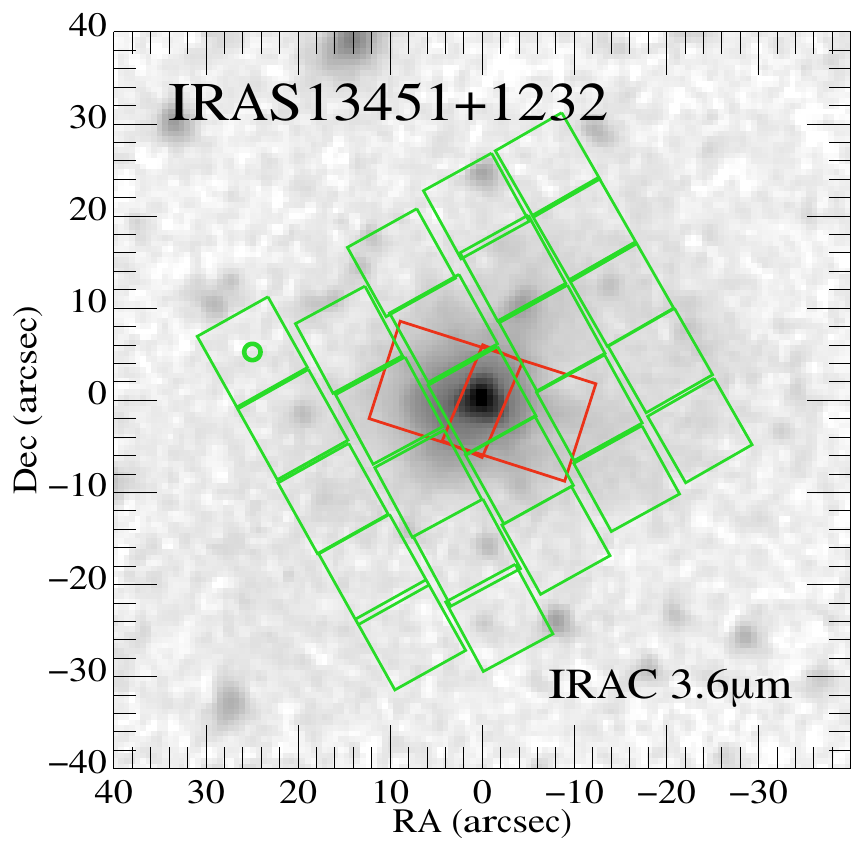}
\centering
\caption{ \scriptsize 3.6$\mu$m images from \spi, superposed to the frame of the PACS spectrometer of 47$\arcsec$ $\times$ 47$\arcsec$, 
the short- and long-wavelengths beams of the SPIRE FTS spectrometer, for the galaxies for which these observations are available, 
and the slit of 22.3$\arcsec$ $\times$ 11.1$\arcsec$ of the long-wavelength high-resolution mode (LH) of the IRS spectrometer of \spi.
1: NGC1056; 2: MRK1066; 3: UGC2608; 4: 3C120; 5: MRK3; 6: UGC5101; 7: NGC3227; 8: NGC3516; 
9: NGC3982; 10: NGC4051; 11: NGC4151; 12: NGC4388; 13: IC3639; 14: NGC5256; 15: IRAS13451+1232; 16: IC4329A.}
 \label{fig:spitzima_1}
 \end{figure*}
  
\clearpage

\begin{figure*}
\includegraphics[width=4.2cm]{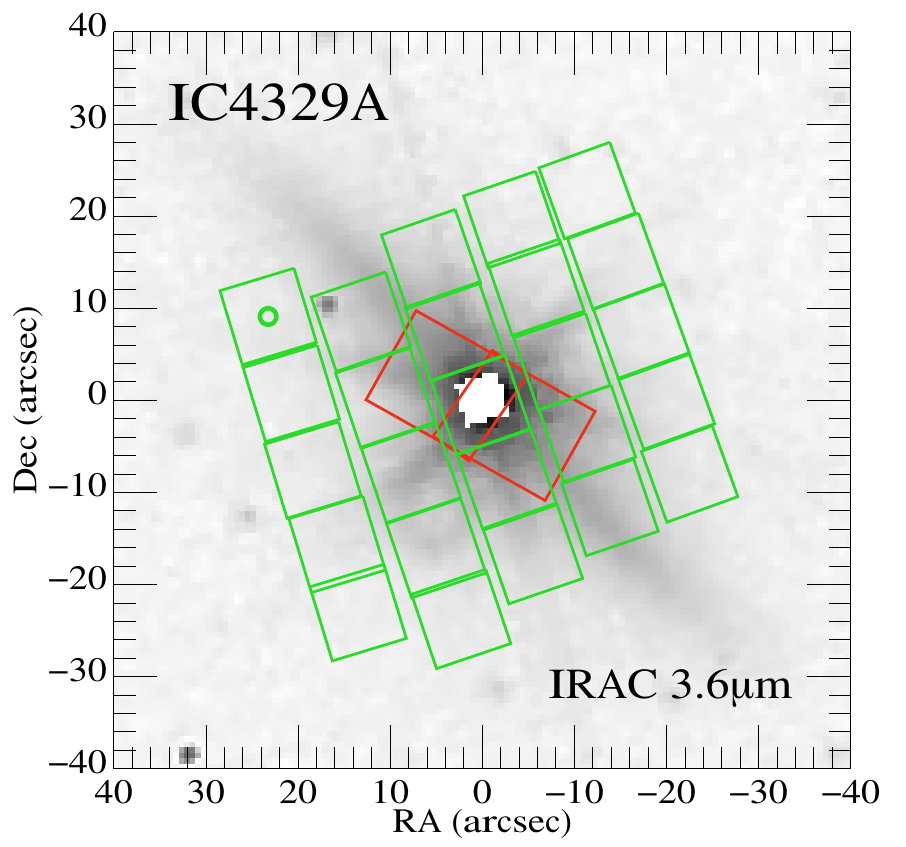}\includegraphics[width=4.3cm]{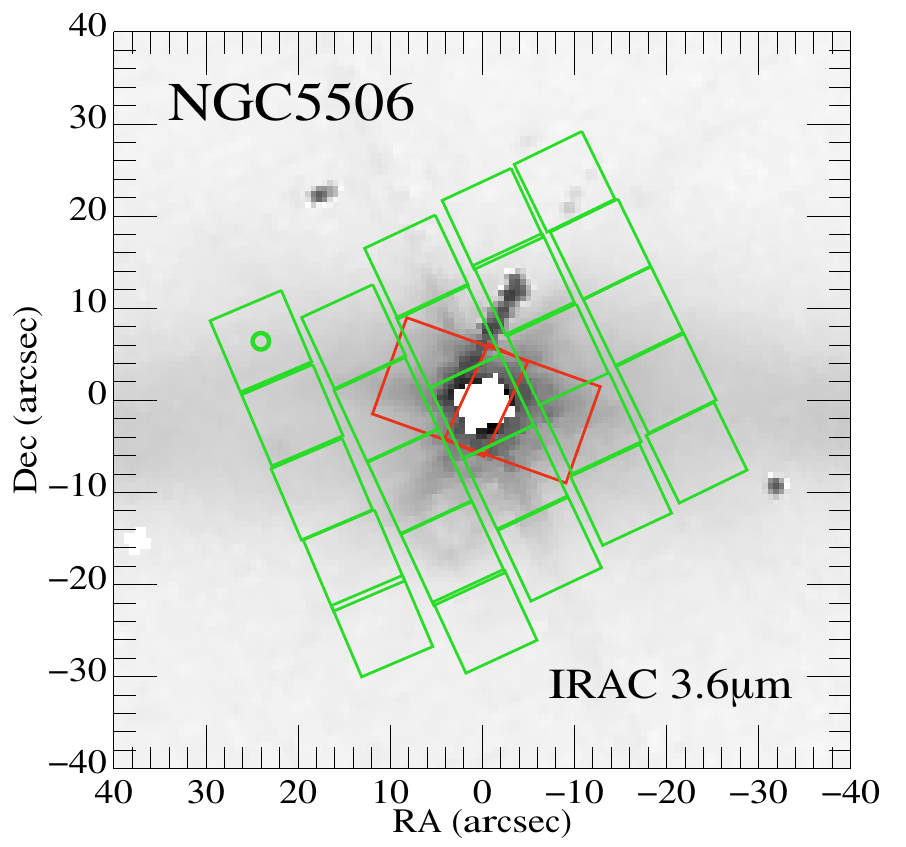}\includegraphics[width=4.2cm]{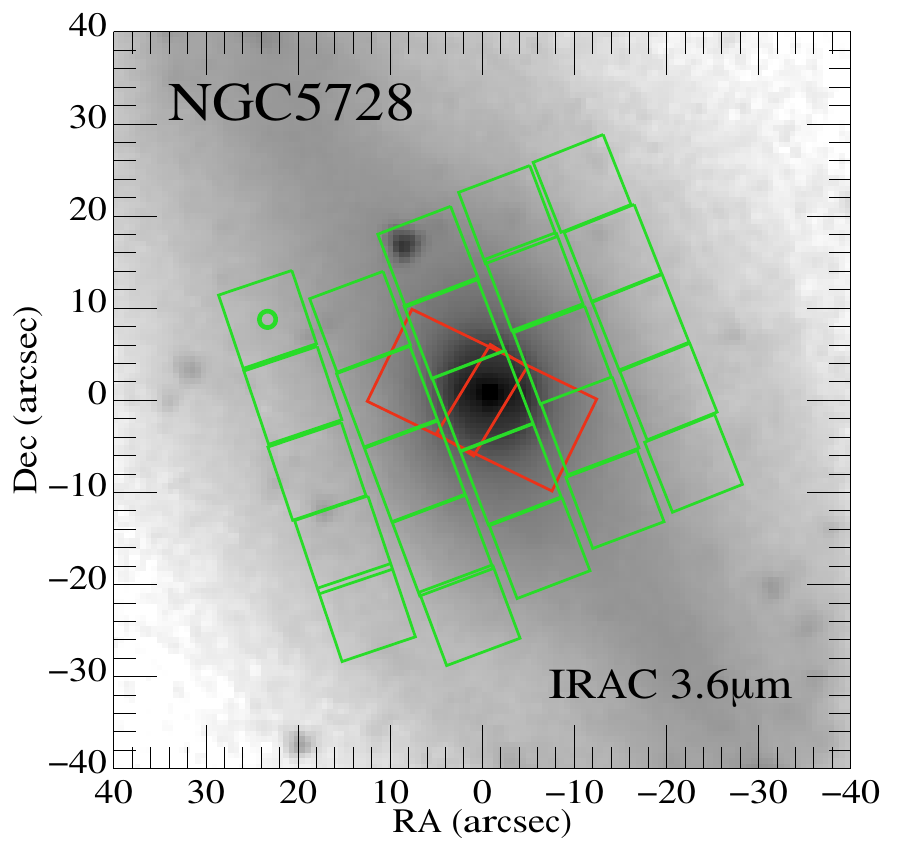}\includegraphics[width=4.2cm]{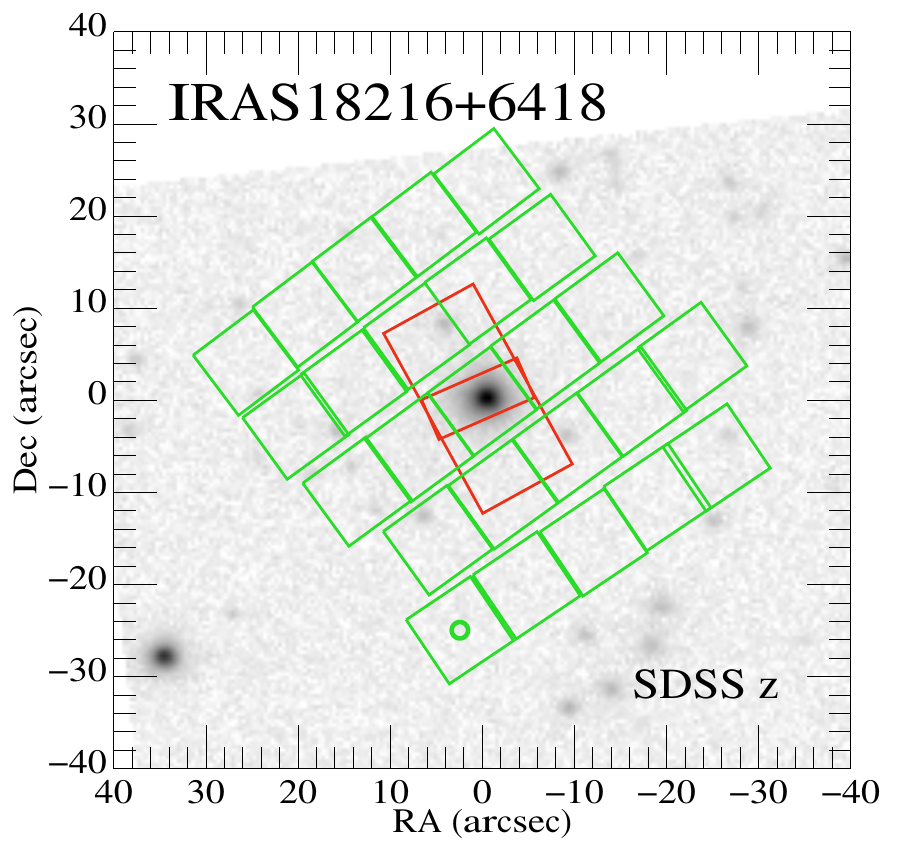}
\includegraphics[width=4.2cm]{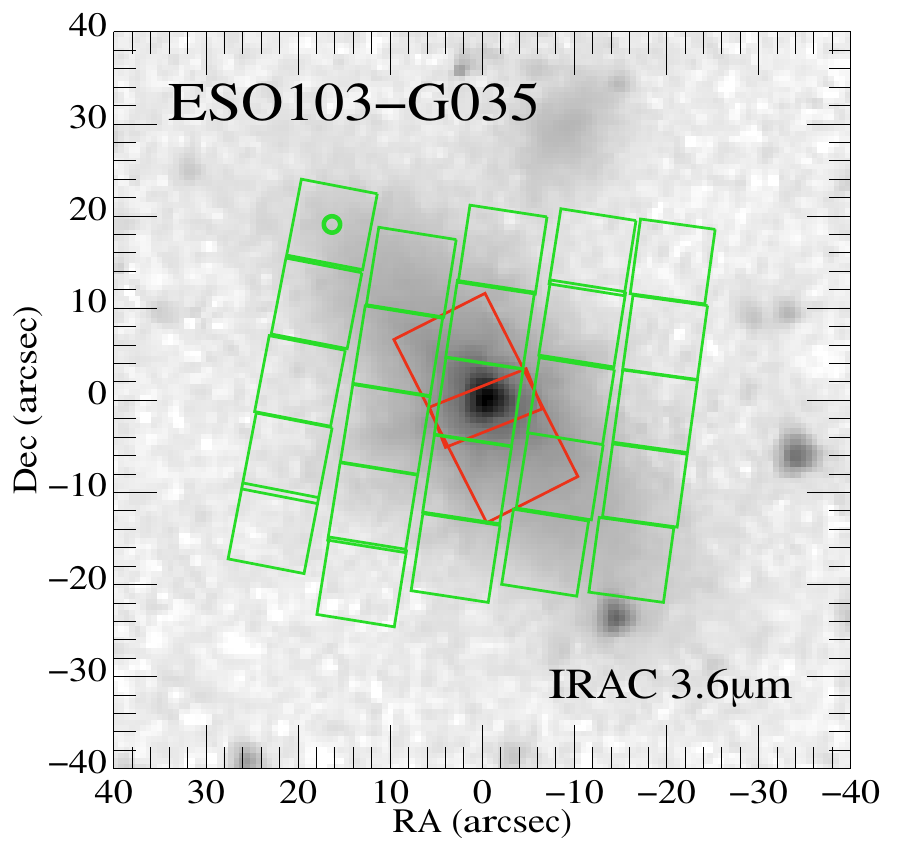}\includegraphics[width=4cm]{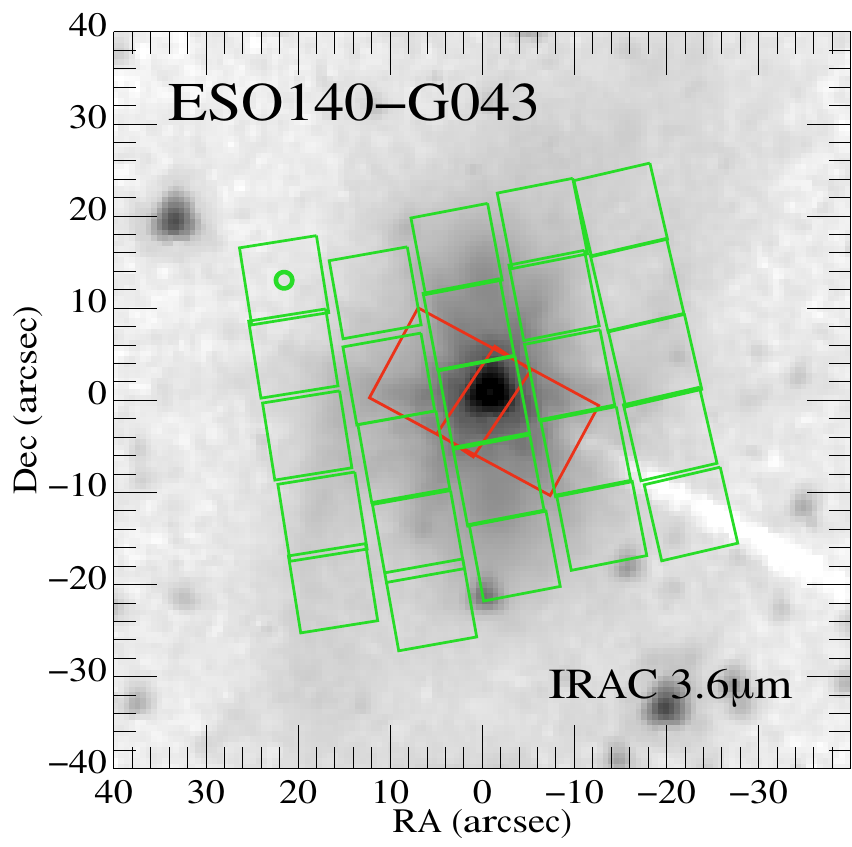}\includegraphics[width=4.3cm]{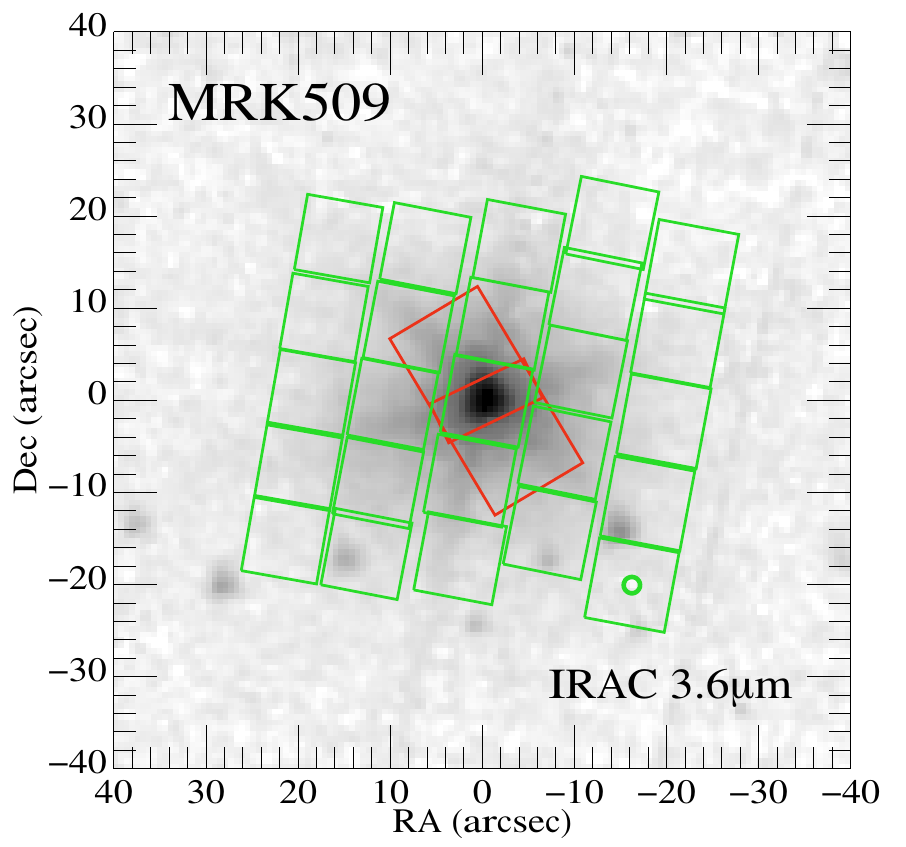}\includegraphics[width=4.2cm]{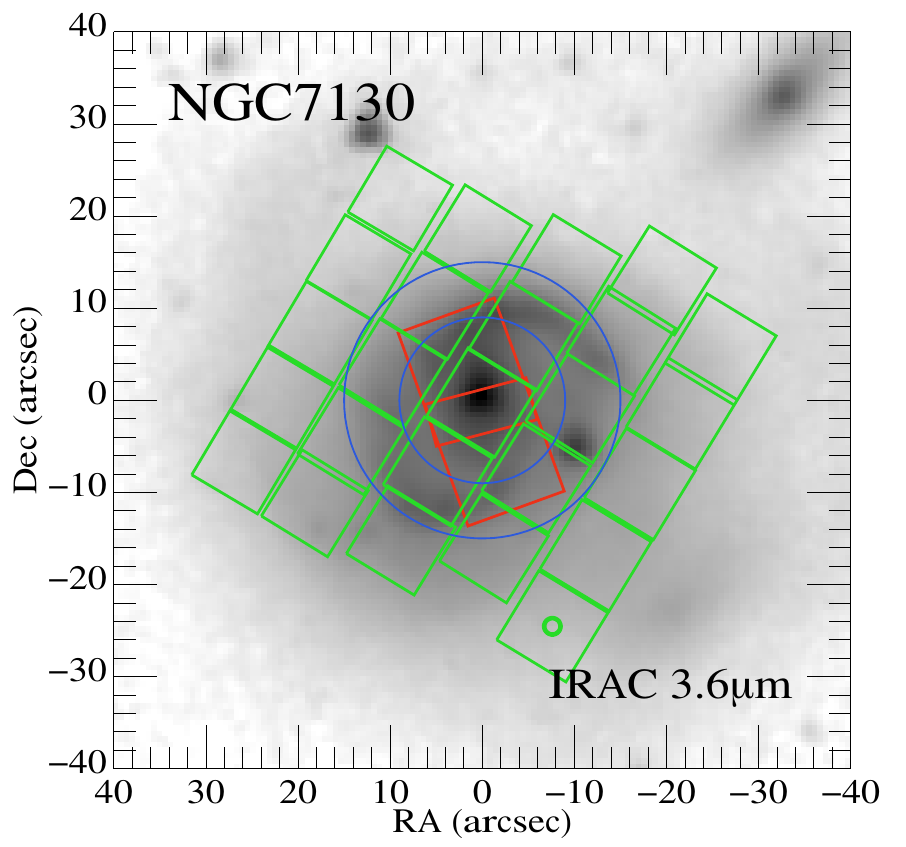}
\includegraphics[width=4.2cm]{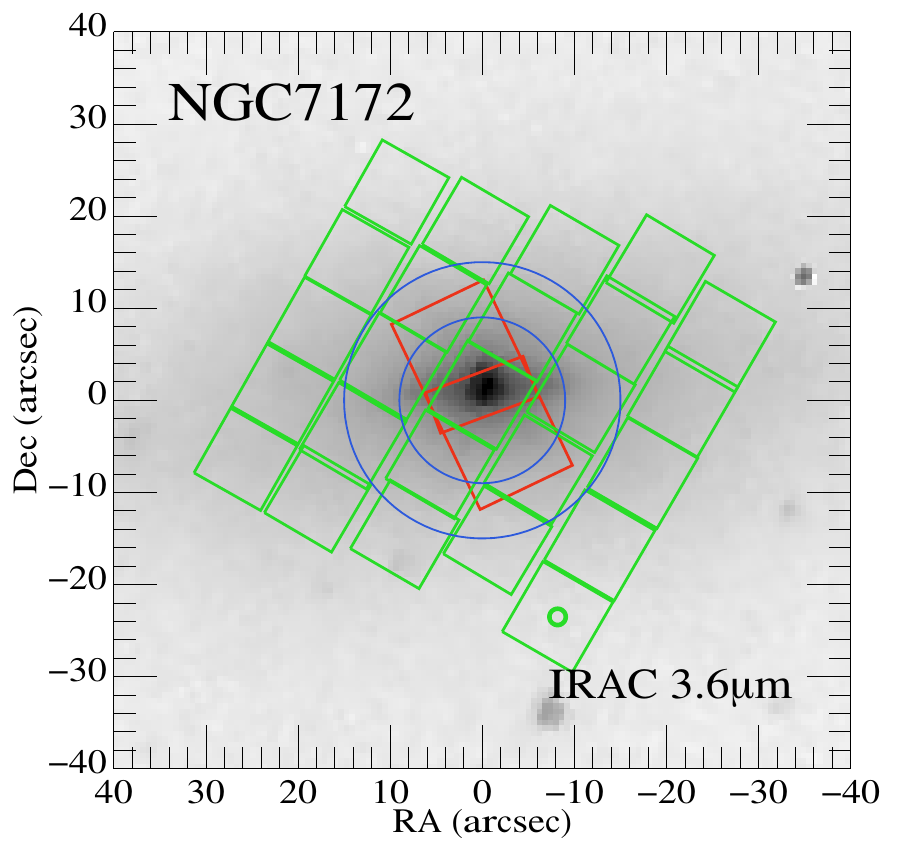}\includegraphics[width=4.2cm]{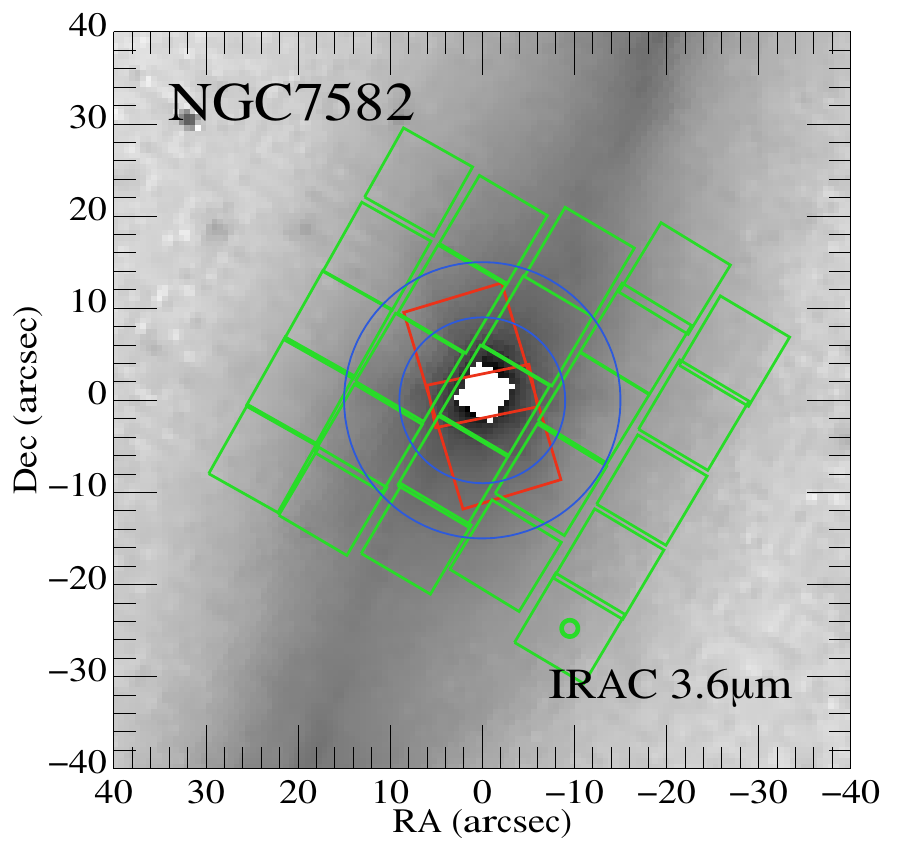}
\caption{ \scriptsize 3.6$\mu$m images from \spi, superposed to the frame of the PACS spectrometer of 47$\arcsec$ $\times$ 47$\arcsec$, 
the short- and long-wavelengths beams of the SPIRE FTS spectrometer, for the galaxies for which these observations are available,  
and the slit of 22.3$\arcsec$ $\times$ 11.1$\arcsec$ of the long-wavelength high-resolution 
mode (LH) of the IRS spectrometer of \spi.
1:  IC4329A; 2: NGC5506; 3: NGC5728; 4: IRAS18216+6418; 5: ESO103-G035; 6: ESO140-G043; 7: MRK509; 8: NGC7130; 9: NGC7172; 10: NGC7582.}
 \label{fig:spitzima_2 }
 \end{figure*}
  
\end{document}